\documentclass{aa}  

\usepackage{txfonts}
\usepackage{graphicx}	% Including figure files
\usepackage{amsmath}	% Advanced maths commands
\usepackage{amssymb}	% Extra maths symbols
\usepackage{pifont}

\usepackage{multicol}        % Multi-column entries in tables
\usepackage{bm}     % Bold maths symbols, including upright Greek
\usepackage{pdflscape}  % Landscape pages
\usepackage{multirow}
\usepackage{textcomp}
\usepackage[T1]{fontenc}
\usepackage{ae,aecompl}
\usepackage{hyperref}
\usepackage{caption}
\usepackage[utf8]{inputenc}
\usepackage{booktabs}
\usepackage{longtable}
\usepackage{threeparttablex}

\usepackage{tablefootnote}

\usepackage{xcolor}

\newcommand{\xmm}{{\em XMM--Newton}}
\newcommand{\BSAX}{{\em Beppo}SAX}

\newcommand{\RXTE}{{\em RXTE}}
\newcommand{\swift}{{\em Swift}}

\newcommand{\INT}{{\em INTEGRAL}}
\newcommand{\nus}{{\em NuSTAR}}

\def\nh {$N_{\rm H}$}
\def\chisq {$\chi^{2}$}

\def\lum {erg\,s$^{-1}$}

\def\cm2{cm$^{-2}$}
\def\arcsec{$^{\prime\prime}$}

\def\deg{$^{\circ}$}

\def\kt{$kT_{\rm BB}$}
\def\r{$R_{\rm BB}$}

\def\ktdisc{$kT_{\rm disc}$}
\def\rdisc{$R_{\rm disc}$}

\def\ktseed{$kT_{\rm seed}$}
\def\kte{$kT_{\rm e}$}

\begin{document} 

\title{The \nus\ view of Ultra-Compact X-ray Binaries}

\titlerunning{\nus\ view of UCXBs}
\authorrunning{}

\author{A.~Borghese\inst{1,2,3}\fnmsep\thanks{ESA Research Fellow} 
          \and
        M.~Armas Padilla\inst{2,3}
            \and
        T.~Mu\~{n}oz-Darias\inst{2,3}
          }

\institute{
European Space Agency (ESA), European Space Astronomy Center (ESAC), Camino Bajo del Castillo s/n, 28692 Villafranca del Castillo, Madrid, Spain \email{alice.borghese@esa.int}
\and
Instituto de Astrofísica de Canarias, E-38205 La Laguna, Tenerife, Spain
\and
Departamento de Astrofísica, Universidad de La Laguna, E-38206 La Laguna, Tenerife, Spain
             }

   \date{Received XXX; accepted XXX}

\abstract{
Ultra-compact X-ray binaries (UCXBs) are a subclass of low-mass X-ray binaries (LMXBs) characterised by tight orbits and hydrogen-poor donor stars. We present a spectral and timing study in the hard X-ray band of 11 of the 20 confirmed UCXBs, based on 37 archival \nus\ observations. Using both X-ray colours and fractional root mean square values, we show that our sample spans the hard, soft, and intermediate X-ray states. Subsequently, we perform an X-ray spectral analysis using, when data allow it, the three-component model -- an approach increasingly adopted for neutron star LMXBs. This work represents the largest LMXB sample analysed to date with this methodology. We focus on the properties of the X-ray continuum and report typical values for each X-ray state. Overall, UCXBs exhibit similar spectral properties to their longer-period counterparts, suggesting no major differences in the innermost regions of X-ray binaries, regardless of disc size or chemical composition. A possible exception is found in the soft-state sample, which shows Comptonisation fractions higher than those typically observed in regular LMXBs, although the statistics remain limited. Finally, we discuss the case of the slow X-ray pulsar 4U\,1626--67, where we report the discovery of a very cold hard state with an electron temperature of $\sim$6~keV -- comparable to those usually observed in soft states of neutron-star LMXBs.
}


\maketitle

\section{Introduction} 

Ultra-compact X-ray binaries (UCXBs) are a group of low-mass X-ray binaries (LMXBs) with shorter orbital periods ($P_{\rm orb}<$60--80\,min) than those of the `classical' LMXBs ($P_{\rm orb}\sim$ hours to tens of days). The tight orbits of these systems imply that the compact object, either a neutron star (NS) or a black hole (BH), is accreting via Roche-lobe overflow from a hydrogen-poor companion \citep
{rappaport82}. UCXBs play a crucial role in different aspects of astrophysics: {\it i)} they are strong, persistent gravitational wave sources for future missions that are sensitive in the low-frequency regime \citep[e.g.,][]{chen2021}; {\it ii)} UCXBs provide important tests of binary evolution theories; {\it iii)} they allow us to investigate the accretion processes under hydrogen-poor conditions \citep[e.g.,][]{nelemans2009}.

At the moment of writing (April 2026), 20 LMXBs are proved to be UCXBs, that is, with confirmed $P_{\rm orb}$ measurements, as reported by \cite{armaspadilla2023}, who built the first comprehensive catalogue of confirmed and candidate UCXBs (UltraCompCAT\footnote{\url{https://research.iac.es/proyecto/compactos/UltraCompCAT/index.php}.}). To pinpoint UCXBs by measuring the $P_{\rm orb}$ can be challenging, thus we often need to rely on indirect methods to identify candidates. In these systems, the accretion disc is small, and thus the disc region responsible for the X-ray to optical reprocessing. Therefore, UCXBs show lower optical-to-X-ray flux ratios than `classical' LMXBs at the same X-ray flux \citep{paradijs1994}. Moreover, persistent LMXBs with low X-ray luminosity ($L_{\rm X,pers}<10^{36}$\,\lum) can hide UCXBs since small discs can be entirely ionised at lower accretion rates \citep{lasota2001,zand2007}. Additional diagnostics are based on spectral data, such as the absence of hydrogen lines in the optical spectra \citep[e.g.,][]{stoop2021} and the presence of X-ray features due to overabundances in the accreted material \citep[e.g.,][]{armaspadilla2019}.  

As for the LMXBs, UCXBs can be persistent sources, always active, or transient, exhibiting outbursts with recurrence times from months to years \citep[see e.g., for a review][]{bahramian23book}. During an outburst, two main accretion states can be identified, labelled hard and soft states. A hard state is observed at the beginning and at the end of an outburst. The X-ray spectrum roughly follows a power-law shape with cutoff energies of $\sim$tens of keV for NS systems and up to $\sim$hundreds of keV for BH sources \citep{burke2017, burkesecond20}. This component arises from a geometrically thick, optically thin inflow, so-called corona, which Comptonises softer X-ray photons coming from the accretion disc and/or boundary layer (for NS systems). Contrastingly, during the soft state, the X-ray emission is dominated by the optically thick, geometrically thin accretion disc that moves inward and becomes hotter, cooling down the corona. In addition to these two states, intermediate states can be found, when the source transitions from the hard to the soft state and vice versa with the soft-to-hard transition occurring at lower luminosities than the hard-to-soft one. 

Several studies have been performed for each of the UCXBs, while global investigations of this class of LMXBs remain very scarce.
For instance, \cite{fiocchi2008} carried out a systematic spectral analysis of a small sample of UCXBs using \INT\, combined with \BSAX\, and \swift\, data whenever possible. They found that these systems spend most of the time in hard states, and all spectra were well described by a two-component model consisting of a disk-blackbody and Comptonised emission. \cite{Koliopanos2021} employed primarily \xmm\ data of 14 confirmed and 2 then candidate UCXBs to constrain the chemical composition of their accretion disc through a homogeneous spectral study focusing on the presence of the iron K$\alpha$ emission line. Recently, \cite{dage2025} presented a global study of the radio counterparts of UCXBs, reporting no strong evidence of a correlation between the radio luminosity and the orbital period. Moreover, they studied the properties of the globular clusters hosting UCXBs with respect to a sample of globular clusters where no known UCXBs reside, noting that the first ones have much higher encounter rates than the second.

In this work, we present for the first time a self-consistent analysis of the X-ray spectral properties and variability of the confirmed UCXBs as a population making use of publicly available\footnote{We considered observations acquired up to 2023 July.} observations of the {\it Nuclear Spectroscopic Telescope Array} \cite[\nus;][]{harrison13}. This sums up to 37 pointings for 11 sources, for a total on-source exposure time of 1.2\,Ms. In Section\,\ref{sec:obs}, we summarise the X-ray data analysis procedure. Then, we describe the details and results of the analysis in Section\,\ref{sec:analysis}. Finally, implications are discussed in Section\,\ref{sec:disc}.

\section{Observations and data reduction} 
\label{sec:obs}

Launched in 2012, \nus\ is the first focusing hard X-ray observatory. It consists of two co-aligned optics focused onto two focal planes, referred to as FPMA and FPMB, observing the sky in the energy range from 3 to 79\,keV. Its effective area peaks at $\sim$900\,cm$^{-2}$ around 10\,keV, adding up the two modules. \nus\ achieves an energy resolution of 400\,eV at 10\,keV and an angular resolution of 18\,\arcsec\ full width at half maximum.

We found 37 \nus\ observations for 11 systems in the {\sc HEASARC} Data Archive. We report a log of the observations included in this work in Appendix\,\ref{sec:log} and detail the properties of the 11 UCXBs in Table\,\ref{tab:properties}, as taken from the catalog UltraCompCAT \citep[][and references therein]{armaspadilla2023}. 

Data reduction was performed using the \nus\ Data Analysis Software ({\sc NuSTARDAS}), which is part of the {\sc HEASoft} package (version 6.32), and the most recent calibration files. We processed the raw data with the tool {\sc nupipeline} to generate cleaned event lists and filtered out passages of the satellite through the South Atlantic Anomaly. We selected the source photons from a circular region with a typical radius of 100--120\arcsec, depending of the source brightness. Noteworthy exceptions are IGR\,J17494$-$3030 and 47\,Tuc\,X$-$9. For the former, we adopted an 80-arcsec radius circle to maximise the source signal-to-noise ratio (it is the UCXB with the lowest net count rate in the 3--15\,keV band,  0.072$\pm$0.001\,counts\,s$^{-1}$). 47\,Tuc\,X$-$9 is located in the core of the globular cluster 47\,Tucanae. To minimise the contamination from other sources of the cluster core, we opted for a circle with a radius of 30\arcsec. In all cases, the background counts were accumulated within a region far from the target and of the same size as that adopted for the source. Swift\,J1756.9--2508 was not detected during the second available observation, carried out when the source returned into quiescence after the 2018 outburst \citep{sanna2018a,li2021}. Stray light contamination, which is unfocused light from bright sources outside the field of view of \nus, is evident and the source region is embedded in it. We derived an upper limit of $\sim$0.34 counts\,s$^{-1}$ on the source count rate in the 3--15\,keV interval, considering only the FPMA event file.

For each observations, we built the source light curve with different timing resolutions to check for the presence of type I X-ray bursts and, if detected, we applied intensity filters to the event lists. 4U\,1916--053 shows periodic dips \citep{walter1982}. Being interested in the persistent emission, we performed a time selection on the data. We created source and background spectra, the corresponding redistribution matrices and ancillary response files using the {\sc nuproducts} pipeline. For 47\,Tuc\,X$-$9, the two available observations were performed only one day apart, thus we merged them to increase the source signal-to-noise ratio. This object is the principal source of X-rays in the cluster above 6\,keV \citep{bahramian2017}. Therefore, we limited our analysis to photons with energies higher than 6\,keV, at odds with all the others sources where we considered energies higher than 3\,keV.

\begin{table*}
\centering
\caption{Main properties of the UCXBs included in this work. The sources are sorted by increasing orbital period. }  
\label{tab:properties}
\footnotesize
\begin{tabular}{@{}lccccc}
\hline
\hline
Source & P/T$^a$ & Accretor$^b$ & $P_{\rm orb}$ & \nh & $d$ \\
       &     &          & (min) & (10$^{22}$\,\cm2) & (kpc) \\
\hline
4U\,1820$-$303 & P & NS & 11.42 & 0.16 & 8.019 \\
4U\,1543$-$624 & P & NS & 18.2 & 0.35 & 9.2 \\
47\,Tuc X$-$9 & P & BH (?) & 28.18 & 0.013 & 4.521 \\
IGR\,J17062$-$6143 & P & NS (AMXP) & 37.9701 & 0.26 & 7.3 \\
4U\,1626$-$67 & P & NS & 41.538 & 0.13 & 8 \\
MAXI\,J0911$-$655 & T & NS (AMXP) & 44.332218 & 0.25 & 10.060 \\
IGR\,J16597$-$3704 & T & NS (AMXP) & 45.97 & 0.82 & 7.242 \\
4U\,1916$-$053 & P & NS & 49.75 & 0.55 & 7.6 \\
4U\,0614+091 & P & NS & 51.3 & 0.3 & 2.59 \\
Swift\,J1756.9$-$2508 & T & NS (AMXP) & 54.7017 & 8.14 & assumed 8 \\
IGR\,J17494$-$3030 & T & NS (AMXP) & 74.9445 & 1.87 & assumed 8 \\
\hline 
\hline
\end{tabular} 
\tablefoot{
The values are taken from the catalog UltraCompCAT \citep{armaspadilla2023}. The reported hydrongen density \nh\ and distance $d$ values are those assumed for the spectral analysis.
\tablefoottext{a}{Type of accretion: persistent (P) or transient (T).}
\tablefoottext{b}{Nature of the accretor: neutron star (NS) or black hole (BH). If the NS is an accreting millisecond X-ray pulsar, AMXP is added. 47\,Tuc X$-$9 has been suggested to host a black hole accretor \citep{bahramian2017}.}}
\end{table*}

\section{Analysis and results}
\label{sec:analysis}
In the following, all uncertainties are quoted at 1$\sigma$ confidence level, unless otherwise specified.

\subsection{Power density spectra and hardness} 
\label{sec:pds}
In order to study the variability of the source, we selected events in the 3--15\,keV energy range, with the exception of 47\,Tuc\,X$-$9 (6--15\,keV). We built the Leahy normalised power density spectra (PDS) using the library {\tt Stingray} \citep{huppenkothen2019} and performing the dead-time correction on each \nus\ PDS by means of the Fourier Amplitude Difference technique \citep{bachetti2018}. For each observation, we averaged the PDS obtained from 50-s long segments with time bins of 5\,ms. For each PDS, we computed the fractional root mean square (RMS) variability in the 0.1--64\,Hz frequency range. Instead of subtracting the Poisson noise contribution, we fitted it with a constant component at high frequencies. For only two sources, 47\,Tuc\,X--9 and IGR\,J17494--3030, the RMS value is not constrained, most likely due to the low statistics. 
Moreover, we derived the hardness as the ratio of the source count rates estimated from the FPMA event file between the 10--16 and 6--10\,keV energy bands. We adopted these energy bands to reduce contamination from the interstellar absorption and the contribution from the soft thermal components. In this way, the hardness can be exploited as an independent tool to trace the Comptonisation fraction. 

We favoured the frequency and energy ranges mentioned above for the RMS and hardness in order to be able to compare our results with those presented by \cite{munozdarias2014}, who performed a systematic analysis of the X-ray colour and fast variability of 50 NS LMXBs monitored by the {\it Rossi X-ray Timing Explorer} (\RXTE). We are aware that \RXTE\ energy band extended down to 2\,keV. We take this information into account when comparing our findings.

\subsection{Spectral analysis}
\label{sec:spec} 
The \nus\ background-subtracted spectra were grouped using the optimal binning scheme of \cite{kaastra2016} by means of the ftool {\sc ftgrouppha}. The spectral analysis was performed using {\sc Xspec} \citep[version 12.13.1;][]{arnaud96} and the \chisq\ statistic. To describe the effects of the photoelectric absorption along the line of sight, we adopted the {\tt tbabs} model with photoionization cross-sections of \citet{verner96} and chemical abundances of \citet{wilms00}. For each observation, we fit simultaneously the spectra for both FPMs, adding a renormalisation constant to account for cross-calibration uncertainties between the two telescopes (we freeze it to 1 for FPMA). In every fit, we fixed the hydrogen column density \nh\ to the value reported in Table\,\ref{tab:properties}. Once the best fit for the adopted model was found, the observed and unabsorbed fluxes were estimated with the convolution model {\tt cflux} in the 0.8--30\,keV energy band.

In the cases of 4U\,1543--624, 4U\,1626--67 (first epoch) and 4U\,1820--303 (all epochs except the first two), the observations provided high-statistical-quality spectra where systematic calibration uncertainties are important. Therefore, we added an energy independent systematic uncertainty of 0.5\% to each spectral channel\footnote{NuSTAR Calibration Update: \url{https://nustarsoc.caltech.edu/NuSTAR_Public/NuSTAROperationSite/CALDB20211020.php}}.

\subsubsection{Spectral models}

For the thermal emission, we tested two models:
\begin{itemize}
    \item the blackbody model {\tt bbodyrad} (BB), parameterised by its temperature \kt\ and normalisation $N_{\rm BB}$, the latter being connected to the blackbody emitting radius through the formula \r=$D_{\rm 10kpc} \sqrt{N_{\rm BB}}$, with $D_{\rm 10kpc}$ the distance of the source in units of 10\,kpc. This model accounts for the NS surface/boundary layer emission;
    \item the multicolor disc model {\tt diskbb} for the accretion disc emission \citep[DISC, e.g.,][]{mitsuda84}, characterised by the temperature at inner disc radius \ktdisc\ and normalisation $N_{\rm disc}$, which can be translated into the inner disc radius \rdisc\ with the relation
    \begin{displaymath}
    R_{\rm disc} = \xi \kappa^2 \left(  \frac{N_{\rm disc}}{\cos{i}}  \right)^{0.5} D_{\rm 10kpc}
    \end{displaymath}
    where $\xi$ is the correction factor for the inner torque-free boundary condition \citep{kubota98}, $\kappa$ is the ratio of colour temperature to the effective temperature \citep{shimura95}, $i$ is the inclination of the system. We assumed $\xi$=0.4 and $\kappa$=1.7. For the inclination, we adopted a value of 43\deg\ for 4U\,1820--303 derived from the modelling of the ultraviolet flux modulation of the system \citep{anderson1997}. The presence of dips and lack of eclipses in the light curve of 4U\,1916--053 constrained the inclination of the system to be $ i \lesssim$ 79\deg\ \citep{smale1988}. Previous works \citep[e.g.,][]{boirin2004} assumed $i$=70\deg, we therefore adopted the same value. The other sources in the sample do not show eclipses/dips, setting an upper limit $ i \lesssim 75^{\circ}$ on the inclination angle \citep[e.g.,][]{degenaar2017, sanna2017, sanna2018a} and we chose a representative value of 70\deg.
    \end{itemize}
    
We applied the {\tt nthcomp} model (NTHCOMP) to take into account the contribution of the Comptonisation corona \citep{zdziarski96, zycki99}. The involved parameters are the power-law photon index $\Gamma$, the electron temperature of the Comptonising medium \kte, the seed photon temperature \ktseed, and the normalisation $N_{\rm comp}$. The seed photons might come from the disc or the NS (boundary layer and/or surface). Therefore, when combining the {\tt nthcomp} model with the BB and/or DISC components, we linked \ktseed\ to \kt\ or \ktdisc\ and changed the seed photon shape parameter {\it inp\_type} accordingly, 0 or 1 for BB or DISC shape, respectively.  

When the Fe K$\alpha$ emission line was observed, a simple Gaussian component ({\tt gauss} in {\sc Xspec}) was included in the model and its width fixed to the values reported in the literature (4U\,1543--624 being the exception with width free). We detected the Fe feature at all epochs for 4U\,1543--624, 4U\,0614+091 and IGR\,J17062--6143, while for 4U\,1820--303 the spectral line was present in ten over eleven observations. The second source with a variable presence of the iron line is 4U\,1626--67. The feature is broad with a central energy of $\sim$6.8\,keV. \cite{dai2017} suggested that the feature is a possible blending of two lines, because the central energy lies in between the energy of the Ly$\alpha$ transitions of the He-like (6.70\,keV) and the H-like (6.97\,keV) iron ions. This system is unique, as it is the only UCXB harbouring a slow-spinning (7.7\,s) NS with a magnetic field of $\sim$(3--4)$\times$10$^{12}$\,G, estimated from a resonant cyclotron scattering feature at $\sim$35--37\,keV \citep{rappaport77, orlandini1998}. We used the {\tt gabs} component to model the resonant cyclotron scattering feature. We found no evidence of the iron emission line for 4U\,1916--053, however the spectrum showed two narrow absorption features at $\sim$8.14\,keV and $\sim$6.77\,keV, which might be the Fe {\sc xxvi} K$\beta$ and Fe {\sc xxv} K$\alpha$ absorption lines. We modeled them using the additive spectral component {\tt gauss} with the width frozen at 20\,eV \citep{gambino2019}. In the spectra of the remaining sources (MAXI\,J0911--655, Swift\,J1756.9--2508, IGR\,J16597--3704, IGR\,J17494--3030 and 47\,Tuc\,X--9), we did not detect any feature.

\subsubsection{Spectral fits} 
\label{sec:SpecFit}

At this stage of the study, we model the spectra regardless of the state of the source. To select the acceptable fits from a statistically point of view, we considered the null hypothesis probability (nhp), which represents the probability that the deviations between the data and the model are due to chance alone. In general, a model can be rejected when the nhp is smaller than 0.05 \citep[see e.g.,][]{armaspadilla2017, armaspadilla2019}. 

For each spectrum, we started by fitting it with the simplest model, meaning one-component model such as a BB, DISC or NTHCOMP. 
The number of required components was evaluated by means of an $F$-test: an additional component was included if it yielded an improvement in the fit of at least 3$\sigma$. When a second component was required, we combined the Comptonised component with a thermal one (i.e., DISC+NTHCOMP or BB+NTHCOMP). While, in the case of a three-component model, we employed the model DISC+BB+NTHCOMP with two possible solutions for the seed photon temperature: \ktseed=\ktdisc\ or \ktseed=\kt. Therefore, hereafter, we will refer to these as the DISC and BB approaches, respectively. Note that for every spectrum, we tested both approaches. 
In the cases where the best fit is provided by a three-component model, uncertainties on the parameters have been computed in two steps, that is, by freezing the thermal components when estimating the error associated with the parameters of the NTHCOMP model and vice versa, given the degeneracy of the model.

Following the above-mentioned criteria (i.e., nhp and $F$-test), we found that for seven systems, the three-component model was the best-fit model at all epochs. For 4U\,1543--624 (one spectrum), the $F$-test showed that the three-component model was significantly better than a two-component one, however the nhp was always lower than the decided threshold (nhp$\sim$0.02--0.03). We retrieved two observations for MAXI\,J0911--655 and Swift\,J1756.9--2508. For both sources, the two-component model suitably fits the spectrum of one of the two epochs, with the other being well modelled by a three-component model. 
In all these cases, both DISC and BB solutions provided statistically equivalent fits, with the residuals not showing any clear or systematic structure. Special instances are IGR\,J17494--3030 (one spectrum) and 47\,Tuc\,X--9 (one spectrum), for which the photon counting statistics was not high enough to allow us to use complex models. For the former, a power law was able to reproduce the data, while for the latter the NTHCOMP model assuming a BB shape for the seed photons resulted in the best-fitting model.

Our spectral analysis revealed the well-known degeneracy issue in the X-ray spectral modelling of NS LMXBs: we obtained statistically acceptable fit by using any combination of the Comptonised component and one or two thermal models regardless of the state of the source. The various models convey different physical implications, like whether the seed photons come from the accretion disc or NS/boundary layer. To deal with the degeneracy problem, we followed the prescription by \cite{lin2007} \citep[see also][]{armaspadilla2017}, who applied multiple physical evaluation criteria to choose a spectral model that may be more suitable than the others. We adopted the following sanity checks:

\begin{enumerate}
\item the Comptonisation fraction is consistent with the PDS, meaning that it should broadly agree with the observed level of aperiodic variability;
\item the NS surface/boundary layer has a higher temperature than the disc \citep[\kt$>$\ktdisc;][]{mitsuda89, popham01}, and therefore the photon seed temperature should not be higher than the NS surface temperature, \ktseed$\leq$\kt; 
\item the inner disc radius \rdisc\ has values higher than/comparable to the size of the NS.
\end{enumerate}

We detailed the three criteria according to our priority list. It is important to bear in mind that the low-energy limit of \nus\ is 3\,keV. 
The lack of coverage at energies $<3$\,keV directly affects the last two sanity checks. For this reason, we did not consider Condition 3 (i.e., \rdisc$\leq$ $R_{\rm NS}$) decisive for the choice of the final model\footnote{Condition 3 is met for all epochs of 4U\,1820$-$303, 4U\,1543$-$624, MAXI\,J0911$-$655 and 4U\,1916$-$053; only for the first epoch of 4U\,J1626$-$67 and Swift\,J1756.9$-$2508; only for the third epoch of IGR\,J17062$-$6143; for all epochs apart from the fourth one of 4U\,0614$+$091. Note that {\it (i)} we did not consider the distance error (not always known) when deriving the uncertainty of $R_{\rm disc}$ and {\it (ii)} $R_{\rm NS}$=10\,km was assumed.} (see Table\,\ref{tab:final_model}). Even when soft X-ray coverage is available, we need to be careful when dealing with radius measurements from the {\tt diskbb} model since it is a Newtonian model. Condition 2 is more relevant for our study since either \kt\ or \ktdisc\ is linked to the seed photon temperature \ktseed, thus affecting the fit through the Comptonisation component. The parameters of NTHCOMP are well constrained, because this component accounts for the hard emission of the spectra ($\gtrsim$ 4--5\,keV). This criterium helped us rule out a few fits. For example, the three-component model with the DISC approach was discarded for the spectrum corresponding to the second epoch of MAXI\,J0911--655 due to the fact that it gave \kt$<$\ktdisc. Finally, Condition 1 is discussed in Section\,\ref{sec:choice_mod} since it helps to reconcile the spectral and timing properties of our sample and affects the nature of the Comptonised emission. 

\subsection{State classification} 
\label{sec:state}
 
\cite{munozdarias2014} showed that the evolution of the fast variability can be adopted to map the different accretion regimes also for NS LMXBs. Like BH X-ray binaries, NS LMXBs at intermediate luminosities, the so-called atolls, display hysteresis loops in the RMS-intensity diagram \citep[RID;][]{munozdarias2011}, where different regions correspond to different states. Hard states are found at RMS $\geq$20\%, while soft states are characterised by less variability, RMS $\leq$5--7\%. 

\begin{figure}
    \centering
    \includegraphics[width=1.\columnwidth]{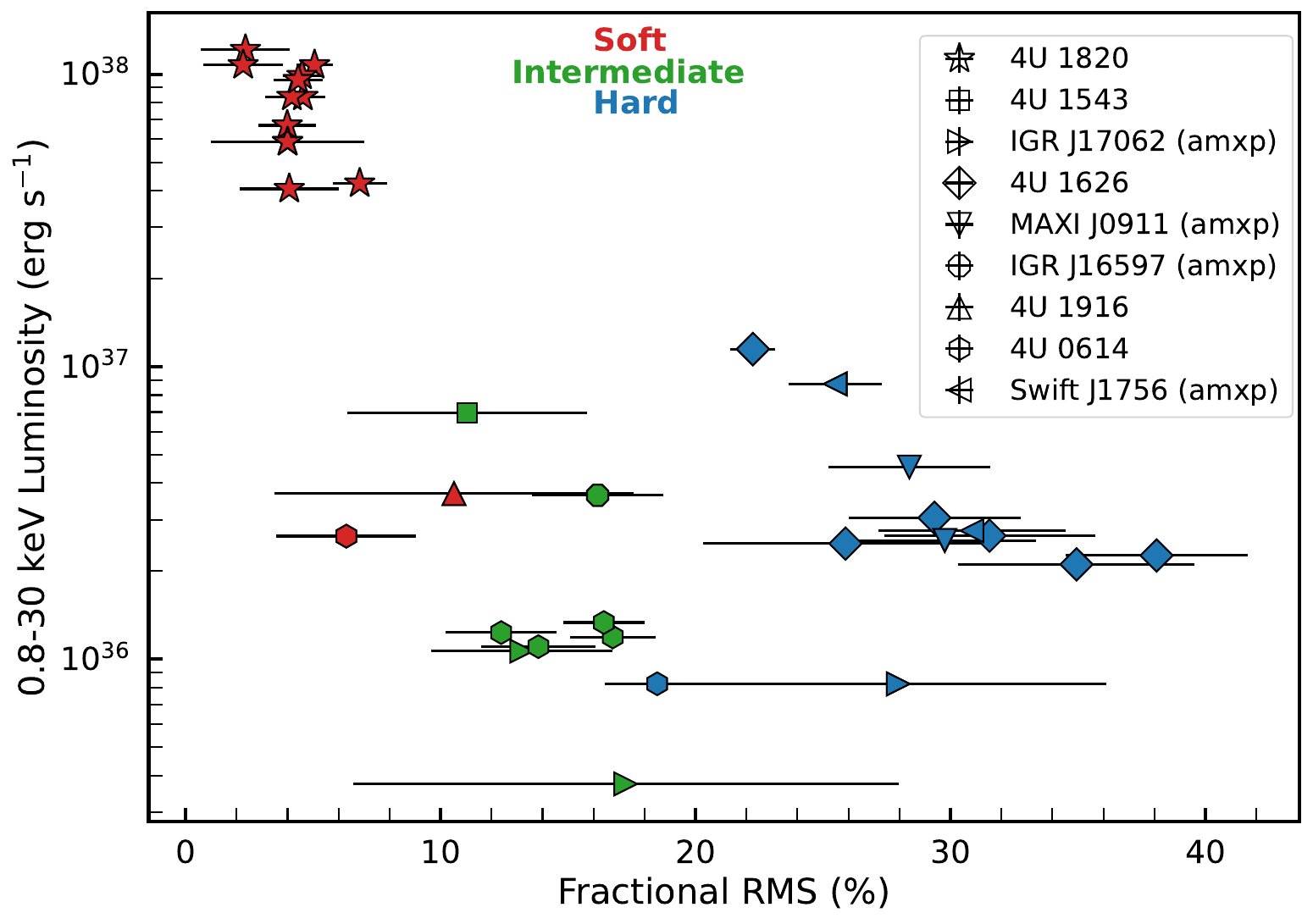}
    \caption{RMS-intensity diagram: 0.8--30\,keV luminosity versus fractional RMS for the systems studied in this work. Luminosity error bars are smaller than the marker size. Red, green, blue indicate the state of the source during each observation: soft, intermediate and hard states, respectively. Same symbol/colour code is used for all the figures.}
    \label{fig:RID}
\end{figure}

We built the RID for our sample by computing the RMS, as explained in Section\,\ref{sec:pds}, and the luminosity in the 0.8--30\,keV energy range from the unbasorbed flux obtained from the best-fitting model and the distance reported in Table\,\ref{tab:properties}. The final plot is exhibited in Figure\,\ref{fig:RID}, which expectedly recalls the results found for atolls by \citet[][a few UCXBs were included in their work]{munozdarias2014}. To define the source state we roughly followed the prescription for the RMS mentioned above: given that \nus\ energy band is slightly harder than that of \RXTE, the boundary between the soft and intermediate states might be at higher RMS. Therefore, we labelled as soft states the observations with RMS $\leq$10\% and as hard states with RMS $\geq$20\%; intermediate states are found in between these limits. Spectral information was used to corroborate our choice for the source states (see Section\,\ref{sec:choice_mod}). An exception to our definition is the fourth epoch of 4U\,0614+091 with RMS=18.5$\pm$2.0\% that we defined as hard state. \cite{munozdarias2014} reported a peculiar behaviour for this system with transitions occurring at similar flux levels in the RID, and found that there are often clumps of observations with RMS in the 15--20\% range at lower luminosities that are hard states, as in our case. Moreover, the spectral results corroborate this classification, being the Comptonisation fraction of the spectrum corresponding to this epoch higher than those of the spectra in intermediate states.

As shown in Fig.\,\ref{fig:RID}, there are a few cases where the RMS value is consistent within 1$\sigma$ confidence level with more than one state. For instance, for the only observation retrieved for 4U\,1916$-$053 the RMS is 11$\pm$7\%, which is crossing the soft-intermediate boundary. We found evidence of two narrow absorption lines in the spectrum that were identified with absorption from Fe {\sc xxv} and Fe {\sc xxvi}. These features are often detected in dipping and, hence, high-inclination sources -- such as 4U\,1916$-$053 -- and are interpreted as the signature of high-ionization absorbers, revealing the presence of disc winds or atmosphere \citep[see for reviews e.g.,][]{diaztrigo2016, munozdarias2026}. These absorption lines are observed only during soft-state epochs, as proved by a thorough spectral study of the NS LMXB EXO\,0748--676 \citep{ponti2014}.
Consequently, we classified the observation of 4U\,1916$-$053 as soft state. 
Moreover, for 4U\,1543--623 (RMS=11$\pm$5\%) and the second epoch of IGR\,J17062--6143 (RMS=17$\pm$11\%) the spectral properties disfavour a soft-state classification, while they point towards an intermediate-state one because of the high values of the electron temperature.

Finally, for 47\,Tuc\,X$-$9 and IGR\,J17494$-$3030, the RMS value is not constrained. For the former, we caught a hint of variability in the 0.1--0.5\,Hz frequency range and calculated a hardness of $\sim$0.9, which favours a hard-state classification. The results of the spectral analysis support this claim, with a photon index $\Gamma$=1.6$^{+0.4}_{-0.2}$ and  luminosity level of $\sim$10$^{34}$\,\lum (0.8--30\,keV). For the latter, we relied on the spectral information. The best-fitting power-law with $\Gamma$=1.92$\pm$0.02 and an inferred 0.8--30\,keV luminosity of $\sim$10$^{35}$\,\lum, together with a hardness of $\sim$0.55, are consistent with a hard state.

\section{Discussion}
\label{sec:disc}

We carried out a systematic study of the physical properties of the confirmed UCXBs in the hard X-rays making use of archival \nus\ data. We analysed 37 archival observations of 11 systems in a consistent way, adopting the same assumptions and criteria.
We determined the state of the sources at a given epoch following prescriptions described in previous works making use of RMS, hardness ratio and spectral signatures such as narrow absorption features \cite[see e.g.,][]{munozdarias2014, ponti2014}. Appendix\,\ref{sec:nus_spec} shows the \nus\ spectra with the best-fitting models.

In the following, we discuss the choice of the spectral models that can reconcile the spectral behaviour with the different states, the evolution of the spectral parameters and the peculiar hard state of the system 4U\,1626--67.

\begin{table*}[h]
\centering
\caption{Summary of the chosen spectral models for the X-ray continuum.}
\label{tab:final_model}
\footnotesize
\begin{tabular}{@{}lccc}
\hline
\hline
Source & Epochs & State & Model \\
\hline
4U\,1820--303 & 11 & soft & DISC+BB+NTHCOMP (\ktseed=\kt) \\
\hline
4U\,1543--624 & 1 & intermediate & DISC+BB+NTHCOMP (\ktseed=\ktdisc) \\
\hline
47\,Tuc\,X--9 & 1 & hard & NTHCOMP ({\it inp\_type}=0) \\
\hline
\multirow{2}{*}{IGR\,J17062--6143} & 1$^{st}$ & hard & DISC+BB+NTHCOMP (\ktseed=\ktdisc) \\
                             & 2$^{nd}$ and 3$^{rd}$ & intermediate & DISC+BB+NTHCOMP (\ktseed=\ktdisc)  \\
\hline
4U\,J1626--67 & 6 & hard & DISC+BB+NTHCOMP (\ktseed=\ktdisc) \\
\hline
\multirow{2}{*}{MAXI\,J0911--655} & 1$^{st}$ & hard & BB+NTHCOMP (\ktseed=\kt) \\
                                  & 2$^{nd}$ & hard & DISC+BB+NTHCOMP (\ktseed=\kt) \\
\hline                             
IGR\,J16597--3704 & 1 & intermediate & DISC+BB+NTHCOMP (\ktseed=\ktdisc) \\
\hline                              
4U\,1916--053 & 1 & soft & DISC+BB+NTHCOMP (\ktseed=\kt) \\
\hline                             
\multirow{4}{*}{4U\,0614+091} & 1$^{st}$ & soft & DISC+BB+NTHCOMP (\ktseed=\ktdisc) \\
                          & 2$^{nd}$ and 3$^{rd}$ & intermediate & DISC+BB+NTHCOMP (\ktseed=\ktdisc) \\
                          & 4$^{th}$ &  hard & DISC+BB+NTHCOMP (\ktseed=\ktdisc) \\
                          & 5$^{th}$ and 6$^{th}$ & intermediate & DISC+BB+NTHCOMP (\ktseed=\ktdisc) \\
\hline                          
\multirow{2}{*}{Swift\,J1756.9--2508} & 1$^{st}$ & hard & DISC+BB+NTHCOMP (\ktseed=\ktdisc) \\
                                      & 2$^{nd}$ & hard & DISC+NTHCOMP (\ktseed=\ktdisc)  \\
\hline
IGR\,J17494--3030 & 1 & hard & POWER LAW \\
\hline 
\hline
\end{tabular} 
\end{table*}

\begin{figure*}[th!]
    \centering
    \includegraphics[width=1.\textwidth]{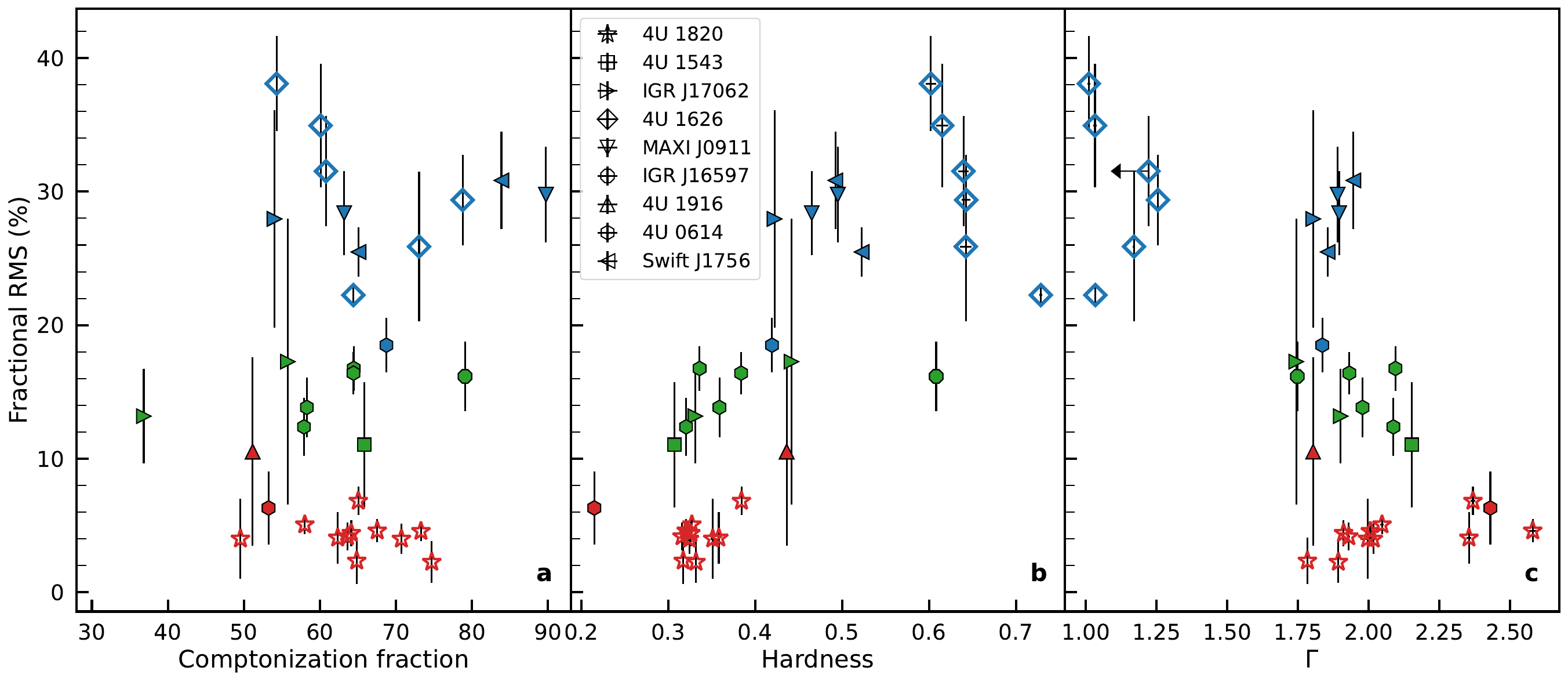}
    \caption{Fractional RMS versus fraction of Comptonised luminosity ({\tt nthcomp}) estimated in the 0.8--30\,keV interval ({\it panel a}), hardness derived as the ratio between the \nus/FPMA count rates in the 10--16\,keV and 6--10\,keV bands ({\it panel b}), and power-law index $\Gamma$ ({\it panel c}).}
    \label{fig:RMS_hardness}
\end{figure*}

\subsection{Nature of the Comptonised emission}
\label{sec:choice_mod}

We found that for most spectra the best-fitting model consists of three components (i.e., {\tt bbodyrad}, {\tt diskbb} and {\tt nthcomp} in {\sc Xspec}). Both the DISC (\ktdisc=\ktseed) and BB (\kt=\ktseed) approaches work, meaning that the Comptonisation seed photons may originate from the NS/boundary layer or from the accretion disc regardless of the source state. Model degeneracy is a long-standing problem that we tackled by applying performance-based criteria, listed in Section\,\ref{sec:SpecFit}, following the recipe designed by \cite{lin2007} \citep[see also][]{armaspadilla2017}. Table\,\ref{tab:final_model} lists the final models that meet these requirements. Conditions 2 and 3 have already been discussed.
Condition 1 can guide us to choose a preferable model considering the Comptonisation fraction and the RMS value (i.e., the state classification).
In general, the Comptonisation fraction traces the X-ray variability, becoming higher in the hard states when accretion drops and the Comptonisation emission becomes dominant with respect to the thermal emission. We estimated the fraction of the Comptonised luminosity in the 0.8--30\,keV energy range for both possible seed photon solutions. When comparing the two approaches, we noticed that the BB solution yielded lower Comptonisation fractions for most spectra in hard and intermediate states than the results from the DISC solution models. In the same way, the soft states remained weakly Comptonised if we adopted the BB solution with respect to the DISC one (with the exception of 4U\,1820--303). Therefore, we used the BB approach for soft states, and the DISC solution for intermediate and hard states, fulfilling Condition 1. The Comptonisation fraction estimated following this prescription as a function of the fractional RMS is shown in Figure\,\ref{fig:RMS_hardness} ({\it panel a}). Within this picture, we assume that the NS surface and/or boundary layer is the dominant supplier of seed photons for the Comptonisation process in the soft states, while the seed photons come from the disc in the other cases. Our findings are in line with the results of previous works on classical NS LMXBs. For example, \citealt{armaspadilla2017} and \citealt{sharma2018} found that for NS LMXBs 4U\,1608--52 and MXB\,1658--289, respectively, in the soft state, the seed photons are provided by the NS surface/boundary layer, while in the hard state either the disc or NS surface is equally favoured. As shown in Table\,\ref{tab:final_model}, there are two exceptions to our modelling: MAXI\,J0911--655 and 4U\,0614+091. The former was observed to be in hard state at both epochs, however only the BB approach yielded results that respect the above-mentioned conditions (see Sec.\,\ref{sec:SpecFit}). We classified the first epoch of the latter as soft state and chose the DISC approach, because the BB best-fitting model gave suspicious (i.e., likely not physically meaningful) parameters for such accretion state, like a power-law photon index $\Gamma$ of $\sim$4. It is important to bear in mind that our models provide a simplified picture being only able to support one component as the dominant source of seed photons for Comptonisation, although they are most likely supplied by both the NS and accretion disc.

In Figure\,\ref{fig:RMS_hardness}, we also show the hardness-RMS diagram \citep[HRD, {\it panel b};][]{belloni2005}. The HRD found in this work resembles that derived by \citet[][Figure\,4]{munozdarias2014}, who found a positive correlation between RMS and hardness until RMS reaches values below 10--5\% when hardness increases as RMS remains constant. This trend marks a hook-shaped track in the HRD, which is a distinctive trait of NS systems. Normal atoll sources (i.e., NS LMXBs accreting below $\sim$30\% of the Eddington luminosity) are observed to follow this peculiar hook-like path in the HRD, while the bright atolls (i.e., persistently bright systems that remain almost permanently in the bright soft states; \citealt{hasinger1989}) show a flat rms-hardness relation, which is similar to what we found for 4U\,1820--303 (empty red stars in Fig.\,\ref{fig:RMS_hardness}). This source is classified as an atoll and it is persistently accreting at a relatively high luminosity, displaying mostly soft states with rare transitions to the hard states and, thus, behaving similar to the bright atolls. As shown in Fig.\,\ref{fig:RID}, 4U\,1820--303 is the most luminous in our sample with luminosity $L_{\rm X}\sim4\times10^{37}-10^{38}$\,\lum\ (0.8--30\,keV), and we always found it in soft states with RMS between 2\% and 8\% and hardness in the 0.3--0.4 range. The distinctive hook shape is recovered by plotting the RMS versus the photon index $\Gamma$ (Fig.\,\ref{fig:RMS_hardness}, {\it panel c}), highlighting an anti-correlation between these two quantities. Generally, the soft states exhibit steeper spectra than the hard states, with 4U\,1820--303 showing spectra as hard as those in the intermediate and hard states ($\Gamma \sim$ 1.8) even if the source is always observed in soft state. This hook-like trend might be caused by the presence of the NS surface/boundary layer contributing significantly and making the spectrum harder, since it has not been seen in BH systems \citep[e.g.,][]{belloni2011}.

Although our present dataset does not allow us to draw firm conclusions, it is worth noting a potential difference between UCXBs and longer‑period NS LMXBs. UCXBs appear not to reach the very low RMS values typically observed in classical LMXBs when they are in soft states. In \cite{munozdarias2014}, 4U\,1820--303 and 4U\,0614+091 rarely fall below the 5\% RMS level, suggesting comparatively harder soft states in UCXBs. In contrast, classical NS LMXB systems exhibit softer soft states, in terms of RMS variability and possibly Comptonisation fraction. For instance, \cite{armaspadilla2017} reported soft-state observations of the LMXB 4U\,1608--52 with RMS $\sim$2\% and Comptonisation fraction $\sim$20\%, while in our sample, the soft states remain highly Compotonised ($\ge$50\%). These differences merit further investigation with a larger set of soft‑state data.

\begin{figure*}[th!]
    \centering
    \includegraphics[width=1.\textwidth]{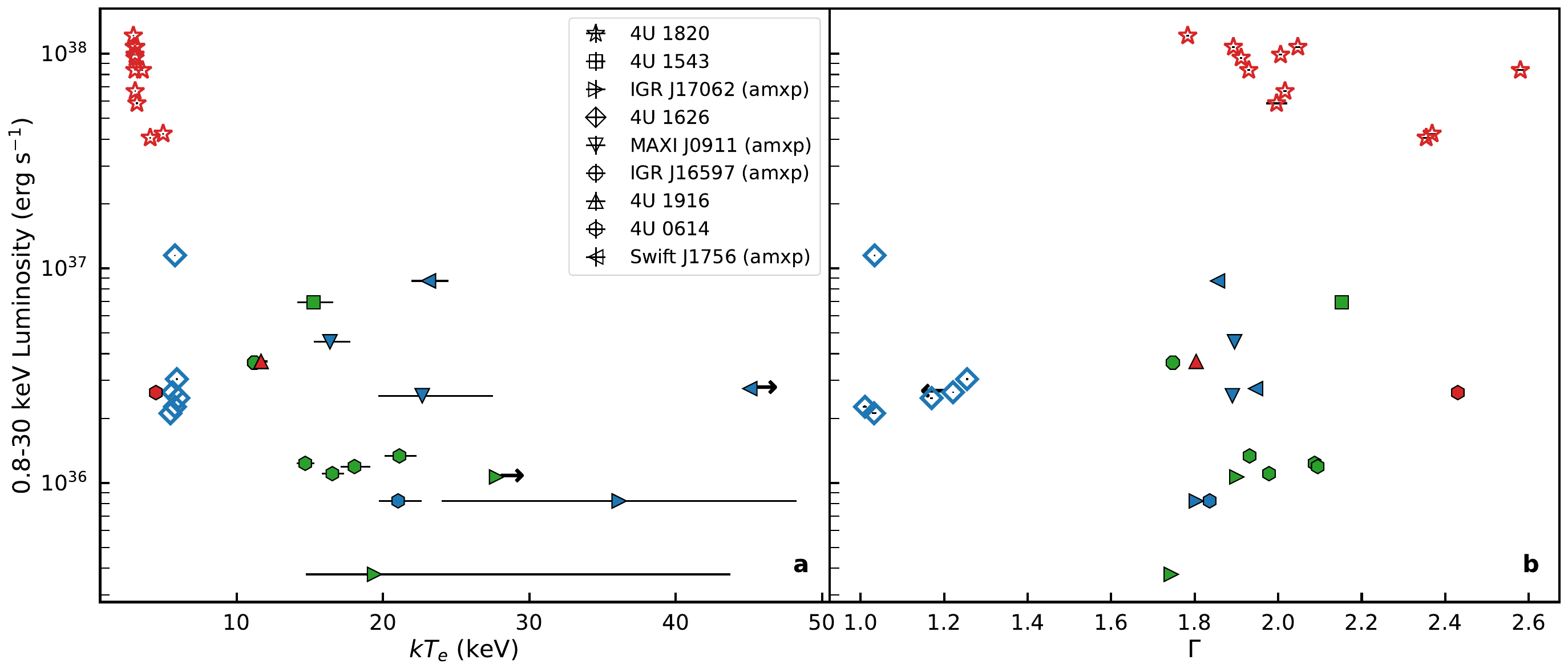}
    \includegraphics[width=1.\textwidth]{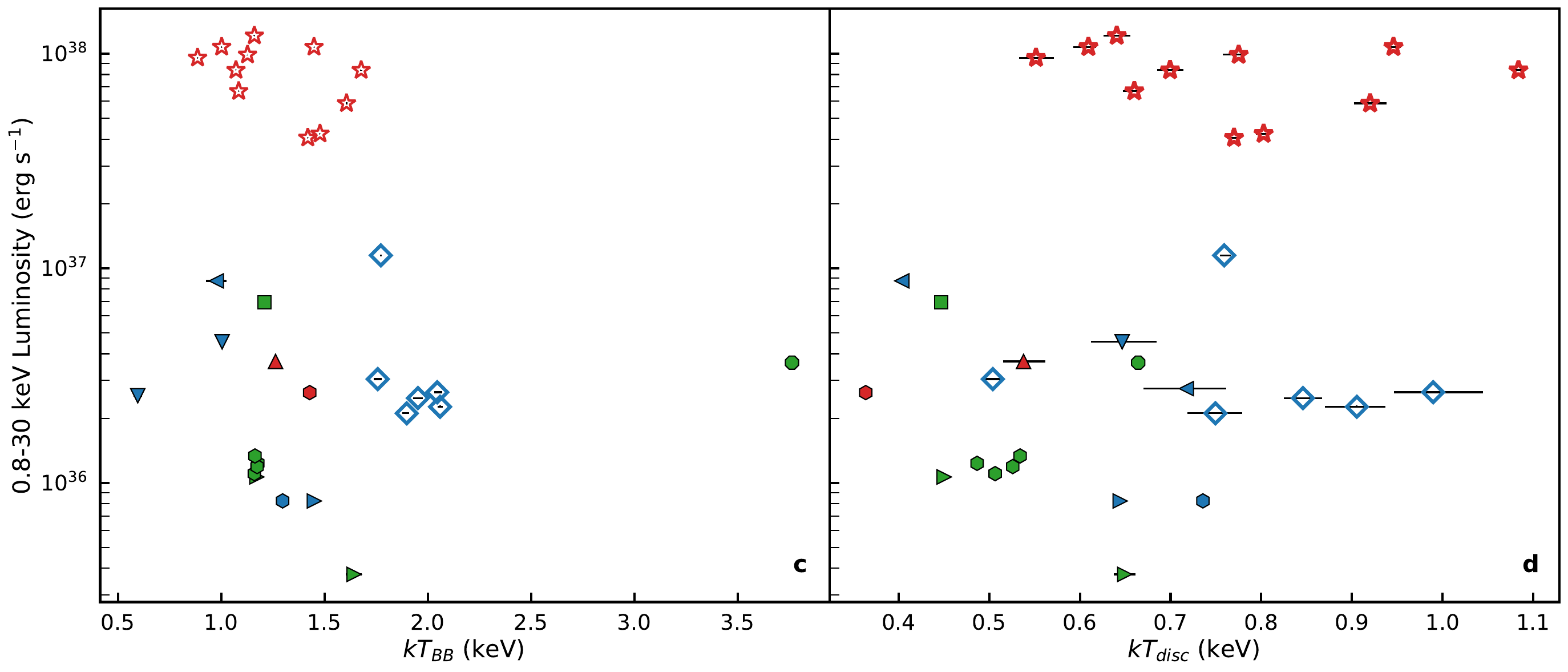}
    \caption{0.8--30\,keV luminosity versus electron temperature of Comptonising medium \kte\ ({\it panel a}), power-law index $\Gamma$ ({\it panel b}), blackbody temperature \kt\ (=\ktseed\ for soft-state spectra; {\it panel c}), and disc temperature \ktdisc\ (=\ktseed\ for hard- and intermediate-state spectra; {\it panel d}).}
    \label{fig:spec_para}
\end{figure*}

\begin{table}[h]
\centering
\caption{Weighted averages of the spectral parameters according to the spectral states.}
\label{tab:spec_res_ave}
\begin{tabular}{@{}lccc}
\hline
\hline
 & Soft & Inter & Hard$^\dagger$ \\
\hline
$kT_{\rm e}$\,(keV) & 3.060$\pm$0.003 & 12.52$\pm$0.19 & 20.24$_{-0.67}^{+0.82}$ \\
$\Gamma$ & 1.969$\pm$0.001 & 1.938$\pm$0.002 & 1.851$\pm$0.002 \\
\kt\,(keV) & 1.128$\pm$0.001$^a$ & 1.225$\pm$0.003 & 1.005$\pm$0.002 \\
\ktdisc\,(keV) & 0.604$\pm$0.002 & 0.512$\pm$0.001$^b$ & 0.511$\pm$0.002$^b$ \\ 
\hline 
\hline
\end{tabular} 
\tablefoot{
\tablefoottext{$\dagger$}{ 4U\,1626--67 was not included to derive the weighted average.}
\tablefoottext{a} {\kt = \ktseed\ for soft-state observations.}
\tablefoottext{b} {\ktdisc = \ktseed\ for intermediate- and hard-state observations.}}
\end{table}

\subsection{X-ray continuum}
\label{sec:para}

Figure\,\ref{fig:spec_para} shows the 0.8--30\,keV luminosity as a function of the spectral parameters $kT_{\rm e}$ ({\it panel a}), $\Gamma$ ({\it panel b}), \kt\ ({\it panel c}), and \ktdisc\ ({\it panel d}), while Table\,\ref{tab:spec_res_ave} reports the weighted averaged values of the spectral parameters according to each spectral state.

Soft states have $kT_{\rm e}$ below $\sim$5\,keV, excluding that of 4U\,1916--053 ($kT_{\rm e}\sim10$\,keV). For hard and intermediate states, $kT_{\rm e}$ covers a larger range, extending from $\sim$10\,keV to $\gtrsim$40\,keV (with the exception of 4U\,1626--67, discussed below). Hard states of LMXBs have been extensively studied, revealing a dichotomy between Comptonisation properties of accreting BHs and NSs \citep[see e.g.,][]{burke2017, burkesecond20}. The latter have generally softer hard state spectra than BHs because of the presence of an extra source of seed photons, that is, the NS surface. \cite{burkesecond20} found that the $kT_{\rm e}$ distribution of NS LMXBs peaks at $\sim$50\,keV with its 80\% quantile being 16--93\,keV. We note that these authors employed the model {\tt compPS} \citep{poutanen1996} to describe the Comptonised emission and restricted $kT_{\rm e}$ to values above 10\,keV, the minimum value for which the numerical method used by the model can be expected to produce reasonable results. For the hard states of the UCXBs in our sample, the $kT_{\rm e}$ values follow the expected spread, which was previously found for classical NS LMXBs, but with a lower average value of $\sim$20\,keV, most likely due to the fact that we did not constrain the lower limit of this spectral parameter. Also the estimated values for $\Gamma$, \kt\ (=\ktseed\ for soft states) and \ktdisc\ (=\ktseed\ for hard and intermediate states) fall within the typical range of such parameters observed in other NS LMXBs at different accretion states \citep[see Table\,3; e.g.,][]{armaspadilla2017, sharma2018, banerjee2024, ludlam2020}. For completeness, we report the average values for $R_{\rm BB}$: 1.066$\pm$0.005\,km, 0.289$\pm$0.002\,km and 0.688$\pm$0.003\,km for soft, intermediate and hard states, respectively. Moreover, we also explored possible (anti-)correlations between the spectral parameters and the orbital period, which is the property that sets UCXBs apart from classical NS LMXBs, without any success. 

We can therefore conclude that the properties of the X-ray continuum in UCXBs are similar to those of their longer-period siblings. This implies that the X-ray spectral behaviour of NS LMXBs with hydrogen-rich, large accretion discs is indistinguishable from that of systems with smaller discs and a different chemical composition.

\subsection{4U\,1626--67: a very cold corona} 
We classified all the epochs of 4U\,1626--67 as hard states given the high values of the RMS (see Fig.\,\ref{fig:RID}). Following our prescription, we chose the three-component model with the DISC approach to describe the spectra. From Fig.\,\ref{fig:spec_para} {\it panel a}, the first striking thing is that the electron temperature values (empty blue diamonds) for this source are as low as those observed in soft states, $kT_{\rm e}$ $\sim$6\,keV. 
4U\,1626--67 is an intriguing source, even within the already exceptional UCXB family, as it is the only UCXB powered by a highly magnetised, slow X-ray pulsar with spin period $P_{\rm spin}$ $\sim$7.7\,s. The strength of the magnetic field is derived to be $\sim$(3--4)$\times$10$^{12}$\,G, based on a cyclotron resonance scattering feature at $\sim$35--37\,keV.

The first epoch included in this work was already analysed by \cite{dai2017}, who modelled the \nus\ dataset jointly with a \swift/XRT and \swift/BAT spectrum, thereby covering a broader energy range (0.5--150\,keV). To describe the continuum, they adopted the physical model {\tt bwmodel}, which is suited for the conditions in the accretion column of an accreting X-ray pulsar, and included an additional thermal ({\tt bbodyrad}) and reflection ({\tt coplrefl}) component. The electron temperature $kT_{\rm e}$, one of the free parameters of the model, was found to lie in the range $\sim$4--4.7\,keV, in good agreement with our results despite the different modelling approaches. 
This result is to be expected on theoretical grounds, as noted by \cite{burke2017}. In the Newtonian approximation, the luminosity of the boundary layer $L_{\rm bl}$ on the NS surface is given by
 \begin{displaymath}
    L_{\rm bl} = \frac{1}{2} \dot{M} \left(  v_{\rm K} - v_{\rm NS}  \right)^{2} 
\end{displaymath}
where $\dot{M}$ is the accretion rate, $v_{\rm K}$ is the Keplerian velocity near the NS surface and $v_{\rm NS}$ is the linear velocity of the NS surface at the equator. The shorter the spin period (i.e. higher $v_{\rm NS}$), the smaller $L_{\rm bl}$. In NS systems, the luminosity of seed photons $L_{\rm seed}$ is approximately equal to $L_{\rm bl}$, therefore shorter spin period systems should have smaller $L_{\rm seed}$. In other words, slowly rotating NSs have a higher surface luminosity and a weaker Comptonisation in the corona: they have a great supply of seed photons, implying a more efficient cooling of the corona electrons and, consequently, leading to colder $kT_{\rm e}$. However, we note that a further study including nine NS-LMXB systems did not find a clear correlation between the NS spin period and the electron temperature $kT_{\rm e}$ \citep{burke2018}.

\section{Conclusions}

We performed a global study of UCXBs using \nus\ observations, which covered all the classical accretion spectral states. We applied a uniform X‑ray spectral analysis based on a three‑component model whenever the data quality allowed it, following an approach that has become increasingly common for NS LMXBs. Our sample is the largest  analysed to date with this methodology. We do not claim that our approach is either a unique solution or an adequate depiction of complexity of nature, however it provides an overall self-consistent picture of UCXBs. Our key findings can be summarised as follows:

$-$ the RMS-diagram of UCXBs is similar to that observed in classical NS LMXBs, including the distinctive hook-like track displayed by brigth atoll sources \citep{munozdarias2014};

$-$ we constrained the nature of the Comptonisation seed photons across the different accretion states, finding a behaviour consistent with that expected from NS LMXBs \citep[e.g.,][]{armaspadilla2017};

$-$ the main properties of the X-ray emission in UCXBs is not significantly different from those of NS LMXBs with longer orbital periods. This suggests that the size and chemical composition of the accretion disc do not affect the properties of the inner regions of the systems, making them uniform across different types of NS X-ray binaries.

\begin{acknowledgements}
We thank the anonymous referee for a careful reading of the paper and insightful comments. This research utilized data, software, and web tools obtained from the High Energy Astrophysics Science Archive Research Center (HEASARC), a service of the Astrophysics Science Division at NASA/GSFC. 
A.B. acknowledges support through the European Space Agency (ESA) research fellowship programme. T.M.D. and M.A.P. acknowledge support by the Spanish Ministry of Science via the Plan de Generacion de conocimiento PID2021-124879NB-I00 and PID2024-161863NB-I00. M.A.P. acknowledges support through the Ramón y Cajal grant RYC2022-035388-I, funded by MCIU/AEI/10.13039/501100011033 and FSE+.

\end{acknowledgements}

\bibliographystyle{aa} 
\bibliography{biblio_UCXB} 

@ARTICLE{sanna2018a,
       author = {{Sanna}, A. and {Pintore}, F. and {Riggio}, A. and {Mazzola}, S.~M. and {Bozzo}, E. and {Di Salvo}, T. and {Ferrigno}, C. and {Gambino}, A.~F. and {Papitto}, A. and {Iaria}, R. and {Burderi}, L.},
        title = "{SWIFT J1756.9-2508: spectral and timing properties of its 2018 outburst}",
      journal = {\mnras},
         year = 2018,
        month = dec,
       volume = {481},
       number = {2},
        pages = {1658-1666},
          doi = {10.1093/mnras/sty2316},
archivePrefix = {arXiv},
       eprint = {1808.06796},
 primaryClass = {astro-ph.HE},
       adsurl = {https://ui.adsabs.harvard.edu/abs/2018MNRAS.481.1658S}
}

@ARTICLE{sanna2018b,
       author = {{Sanna}, A. and {Bahramian}, A. and {Bozzo}, E. and {Heinke}, C. and {Altamirano}, D. and {Wijnands}, R. and {Degenaar}, N. and {Maccarone}, T. and {Riggio}, A. and {Di Salvo}, T. and {Iaria}, R. and {Burgay}, M. and {Possenti}, A. and {Ferrigno}, C. and {Papitto}, A. and {Sivakoff}, G.~R. and {D'Amico}, N. and {Burderi}, L.},
        title = "{Discovery of 105 Hz coherent pulsations in the ultracompact binary IGR J16597-3704}",
      journal = {\aap},
         year = 2018,
        month = feb,
       volume = {610},
          eid = {L2},
        pages = {L2},
          doi = {10.1051/0004-6361/201732262},
archivePrefix = {arXiv},
       eprint = {1711.03092},
 primaryClass = {astro-ph.HE},
       adsurl = {https://ui.adsabs.harvard.edu/abs/2018A&A...610L...2S}
}

@ARTICLE{li2021,
       author = {{Li}, Z.~S. and {Kuiper}, L. and {Falanga}, M. and {Poutanen}, J. and {Tsygankov}, S.~S. and {Galloway}, D.~K. and {Bozzo}, E. and {Pan}, Y.~Y. and {Huang}, Y. and {Zhang}, S.~N. and {Zhang}, S.},
        title = "{Broadband X-ray spectra and timing of the accreting millisecond pulsar Swift J1756.9-2508 during its 2018 and 2019 outbursts}",
      journal = {\aap},
         year = 2021,
        month = may,
       volume = {649},
          eid = {A76},
        pages = {A76},
          doi = {10.1051/0004-6361/202140360},
archivePrefix = {arXiv},
       eprint = {2102.11687},
 primaryClass = {astro-ph.HE},
       adsurl = {https://ui.adsabs.harvard.edu/abs/2021A&A...649A..76L}
}

@ARTICLE{boirin2004,
       author = {{Boirin}, L. and {Parmar}, A.~N. and {Barret}, D. and {Paltani}, S. and {Grindlay}, J.~E.},
        title = "{Discovery of X-ray absorption features from the dipping low-mass X-ray binary XB 1916-053 with XMM-Newton}",
      journal = {\aap},
         year = 2004,
        month = may,
       volume = {418},
        pages = {1061-1072},
          doi = {10.1051/0004-6361:20034550},
archivePrefix = {arXiv},
       eprint = {astro-ph/0402294},
 primaryClass = {astro-ph},
       adsurl = {https://ui.adsabs.harvard.edu/abs/2004A&A...418.1061B}
}

@ARTICLE{gambino2019,
       author = {{Gambino}, A.~F. and {Iaria}, R. and {Di Salvo}, T. and {Mazzola}, S.~M. and {Marino}, A. and {Burderi}, L. and {Riggio}, A. and {Sanna}, A. and {D'Amico}, N.},
        title = "{Spectral analysis of the dipping LMXB system XB 1916-053}",
      journal = {\aap},
         year = 2019,
        month = may,
       volume = {625},
          eid = {A92},
        pages = {A92},
          doi = {10.1051/0004-6361/201832984},
archivePrefix = {arXiv},
       eprint = {1904.05770},
 primaryClass = {astro-ph.HE},
       adsurl = {https://ui.adsabs.harvard.edu/abs/2019A&A...625A..92G}
}

@ARTICLE{orlandini1998,
       author = {{Orlandini}, M. and {Dal Fiume}, D. and {Frontera}, F. and {Del Sordo}, S. and {Piraino}, S. and {Santangelo}, A. and {Segreto}, A. and {Oosterbroek}, T. and {Parmar}, A.~N.},
        title = "{BEPPOSAX Observation of 4U 1626-67: Discovery of an Absorption Cyclotron Resonance Feature}",
      journal = {\apjl},
         year = 1998,
        month = jun,
       volume = {500},
       number = {2},
        pages = {L163-L166},
          doi = {10.1086/311404},
archivePrefix = {arXiv},
       eprint = {astro-ph/9804241},
 primaryClass = {astro-ph},
       adsurl = {https://ui.adsabs.harvard.edu/abs/1998ApJ...500L.163O}
}

@ARTICLE{Mondal2016,
       author = {{Mondal}, Aditya S. and {Dewangan}, G.~C. and {Pahari}, M. and {Misra}, R. and {Kembhavi}, A.~K. and {Raychaudhuri}, B.},
        title = "{Broad-band X-ray emission and the reality of the broad iron line from the neutron star-white dwarf X-ray binary 4U 1820-30}",
      journal = {\mnras},
         year = 2016,
        month = sep,
       volume = {461},
       number = {2},
        pages = {1917-1926},
          doi = {10.1093/mnras/stw1464},
archivePrefix = {arXiv},
       eprint = {1606.05307},
 primaryClass = {astro-ph.HE},
       adsurl = {https://ui.adsabs.harvard.edu/abs/2016MNRAS.461.1917M}
}

@ARTICLE{Koliopanos2021,
       author = {{Koliopanos}, Filippos and {P{\'e}ault}, Mathias and {Vasilopoulos}, Georgios and {Webb}, Natalie},
        title = "{The chemical composition of the accretion disc and donor star in ultra-compact X-ray binaries: A comprehensive X-ray analysis}",
      journal = {\mnras},
         year = 2021,
        month = feb,
       volume = {501},
       number = {1},
        pages = {548-563},
          doi = {10.1093/mnras/staa3474},
archivePrefix = {arXiv},
       eprint = {2001.00716},
 primaryClass = {astro-ph.HE},
       adsurl = {https://ui.adsabs.harvard.edu/abs/2021MNRAS.501..548K}
}

@ARTICLE{armaspadilla2017,
       author = {{Armas Padilla}, M. and {Ueda}, Y. and {Hori}, T. and {Shidatsu}, M. and {Mu{\~n}oz-Darias}, T.},
        title = "{Suzaku spectroscopy of the neutron star transient 4U 1608-52 during its outburst decay.}",
      journal = {\mnras},
         year = 2017,
        month = may,
       volume = {467},
       number = {1},
        pages = {290-297},
          doi = {10.1093/mnras/stx020},
archivePrefix = {arXiv},
       eprint = {1701.02728},
 primaryClass = {astro-ph.HE},
       adsurl = {https://ui.adsabs.harvard.edu/abs/2017MNRAS.467..290A}
}

@ARTICLE{armaspadilla2019,
       author = {{Armas Padilla}, M. and {L{\'o}pez-Navas}, E.},
        title = "{On the ultra-compact nature of the neutron star system 1RXS J170854.4-321857: insights from X-ray spectroscopy}",
      journal = {\mnras},
         year = 2019,
        month = oct,
       volume = {488},
       number = {4},
        pages = {5014-5019},
          doi = {10.1093/mnras/stz2004},
archivePrefix = {arXiv},
       eprint = {1907.10631},
 primaryClass = {astro-ph.HE},
       adsurl = {https://ui.adsabs.harvard.edu/abs/2019MNRAS.488.5014A}
}

@ARTICLE{armaspadilla2023,
       author = {{Armas Padilla}, M. and {Corral-Santana}, J.~M. and {Borghese}, A. and {C{\'u}neo}, V.~A. and {Mu{\~n}oz-Darias}, T. and {Casares}, J. and {Torres}, M.~A.~P.},
        title = "{UltraCompCAT: A comprehensive catalogue of ultra-compact and short orbital period X-ray binaries}",
      journal = {\aap},
         year = 2023,
        month = sep,
       volume = {677},
          eid = {A186},
        pages = {A186},
          doi = {10.1051/0004-6361/202346797},
archivePrefix = {arXiv},
       eprint = {2305.07691},
 primaryClass = {astro-ph.HE},
       adsurl = {https://ui.adsabs.harvard.edu/abs/2023A&A...677A.186A}
}

@ARTICLE{lin2007,
       author = {{Lin}, Dacheng and {Remillard}, Ronald A. and {Homan}, Jeroen},
        title = "{Evaluating Spectral Models and the X-Ray States of Neutron Star X-Ray Transients}",
      journal = {\apj},
         year = 2007,
        month = oct,
       volume = {667},
       number = {2},
        pages = {1073-1086},
          doi = {10.1086/521181},
archivePrefix = {arXiv},
       eprint = {astro-ph/0702089},
 primaryClass = {astro-ph},
       adsurl = {https://ui.adsabs.harvard.edu/abs/2007ApJ...667.1073L}
}

@ARTICLE{ludlam2019,
       author = {{Ludlam}, R.~M. and {Miller}, J.~M. and {Barret}, D. and {Cackett}, E.~M. and {Coughenour}, B.~M. and {Dauser}, T. and {Degenaar}, N. and {Garc{\'\i}a}, J.~A. and {Harrison}, F.~A. and {Paerels}, F.},
        title = "{NuSTAR Observations of the Accreting Atolls GX 3+1, 4U 1702-429, 4U 0614+091, and 4U 1746-371}",
      journal = {\apj},
         year = 2019,
        month = mar,
       volume = {873},
       number = {1},
          eid = {99},
        pages = {99},
          doi = {10.3847/1538-4357/ab0414},
archivePrefix = {arXiv},
       eprint = {1902.00520},
 primaryClass = {astro-ph.HE},
       adsurl = {https://ui.adsabs.harvard.edu/abs/2019ApJ...873...99L}
}

@ARTICLE{ludlam2021,
       author = {{Ludlam}, R.~M. and {Jaodand}, A.~D. and {Garc{\'\i}a}, J.~A. and {Degenaar}, N. and {Tomsick}, J.~A. and {Cackett}, E.~M. and {Fabian}, A.~C. and {Gandhi}, P. and {Buisson}, D.~J.~K. and {Shaw}, A.~W. and {Chakrabarty}, D.},
        title = "{Simultaneous NICER and NuSTAR Observations of the Ultracompact X-Ray Binary 4U 1543-624}",
      journal = {\apj},
         year = 2021,
        month = apr,
       volume = {911},
       number = {2},
          eid = {123},
        pages = {123},
          doi = {10.3847/1538-4357/abedb0},
archivePrefix = {arXiv},
       eprint = {2012.10461},
 primaryClass = {astro-ph.HE},
       adsurl = {https://ui.adsabs.harvard.edu/abs/2021ApJ...911..123L}
}

@ARTICLE{VanDenEijnden2018,
       author = {{van den Eijnden}, J. and {Degenaar}, N. and {Pinto}, C. and {Patruno}, A. and {Wette}, K. and {Messenger}, C. and {Hern{\'a}ndez Santisteban}, J.~V. and {Wijnands}, R. and {Miller}, J.~M. and {Altamirano}, D. and {Paerels}, F. and {Chakrabarty}, D. and {Fabian}, A.~C.},
        title = "{The very faint X-ray binary IGR J17062-6143: a truncated disc, no pulsations, and a possible outflow}",
      journal = {\mnras},
         year = 2018,
        month = apr,
       volume = {475},
       number = {2},
        pages = {2027-2044},
          doi = {10.1093/mnras/stx3224},
       adsurl = {https://ui.adsabs.harvard.edu/abs/2018MNRAS.475.2027V}
}

@ARTICLE{degenaar2017,
       author = {{Degenaar}, N. and {Pinto}, C. and {Miller}, J.~M. and {Wijnands}, R. and {Altamirano}, D. and {Paerels}, F. and {Fabian}, A.~C. and {Chakrabarty}, D.},
        title = "{An in-depth study of a neutron star accreting at low Eddington rate: on the possibility of a truncated disc and an outflow}",
      journal = {\mnras},
         year = 2017,
        month = jan,
       volume = {464},
       number = {1},
        pages = {398-409},
          doi = {10.1093/mnras/stw2355},
archivePrefix = {arXiv},
       eprint = {1609.04816},
 primaryClass = {astro-ph.HE},
       adsurl = {https://ui.adsabs.harvard.edu/abs/2017MNRAS.464..398D}
}

@ARTICLE{dai2017,
       author = {{D'A{\`\i}}, A. and {Cusumano}, G. and {Del Santo}, M. and {La Parola}, V. and {Segreto}, A.},
        title = "{A broad-band self-consistent modelling of the X-ray spectrum of 4U 1626-67}",
      journal = {\mnras},
         year = 2017,
        month = sep,
       volume = {470},
       number = {2},
        pages = {2457-2468},
          doi = {10.1093/mnras/stx1146},
archivePrefix = {arXiv},
       eprint = {1705.03404},
 primaryClass = {astro-ph.HE},
       adsurl = {https://ui.adsabs.harvard.edu/abs/2017MNRAS.470.2457D}
}

@ARTICLE{sanna2017,
       author = {{Sanna}, A. and {Papitto}, A. and {Burderi}, L. and {Bozzo}, E. and {Riggio}, A. and {Di Salvo}, T. and {Ferrigno}, C. and {Rea}, N. and {Iaria}, R.},
        title = "{Discovery of a new accreting millisecond X-ray pulsar in the globular cluster NGC 2808}",
      journal = {\aap},
         year = 2017,
        month = feb,
       volume = {598},
          eid = {A34},
        pages = {A34},
          doi = {10.1051/0004-6361/201629406},
archivePrefix = {arXiv},
       eprint = {1611.02995},
 primaryClass = {astro-ph.HE},
       adsurl = {https://ui.adsabs.harvard.edu/abs/2017A&A...598A..34S}
}

@ARTICLE{moutard2023,
       author = {{Moutard}, David and {Ludlam}, Renee and {Garc{\'\i}a}, Javier A. and {Altamirano}, Diego and {Buisson}, Douglas J.~K. and {Cackett}, Edward M. and {Chenevez}, J{\'e}r{\^o}me and {Degenaar}, Nathalie and {Fabian}, Andrew C. and {Homan}, Jeroen and {Jaodand}, Amruta and {Pike}, Sean N. and {Shaw}, Aarran W. and {Strohmayer}, Tod E. and {Tomsick}, John A. and {Coughenour}, Benjamin M.},
        title = "{Simultaneous NICER and NuSTAR Observations of the Ultra-compact X-ray Binary 4U 0614+091}",
      journal = {arXiv e-prints},
         year = 2023,
        month = aug,
          eid = {arXiv:2308.15581},
        pages = {arXiv:2308.15581},
          doi = {10.48550/arXiv.2308.15581},
archivePrefix = {arXiv},
       eprint = {2308.15581},
 primaryClass = {astro-ph.HE},
       adsurl = {https://ui.adsabs.harvard.edu/abs/2023arXiv230815581M}
}

@ARTICLE{munozdarias2014,
       author = {{Mu{\~n}oz-Darias}, T. and {Fender}, R.~P. and {Motta}, S.~E. and {Belloni}, T.~M.},
        title = "{Black hole-like hysteresis and accretion states in neutron star low-mass X-ray binaries}",
      journal = {\mnras},
         year = 2014,
        month = oct,
       volume = {443},
       number = {4},
        pages = {3270-3283},
          doi = {10.1093/mnras/stu1334},
archivePrefix = {arXiv},
       eprint = {1407.1318},
 primaryClass = {astro-ph.HE},
       adsurl = {https://ui.adsabs.harvard.edu/abs/2014MNRAS.443.3270M}
}

@ARTICLE{marino2023,
       author = {{Marino}, A. and {Russell}, T.~D. and {Del Santo}, M. and {Beri}, A. and {Sanna}, A. and {Coti Zelati}, F. and {Degenaar}, N. and {Altamirano}, D. and {Ambrosi}, E. and {Anitra}, A. and {Carotenuto}, F. and {D'A{\`\i}}, A. and {Di Salvo}, T. and {Manca}, A. and {Motta}, S.~E. and {Pinto}, C. and {Pintore}, F. and {Rea}, N. and {van den Eijnden}, J.},
        title = "{The accretion/ejection link in the neutron star X-ray binary 4U 1820-30 I: a boundary layer-jet coupling?}",
      journal = {\mnras},
         year = 2023,
        month = oct,
       volume = {525},
       number = {2},
        pages = {2366-2379},
          doi = {10.1093/mnras/stad2386},
archivePrefix = {arXiv},
       eprint = {2307.16566},
 primaryClass = {astro-ph.HE},
       adsurl = {https://ui.adsabs.harvard.edu/abs/2023MNRAS.525.2366M}
}

@ARTICLE{anderson1997,
       author = {{Anderson}, Scott F. and {Margon}, Bruce and {Deutsch}, Eric W. and {Downes}, Ronald A. and {Allen}, Richard G.},
        title = "{Time-Resolved Ultraviolet Observations of the Globular Cluster X-Ray Source in NGC 6624: The Shortest Known Period Binary System}",
      journal = {\apjl},
         year = 1997,
        month = jun,
       volume = {482},
       number = {1},
        pages = {L69-L72},
          doi = {10.1086/310672},
archivePrefix = {arXiv},
       eprint = {astro-ph/9703180},
 primaryClass = {astro-ph},
       adsurl = {https://ui.adsabs.harvard.edu/abs/1997ApJ...482L..69A}
}

@ARTICLE{dimarco2023,
       author = {{Di Marco}, Alessandro and {La Monaca}, Fabio and {Poutanen}, Juri and {Russell}, Thomas D. and {Anitra}, Alessio and {Farinelli}, Ruben and {Mastroserio}, Guglielmo and {Muleri}, Fabio and {Xie}, Fei and {Bachetti}, Matteo and {Burderi}, Luciano and {Carotenuto}, Francesco and {Del Santo}, Melania and {Di Salvo}, Tiziana and {Dov{\v{c}}iak}, Michal and {Gnarini}, Andrea and {Iaria}, Rosario and {Kajava}, Jari J.~E. and {Liu}, Kuan and {Middei}, Riccardo and {O'Dell}, Stephen L. and {Pilia}, Maura and {Rankin}, John and {Sanna}, Andrea and {Eijnden}, Jakob van den and {Weisskopf}, Martin C. and {Bobrikova}, Anna and {Capitanio}, Fiamma and {Costa}, Enrico and {Kaaret}, Philip and {Marino}, Alessio and {Soffitta}, Paolo and {Ursini}, Francesco and {Ambrosino}, Filippo and {Cocchi}, Massimo and {Fabiani}, Sergio and {Marshall}, Herman L. and {Matt}, Giorgio and {Motta}, Sara Elisa and {Papitto}, Alessandro and {Stella}, Luigi and {Tarana}, Antonella and {Zane}, Silvia and {Agudo}, Iv{\'a}n and {Antonelli}, Lucio A. and {Baldini}, Luca and {Baumgartner}, Wayne H. and {Bellazzini}, Ronaldo and {Bianchi}, Stefano and {Bongiorno}, Stephen D. and {Bonino}, Raffaella and {Brez}, Alessandro and {Bucciantini}, Niccol{\`o} and {Castellano}, Simone and {Cavazzuti}, Elisabetta and {Chen}, Chien-Ting and {Ciprini}, Stefano and {De Rosa}, Alessandra and {Del Monte}, Ettore and {Di Gesu}, Laura and {Di Lalla}, Niccol{\`o} and {Donnarumma}, Immacolata and {Doroshenko}, Victor and {Ehlert}, Steven R. and {Enoto}, Teruaki and {Evangelista}, Yuri and {Ferrazzoli}, Riccardo and {Garcia}, Javier A. and {Gunji}, Shuichi and {Hayashida}, Kiyoshi and {Heyl}, Jeremy and {Iwakiri}, Wataru and {Jorstad}, Svetlana G. and {Karas}, Vladimir and {Kislat}, Fabian and {Kitaguchi}, Takao and {Kolodziejczak}, Jeffery J. and {Krawczynski}, Henric and {Latronico}, Luca and {Liodakis}, Ioannis and {Maldera}, Simone and {Manfreda}, Alberto and {Marin}, Fr{\'e}d{\'e}ric and {Marinucci}, Andrea and {Marscher}, Alan P. and {Massaro}, Francesco and {Mitsuishi}, Ikuyuki and {Mizuno}, Tsunefumi and {Negro}, Michela and {Ng}, Chi-Yung and {Omodei}, Nicola and {Oppedisano}, Chiara and {Pavlov}, George G. and {Peirson}, Abel L. and {Perri}, Matteo and {Pesce-Rollins}, Melissa and {Petrucci}, Pierre-Olivier and {Possenti}, Andrea and {Puccetti}, Simonetta and {Ramsey}, Brian D. and {Ratheesh}, Ajay and {Roberts}, Oliver J. and {Romani}, Roger W. and {Sgr{\`o}}, Carmelo and {Slane}, Patrick and {Spandre}, Gloria and {Swartz}, Douglas A. and {Tamagawa}, Toru and {Tavecchio}, Fabrizio and {Taverna}, Roberto and {Tawara}, Yuzuru and {Tennant}, Allyn F. and {Thomas}, Nicholas E. and {Tombesi}, Francesco and {Trois}, Alessio and {Tsygankov}, Sergey S. and {Turolla}, Roberto and {Vink}, Jacco and {Wu}, Kinwah and {IXPE Collaboration}},
        title = "{First Detection of X-Ray Polarization from the Accreting Neutron Star 4U 1820-303}",
      journal = {\apjl},
         year = 2023,
        month = aug,
       volume = {953},
       number = {2},
          eid = {L22},
        pages = {L22},
          doi = {10.3847/2041-8213/acec6e},
archivePrefix = {arXiv},
       eprint = {2306.08476},
 primaryClass = {astro-ph.HE},
       adsurl = {https://ui.adsabs.harvard.edu/abs/2023ApJ...953L..22D}
}

@ARTICLE{sharma2023,
       author = {{Sharma}, Rahul and {Jain}, Chetana and {Paul}, Biswajit},
        title = "{4U 1626-67 returns to spin-down: timing features toe the line}",
      journal = {\mnras},
         year = 2023,
        month = nov,
       volume = {526},
       number = {1},
        pages = {L35-L40},
          doi = {10.1093/mnrasl/slad110},
archivePrefix = {arXiv},
       eprint = {2307.16599},
 primaryClass = {astro-ph.HE},
       adsurl = {https://ui.adsabs.harvard.edu/abs/2023MNRAS.526L..35S}
}

@ARTICLE{rappaport77,
       author = {{Rappaport}, S. and {Markert}, T. and {Li}, F.~K. and {Clark}, G.~W. and {Jernigan}, J.~G. and {McClintock}, J.~E.},
        title = "{Discovery of a 7.68 second X-ray periodicity in 3U 1626-67.}",
      journal = {\apjl},
         year = 1977,
        month = oct,
       volume = {217},
        pages = {L29-L33},
          doi = {10.1086/182532},
       adsurl = {https://ui.adsabs.harvard.edu/abs/1977ApJ...217L..29R}
}

@ARTICLE{rappaport82,
       author = {{Rappaport}, S. and {Joss}, P.~C. and {Webbink}, R.~F.},
        title = "{The evolution of highly compact binary stellar systems.}",
      journal = {\apj},
         year = 1982,
        month = mar,
       volume = {254},
        pages = {616-640},
          doi = {10.1086/159772},
       adsurl = {https://ui.adsabs.harvard.edu/abs/1982ApJ...254..616R}
}

@ARTICLE{chen2021,
       author = {{Chen}, Hai-Liang and {Tauris}, Thomas M. and {Han}, Zhanwen and {Chen}, Xuefei},
        title = "{Formation of millisecond pulsars with helium white dwarfs, ultra-compact X-ray binaries, and gravitational wave sources}",
      journal = {\mnras},
         year = 2021,
        month = may,
       volume = {503},
       number = {3},
        pages = {3540-3551},
          doi = {10.1093/mnras/stab670},
archivePrefix = {arXiv},
       eprint = {2103.02931},
 primaryClass = {astro-ph.SR},
       adsurl = {https://ui.adsabs.harvard.edu/abs/2021MNRAS.503.3540C}
}

@INPROCEEDINGS{nelemans2009,
       author = {{Nelemans}, G. and {Wood}, M. and {Groot}, P. and {Anderson}, S. and {Belczynski}, K. and {Benacquista}, M. and {Charles}, P. and {Cumming}, A. and {Deloye}, C. and {Jonker}, P. and {Kalogera}, V. and {Knigge}, C. and {Marsh}, T. and {Motl}, P. and {Napiwotzki}, R. and {O'Brien}, K. and {Phinney}, E.~S. and {Ramsay}, G. and {Shahbaz}, T. and {Solheim}, J. -E. and {Steeghs}, D. and {van der Sluys}, M. and {Verbunt}, F. and {Warner}, B. and {Werner}, K. and {Wu}, K. and {Yungelson}, L.~R.},
        title = "{The astrophysics of ultra-compact binaries}",
    booktitle = {astro2010: The Astronomy and Astrophysics Decadal Survey},
         year = 2009,
       volume = {2010},
        month = jan,
        pages = {221},
          doi = {10.48550/arXiv.0902.2923},
archivePrefix = {arXiv},
       eprint = {0902.2923},
 primaryClass = {astro-ph.SR},
       adsurl = {https://ui.adsabs.harvard.edu/abs/2009astro2010S.221N}
}

@ARTICLE{paradijs1994,
       author = {{van Paradijs}, J. and {McClintock}, J.~E.},
        title = "{Absolute visual magnitudes of low-mass X-ray binaries.}",
      journal = {\aap},
         year = 1994,
        month = oct,
       volume = {290},
        pages = {133-136},
       adsurl = {https://ui.adsabs.harvard.edu/abs/1994A&A...290..133V}
}

@ARTICLE{zand2007,
       author = {{in't Zand}, J.~J.~M. and {Jonker}, P.~G. and {Markwardt}, C.~B.},
        title = "{Six new candidate ultracompact X-ray binaries}",
      journal = {\aap},
         year = 2007,
        month = apr,
       volume = {465},
       number = {3},
        pages = {953-963},
          doi = {10.1051/0004-6361:20066678},
archivePrefix = {arXiv},
       eprint = {astro-ph/0701810},
 primaryClass = {astro-ph},
       adsurl = {https://ui.adsabs.harvard.edu/abs/2007A&A...465..953I}
}

@ARTICLE{lasota2001,
       author = {{Lasota}, Jean-Pierre},
        title = "{The disc instability model of dwarf novae and low-mass X-ray binary transients}",
      journal = {\nar},
         year = 2001,
        month = jun,
       volume = {45},
       number = {7},
        pages = {449-508},
          doi = {10.1016/S1387-6473(01)00112-9},
archivePrefix = {arXiv},
       eprint = {astro-ph/0102072},
 primaryClass = {astro-ph},
       adsurl = {https://ui.adsabs.harvard.edu/abs/2001NewAR..45..449L}
}

@ARTICLE{stoop2021,
       author = {{Stoop}, M. and {van den Eijnden}, J. and {Degenaar}, N. and {Bahramian}, A. and {Swihart}, S.~J. and {Strader}, J. and {Jim{\'e}nez-Ibarra}, F. and {Mu{\~n}oz-Darias}, T. and {Armas Padilla}, M. and {Shaw}, A.~W. and {Maccarone}, T.~J. and {Wijnands}, R. and {Russell}, T.~D. and {Hern{\'a}ndez Santisteban}, J.~V. and {Miller-Jones}, J.~C.~A. and {Russell}, D.~M. and {Maitra}, D. and {Heinke}, C.~O. and {Sivakoff}, G.~R. and {Lewis}, F. and {Bramich}, D.~M.},
        title = "{Multiwavelength observations reveal a faint candidate black hole X-ray binary in IGR J17285-2922}",
      journal = {\mnras},
         year = 2021,
        month = oct,
       volume = {507},
       number = {1},
        pages = {330-349},
          doi = {10.1093/mnras/stab2127},
       adsurl = {https://ui.adsabs.harvard.edu/abs/2021MNRAS.507..330S}
}

@article{harrison13,
	Adsurl = {http://adsabs.harvard.edu/abs/2013ApJ...770..103H},
	Archiveprefix = {arXiv},
	Author = {{Harrison}, F.~A. and {Craig}, W.~W. and {Christensen}, F.~E. and {Hailey}, C.~J. and {Zhang}, W.~W. and {Boggs}, S.~E. and {Stern}, D. and {Cook}, W.~R. and {Forster}, K. and {Giommi}, P. and {Grefenstette}, B.~W. and {Kim}, Y. and {Kitaguchi}, T. and {Koglin}, J.~E. and {Madsen}, K.~K. and {Mao}, P.~H. and {Miyasaka}, H. and {Mori}, K. and {Perri}, M. and {Pivovaroff}, M.~J. and {Puccetti}, S. and {Rana}, V.~R. and {Westergaard}, N.~J. and {Willis}, J. and {Zoglauer}, A. and {An}, H. and {Bachetti}, M. and {Barri{\`e}re}, N.~M. and {Bellm}, E.~C. and {Bhalerao}, V. and {Brejnholt}, N.~F. and {Fuerst}, F. and {Liebe}, C.~C. and {Markwardt}, C.~B. and {Nynka}, M. and {Vogel}, J.~K. and {Walton}, D.~J. and {Wik}, D.~R. and {Alexander}, D.~M. and {Cominsky}, L.~R. and {Hornschemeier}, A.~E. and {Hornstrup}, A. and {Kaspi}, V.~M. and {Madejski}, G.~M. and {Matt}, G. and {Molendi}, S. and {Smith}, D.~M. and {Tomsick}, J.~A. and {Ajello}, M. and {Ballantyne}, D.~R. and {Balokovi{\'c}}, M. and {Barret}, D. and {Bauer}, F.~E. and {Blandford}, R.~D. and {Brandt}, W.~N. and {Brenneman}, L.~W. and {Chiang}, J. and {Chakrabarty}, D. and {Chenevez}, J. and {Comastri}, A. and {Dufour}, F. and {Elvis}, M. and {Fabian}, A.~C. and {Farrah}, D. and {Fryer}, C.~L. and {Gotthelf}, E.~V. and {Grindlay}, J.~E. and {Helfand}, D.~J. and {Krivonos}, R. and {Meier}, D.~L. and {Miller}, J.~M. and {Natalucci}, L. and {Ogle}, P. and {Ofek}, E.~O. and {Ptak}, A. and {Reynolds}, S.~P. and {Rigby}, J.~R. and {Tagliaferri}, G. and {Thorsett}, S.~E. and {Treister}, E. and {Urry}, C.~M.},
	Eid = {103},
	Journal = {\apj},
	Month = jun,
	Pages = {103},
	Primaryclass = {astro-ph.IM},
	Title = {{The Nuclear Spectroscopic Telescope Array (NuSTAR) High-energy X-Ray Mission}},
	Volume = 770,
	Year = 2013}

@ARTICLE{bahramian2017,
       author = {{Bahramian}, Arash and {Heinke}, Craig O. and {Tudor}, Vlad and {Miller-Jones}, James C.~A. and {Bogdanov}, Slavko and {Maccarone}, Thomas J. and {Knigge}, Christian and {Sivakoff}, Gregory R. and {Chomiuk}, Laura and {Strader}, Jay and {Garcia}, Javier A. and {Kallman}, Timothy},
        title = "{The ultracompact nature of the black hole candidate X-ray binary 47 Tuc X9}",
      journal = {\mnras},
         year = 2017,
        month = may,
       volume = {467},
       number = {2},
        pages = {2199-2216},
          doi = {10.1093/mnras/stx166},
archivePrefix = {arXiv},
       eprint = {1702.02167},
 primaryClass = {astro-ph.HE},
       adsurl = {https://ui.adsabs.harvard.edu/abs/2017MNRAS.467.2199B}
}

@ARTICLE{kaastra2016,
       author = {{Kaastra}, J.~S. and {Bleeker}, J.~A.~M.},
        title = "{Optimal binning of X-ray spectra and response matrix design}",
      journal = {\aap},
         year = 2016,
        month = mar,
       volume = {587},
          eid = {A151},
        pages = {A151},
          doi = {10.1051/0004-6361/201527395},
archivePrefix = {arXiv},
       eprint = {1601.05309},
 primaryClass = {astro-ph.IM},
       adsurl = {https://ui.adsabs.harvard.edu/abs/2016A&A...587A.151K}
}

@inproceedings{arnaud96,
	Adsurl = {http://adsabs.harvard.edu/abs/1996ASPC..101...17A},
	Author = {{Arnaud}, K.~A.},
	Booktitle = {XSPEC: The First Ten Years},
	Editor = {{Jacoby}, G.~H. and {Barnes}, J.},
	Pages = {17-20},
	Publisher = {ASP, San Francisco},
	Series = {Astronomical Data Analysis Software and Systems V},
	Title = {{XSPEC: The First Ten Years}},
	Volume = 101,
	Year = 1996}

@ARTICLE{verner96,
       author = {{Verner}, D.~A. and {Ferland}, G.~J. and {Korista}, K.~T. and
         {Yakovlev}, D.~G.},
        title = "{Atomic Data for Astrophysics. II. New Analytic FITS for Photoionization Cross Sections of Atoms and Ions}",
      journal = {\apj},
         year = 1996,
        month = jul,
       volume = {465},
        pages = {487},
       adsurl = {https://ui.adsabs.harvard.edu/abs/1996ApJ...465..487V}
}

@article{wilms00,
	Adsurl = {http://adsabs.harvard.edu/abs/2000ApJ...542..914W},
	Author = {{Wilms}, J. and {Allen}, A. and {McCray}, R.},
	Journal = {\apj},
	Month = oct,
	Pages = {914-924},
	Title = {{On the Absorption of X-Rays in the Interstellar Medium}},
	Volume = 542,
	Year = 2000}

@ARTICLE{huppenkothen2019,
       author = {{Huppenkothen}, Daniela and {Bachetti}, Matteo and {Stevens}, Abigail L. and {Migliari}, Simone and {Balm}, Paul and {Hammad}, Omar and {Khan}, Usman Mahmood and {Mishra}, Himanshu and {Rashid}, Haroon and {Sharma}, Swapnil and {Martinez Ribeiro}, Evandro and {Valles Blanco}, Ricardo},
        title = "{Stingray: A Modern Python Library for Spectral Timing}",
      journal = {\apj},
         year = 2019,
        month = aug,
       volume = {881},
       number = {1},
          eid = {39},
        pages = {39},
          doi = {10.3847/1538-4357/ab258d},
archivePrefix = {arXiv},
       eprint = {1901.07681},
 primaryClass = {astro-ph.IM},
       adsurl = {https://ui.adsabs.harvard.edu/abs/2019ApJ...881...39H}
}

@ARTICLE{bachetti2018,
       author = {{Bachetti}, Matteo and {Huppenkothen}, Daniela},
        title = "{No Time for Dead Time: Use the Fourier Amplitude Differences to Normalize Dead-time-affected Periodograms}",
      journal = {\apjl},
         year = 2018,
        month = feb,
       volume = {853},
       number = {2},
          eid = {L21},
        pages = {L21},
          doi = {10.3847/2041-8213/aaa83b},
archivePrefix = {arXiv},
       eprint = {1709.09700},
 primaryClass = {astro-ph.HE},
       adsurl = {https://ui.adsabs.harvard.edu/abs/2018ApJ...853L..21B}
}

@ARTICLE{burke2017,
       author = {{Burke}, M.~J. and {Gilfanov}, M. and {Sunyaev}, R.},
        title = "{A dichotomy between the hard state spectral properties of black hole and neutron star X-ray binaries}",
      journal = {\mnras},
         year = 2017,
        month = apr,
       volume = {466},
       number = {1},
        pages = {194-212},
          doi = {10.1093/mnras/stw2514},
archivePrefix = {arXiv},
       eprint = {1609.09511},
 primaryClass = {astro-ph.HE},
       adsurl = {https://ui.adsabs.harvard.edu/abs/2017MNRAS.466..194B}
}

@ARTICLE{walter1982,
       author = {{Walter}, F.~M. and {Mason}, K.~O. and {Clarke}, J.~T. and {Halpern}, J. and {Grindlay}, J.~E. and {Bowyer}, S. and {Henry}, J.~P.},
        title = "{Discovery of a 50 MN binary period and a likely 22 magnitude optical counterpart for the X-ray burster 4U 1915-05.}",
      journal = {\apjl},
         year = 1982,
        month = feb,
       volume = {253},
        pages = {L67-L71},
          doi = {10.1086/183738},
       adsurl = {https://ui.adsabs.harvard.edu/abs/1982ApJ...253L..67W}
}

@ARTICLE{mitsuda84,
       author = {{Mitsuda}, K. and {Inoue}, H. and {Koyama}, K. and {Makishima}, K. and {Matsuoka}, M. and {Ogawara}, Y. and {Shibazaki}, N. and {Suzuki}, K. and {Tanaka}, Y. and {Hirano}, T.},
        title = "{Energy spectra of low-mass binary X-ray sources observed from Tenma.}",
      journal = {\pasj},
         year = 1984,
        month = jan,
       volume = {36},
        pages = {741-759},
       adsurl = {https://ui.adsabs.harvard.edu/abs/1984PASJ...36..741M}
}

@ARTICLE{mitsuda89,
       author = {{Mitsuda}, Kazuhisa and {Inoue}, Hajime and {Nakamura}, Norio and {Tanaka}, Yasuo},
        title = "{Luminosity-related changes of the energy spectrum of X 1608-522.}",
      journal = {\pasj},
         year = 1989,
        month = jan,
       volume = {41},
        pages = {97-111},
       adsurl = {https://ui.adsabs.harvard.edu/abs/1989PASJ...41...97M}
}

@ARTICLE{kubota98,
       author = {{Kubota}, Aya and {Tanaka}, Yasuo and {Makishima}, Kazuo and {Ueda}, Yoshihiro and {Dotani}, Tadayasu and {Inoue}, Hajime and {Yamaoka}, Kazutaka},
        title = "{Evidence for a Black Hole in the X-Ray Transient GRS 1009-45}",
      journal = {\pasj},
         year = 1998,
        month = dec,
       volume = {50},
        pages = {667-673},
          doi = {10.1093/pasj/50.6.667},
       adsurl = {https://ui.adsabs.harvard.edu/abs/1998PASJ...50..667K}
}

@ARTICLE{shimura95,
       author = {{Shimura}, Toshiya and {Takahara}, Fumio},
        title = "{On the Spectral Hardening Factor of the X-Ray Emission from Accretion Disks in Black Hole Candidates}",
      journal = {\apj},
         year = 1995,
        month = jun,
       volume = {445},
        pages = {780},
          doi = {10.1086/175740},
       adsurl = {https://ui.adsabs.harvard.edu/abs/1995ApJ...445..780S}
}

@ARTICLE{zdziarski96,
       author = {{Zdziarski}, A.~A. and {Johnson}, W.~N. and {Magdziarz}, P.},
        title = "{Broad-band {\ensuremath{\gamma}}-ray and X-ray spectra of NGC 4151 and their implications for physical processes and geometry.}",
      journal = {\mnras},
         year = 1996,
        month = nov,
       volume = {283},
       number = {1},
        pages = {193-206},
          doi = {10.1093/mnras/283.1.193},
archivePrefix = {arXiv},
       eprint = {astro-ph/9607015},
 primaryClass = {astro-ph},
       adsurl = {https://ui.adsabs.harvard.edu/abs/1996MNRAS.283..193Z}
}

@ARTICLE{zycki99,
       author = {{{\.Z}ycki}, Piotr T. and {Done}, Chris and {Smith}, David A.},
        title = "{The 1989 May outburst of the soft X-ray transient GS 2023+338 (V404 Cyg)}",
      journal = {\mnras},
         year = 1999,
        month = nov,
       volume = {309},
       number = {3},
        pages = {561-575},
          doi = {10.1046/j.1365-8711.1999.02885.x},
archivePrefix = {arXiv},
       eprint = {astro-ph/9904304},
 primaryClass = {astro-ph},
       adsurl = {https://ui.adsabs.harvard.edu/abs/1999MNRAS.309..561Z}
}

@ARTICLE{popham01,
       author = {{Popham}, Robert and {Sunyaev}, Rashid},
        title = "{Accretion Disk Boundary Layers around Neutron Stars: X-Ray Production in Low-Mass X-Ray Binaries}",
      journal = {\apj},
         year = 2001,
        month = jan,
       volume = {547},
       number = {1},
        pages = {355-383},
          doi = {10.1086/318336},
archivePrefix = {arXiv},
       eprint = {astro-ph/0004017},
 primaryClass = {astro-ph},
       adsurl = {https://ui.adsabs.harvard.edu/abs/2001ApJ...547..355P}
}

@ARTICLE{burkesecond20,
       author = {{Banerjee}, Srimanta and {Gilfanov}, Marat and {Bhattacharyya}, Sudip and {Sunyaev}, Rashid},
        title = "{Observing imprints of black hole event horizon on X-ray spectra}",
      journal = {\mnras},
         year = 2020,
        month = nov,
       volume = {498},
       number = {4},
        pages = {5353-5360},
          doi = {10.1093/mnras/staa2788},
archivePrefix = {arXiv},
       eprint = {2009.07222},
 primaryClass = {astro-ph.HE},
       adsurl = {https://ui.adsabs.harvard.edu/abs/2020MNRAS.498.5353B}
}

@INCOLLECTION{bahramian23book,
       author = {{Bahramian}, Arash and {Degenaar}, Nathalie},
        title = "{Low-Mass X-ray Binaries}",
    booktitle = {Handbook of X-ray and Gamma-ray Astrophysics. Edited by Cosimo Bambi and Andrea Santangelo},
         year = 2023,
          eid = {120},
        pages = {120},
          doi = {10.1007/978-981-16-4544-0_94-1},
       adsurl = {https://ui.adsabs.harvard.edu/abs/2023hxga.book..120B}
}

@ARTICLE{fiocchi2008,
       author = {{Fiocchi}, M. and {Bazzano}, A. and {Ubertini}, P. and {Bird}, A.~J. and {Natalucci}, L. and {Sguera}, V.},
        title = "{The INTEGRAL long monitoring of persistent ultra compact X-ray bursters}",
      journal = {\aap},
         year = 2008,
        month = dec,
       volume = {492},
       number = {2},
        pages = {557-563},
          doi = {10.1051/0004-6361:200809715},
archivePrefix = {arXiv},
       eprint = {0810.1878},
 primaryClass = {astro-ph},
       adsurl = {https://ui.adsabs.harvard.edu/abs/2008A&A...492..557F}
}

@ARTICLE{diaztrigo2016,
       author = {{D{\'\i}az Trigo}, M. and {Boirin}, L.},
        title = "{Accretion disc atmospheres and winds in low-mass X-ray binaries}",
      journal = {Astronomische Nachrichten},
         year = 2016,
        month = may,
       volume = {337},
       number = {4-5},
        pages = {368},
          doi = {10.1002/asna.201612315},
archivePrefix = {arXiv},
       eprint = {1510.03576},
 primaryClass = {astro-ph.HE},
       adsurl = {https://ui.adsabs.harvard.edu/abs/2016AN....337..368D}
}

@ARTICLE{ponti2014,
       author = {{Ponti}, Gabriele and {Mu{\~n}oz-Darias}, Teodoro and {Fender}, Robert P.},
        title = "{A connection between accretion state and Fe K absorption in an accreting neutron star: black hole-like soft-state winds?}",
      journal = {\mnras},
         year = 2014,
        month = oct,
       volume = {444},
       number = {2},
        pages = {1829-1834},
          doi = {10.1093/mnras/stu1742},
archivePrefix = {arXiv},
       eprint = {1407.4468},
 primaryClass = {astro-ph.HE},
       adsurl = {https://ui.adsabs.harvard.edu/abs/2014MNRAS.444.1829P}
}

@ARTICLE{burke2018,
       author = {{Burke}, M.~J. and {Gilfanov}, M. and {Sunyaev}, R.},
        title = "{The impact of neutron star spin on X-ray spectra}",
      journal = {\mnras},
         year = 2018,
        month = feb,
       volume = {474},
       number = {1},
        pages = {760-769},
          doi = {10.1093/mnras/stx2821},
archivePrefix = {arXiv},
       eprint = {1707.07742},
 primaryClass = {astro-ph.HE},
       adsurl = {https://ui.adsabs.harvard.edu/abs/2018MNRAS.474..760B}
}

@ARTICLE{hasinger1989,
       author = {{Hasinger}, G. and {van der Klis}, M.},
        title = "{Two patterns of correlated X-ray timing and spectral behaviour in low-mass X-ray binaries.}",
      journal = {\aap},
         year = 1989,
        month = nov,
       volume = {225},
        pages = {79-96},
       adsurl = {https://ui.adsabs.harvard.edu/abs/1989A&A...225...79H}
}

@ARTICLE{sharma2018,
       author = {{Sharma}, Rahul and {Jaleel}, Abdul and {Jain}, Chetana and {Pandey}, Jeewan C. and {Paul}, Biswajit and {Dutta}, Anjan},
        title = "{Spectral properties of MXB 1658-298 in the low/hard and high/soft state}",
      journal = {\mnras},
         year = 2018,
        month = dec,
       volume = {481},
       number = {4},
        pages = {5560-5569},
          doi = {10.1093/mnras/sty2678},
archivePrefix = {arXiv},
       eprint = {1810.01827},
 primaryClass = {astro-ph.HE},
       adsurl = {https://ui.adsabs.harvard.edu/abs/2018MNRAS.481.5560S}
}

@ARTICLE{banerjee2024,
       author = {{Banerjee}, Srimanta and {Homan}, Jeroen},
        title = "{Probing the accretion geometry of the atoll source 4U 1702-429 in different spectral states with NICER, NuSTAR, and AstroSat}",
      journal = {\mnras},
         year = 2024,
        month = apr,
       volume = {529},
       number = {4},
        pages = {4311-4324},
          doi = {10.1093/mnras/stae541},
archivePrefix = {arXiv},
       eprint = {2402.11844},
 primaryClass = {astro-ph.HE},
       adsurl = {https://ui.adsabs.harvard.edu/abs/2024MNRAS.529.4311B}
}

@ARTICLE{ludlam2020,
       author = {{Ludlam}, R.~M. and {Cackett}, E.~M. and {Garc{\'\i}a}, J.~A. and {Miller}, J.~M. and {Bult}, P.~M. and {Strohmayer}, T.~E. and {Guillot}, S. and {Jaisawal}, G.~K. and {Malacaria}, C. and {Fabian}, A.~C. and {Markwardt}, C.~B.},
        title = "{NICER-NuSTAR Observations of the Neutron Star Low-mass X-Ray Binary 4U 1735-44}",
      journal = {\apj},
         year = 2020,
        month = may,
       volume = {895},
       number = {1},
          eid = {45},
        pages = {45},
          doi = {10.3847/1538-4357/ab89a6},
archivePrefix = {arXiv},
       eprint = {2004.06723},
 primaryClass = {astro-ph.HE},
       adsurl = {https://ui.adsabs.harvard.edu/abs/2020ApJ...895...45L}
}

@ARTICLE{munozdarias2011,
       author = {{Mu{\~n}oz-Darias}, T. and {Motta}, S. and {Belloni}, T.~M.},
        title = "{Fast variability as a tracer of accretion regimes in black hole transients}",
      journal = {\mnras},
         year = 2011,
        month = jan,
       volume = {410},
       number = {1},
        pages = {679-684},
          doi = {10.1111/j.1365-2966.2010.17476.x},
archivePrefix = {arXiv},
       eprint = {1008.0558},
 primaryClass = {astro-ph.HE},
       adsurl = {https://ui.adsabs.harvard.edu/abs/2011MNRAS.410..679M}
}

@ARTICLE{munozdarias2026,
       author = {{Mu{\~n}oz-Darias}, Teo and {D{\'\i}az Trigo}, Mar{\'\i}a and {Done}, Chris and {Ponti}, Gabriele and {Tomaru}, Ryota},
        title = "{Accretion disc winds in X-ray binaries}",
      journal = {arXiv e-prints},
         year = 2026,
        month = jan,
          eid = {arXiv:2601.05319},
        pages = {arXiv:2601.05319},
          doi = {10.48550/arXiv.2601.05319},
archivePrefix = {arXiv},
       eprint = {2601.05319},
 primaryClass = {astro-ph.HE},
       adsurl = {https://ui.adsabs.harvard.edu/abs/2026arXiv260105319M}
}

@ARTICLE{belloni2005,
       author = {{Belloni}, T. and {Homan}, J. and {Casella}, P. and {van der Klis}, M. and {Nespoli}, E. and {Lewin}, W.~H.~G. and {Miller}, J.~M. and {M{\'e}ndez}, M.},
        title = "{The evolution of the timing properties of the black-hole transient GX 339-4 during its 2002/2003 outburst}",
      journal = {\aap},
         year = 2005,
        month = sep,
       volume = {440},
       number = {1},
        pages = {207-222},
          doi = {10.1051/0004-6361:20042457},
archivePrefix = {arXiv},
       eprint = {astro-ph/0504577},
 primaryClass = {astro-ph},
       adsurl = {https://ui.adsabs.harvard.edu/abs/2005A&A...440..207B}
}

@ARTICLE{belloni2011,
       author = {{Belloni}, T.~M. and {Motta}, S.~E. and {Mu{\~n}oz-Darias}, T.},
        title = "{Black hole transients}",
      journal = {Bulletin of the Astronomical Society of India},
         year = 2011,
        month = sep,
       volume = {39},
       number = {3},
        pages = {409-428},
          doi = {10.48550/arXiv.1109.3388},
archivePrefix = {arXiv},
       eprint = {1109.3388},
 primaryClass = {astro-ph.HE},
       adsurl = {https://ui.adsabs.harvard.edu/abs/2011BASI...39..409B}
}

@ARTICLE{poutanen1996,
       author = {{Poutanen}, Juri and {Svensson}, Roland},
        title = "{The Two-Phase Pair Corona Model for Active Galactic Nuclei and X-Ray Binaries: How to Obtain Exact Solutions}",
      journal = {\apj},
         year = 1996,
        month = oct,
       volume = {470},
        pages = {249},
          doi = {10.1086/177865},
archivePrefix = {arXiv},
       eprint = {astro-ph/9605073},
 primaryClass = {astro-ph},
       adsurl = {https://ui.adsabs.harvard.edu/abs/1996ApJ...470..249P}
}

@ARTICLE{dage2025,
       author = {{Dage}, Kristen C. and {Panurach}, Teresa and {Oh}, Kwangmin and {Sudha}, Malu and {Armas Padilla}, Montserrat and {Bahramian}, Arash and {Cackett}, Edward M. and {Galvin}, Timothy J. and {Heinke}, Craig O. and {Ludlam}, Renee and {Mahida}, Angiraben D. and {Plotkin}, Richard M. and {Russell}, Thomas D. and {Sett}, Susmita and {Saikia}, Payaswini and {Shaw}, Aarran W. and {Tetarenko}, Alexandra J.},
        title = "{Radio Continuum Studies of Ultracompact and Short Orbital Period X-Ray Binaries}",
      journal = {\apj},
         year = 2025,
        month = jul,
       volume = {988},
       number = {1},
          eid = {131},
        pages = {131},
          doi = {10.3847/1538-4357/adea41},
archivePrefix = {arXiv},
       eprint = {2507.00345},
 primaryClass = {astro-ph.HE},
       adsurl = {https://ui.adsabs.harvard.edu/abs/2025ApJ...988..131D}
}

@ARTICLE{sharma2025,
       author = {{Sharma}, Rahul and {Jain}, Chetana and {Paul}, Biswajit and {Beri}, Aru},
        title = "{Sidebands to mHz QPOs in 4U 1626-67 in the second spin-down state}",
      journal = {\mnras},
         year = 2025,
        month = apr,
       volume = {538},
       number = {2},
        pages = {1046-1054},
          doi = {10.1093/mnras/staf379},
archivePrefix = {arXiv},
       eprint = {2503.05444},
 primaryClass = {astro-ph.HE},
       adsurl = {https://ui.adsabs.harvard.edu/abs/2025MNRAS.538.1046S}
}

@ARTICLE{mohammed2024,
       author = {{Tobrej}, Mohammed and {Tamang}, Ruchi and {Rai}, Binay and {Ghising}, Manoj and {Paul}, Bikash Chandra},
        title = "{The ongoing spin-down episode of 4U 1626-67}",
      journal = {\mnras},
         year = 2024,
        month = feb,
       volume = {528},
       number = {2},
        pages = {3550-3558},
          doi = {10.1093/mnras/stae256},
archivePrefix = {arXiv},}

@ARTICLE{bult2017,
       author = {{Bult}, Peter},
        title = "{The Stochastic X-Ray Variability of the Accreting Millisecond Pulsar MAXI J0911-655}",
      journal = {\apj},
         year = 2017,
        month = mar,
       volume = {837},
       number = {1},
          eid = {61},
        pages = {61},
          doi = {10.3847/1538-4357/aa607f},
archivePrefix = {arXiv},
       eprint = {1702.04182},
 primaryClass = {astro-ph.HE},
       adsurl = {https://ui.adsabs.harvard.edu/abs/2017ApJ...837...61B}
}

@ARTICLE{sharma2025_4u1916,
       author = {{Sharma}, Rahul},
        title = "{Broadband spectral and emission geometry analysis of XB 1916{\textendash}053 with Chandra and NuSTAR}",
      journal = {Journal of High Energy Astrophysics},
         year = 2025,
        month = jul,
       volume = {47},
          eid = {100376},
        pages = {100376},
          doi = {10.1016/j.jheap.2025.100376},
       adsurl = {https://ui.adsabs.harvard.edu/abs/2025JHEAp..4700376S}
}

@ARTICLE{smale1988,
       author = {{Smale}, Alan P. and {Mason}, Keith O. and {White}, Nick E. and {Gottwald}, Manfred},
        title = "{X-ray observations of the 50-min dipping source XB 1916-053.}",
      journal = {\mnras},
         year = 1988,
        month = jun,
       volume = {232},
        pages = {647-660},
          doi = {10.1093/mnras/232.3.647},
       adsurl = {https://ui.adsabs.harvard.edu/abs/1988MNRAS.232..647S}
}

\onecolumn
\begin{appendix}

\section{Log of observations}
\label{sec:log}
\footnotesize
This section provides the log of the \nus\ observations, that were analysed in this work. Each table lists: \\
-- the observation start time;\\
-- the exposure time for FPMA after filtering for type-I X-ray bursts and background flaring events;\\
-- the source background-subtracted count rate in the 3--15\,keV energy band for FPMA. We chose this energy band because all the sources are detected between 3 and 15\,keV. The only exception is 47\,Tuc X$-$9: this object is the dominant source of X-rays above 6\,keV in the cluster, therefore we considered the 6--15\,keV interval to minimise the contamination \citep{bahramian2017}; \\
-- 0.8-30\,keV luminosity from the unabsorbed flux obtained from the best-fitting model and distance reported in Table\,\ref{tab:properties};\\ 
-- references to the papers where the observation was analysed for the first time for either spectral or timing studies.

\begin{table}[th!]
\centering
\caption{X-ray observation log.}
\label{tab:obsX}
\footnotesize
\begin{tabular}{cccccc}
\hline
\hline
Obs.ID & Start day (TT)  & Exp. & Source count rate & $L_{\rm X,0.8-30keV}$ & Reference for first use  \\
   &  (YYYY-MM-DD hh:mm:ss)  & (ks) & (counts s$^{-1}$) & (\lum) & of NuSTAR observation \\
\hline
\multicolumn{6}{c}{4U\,1820-303} \\
80001011002 & 2013-07-08 06:06:07 & 1.8 & 100.8$\pm$0.2 & (5.87$\pm$0.07)$\times$10$^{37}$ & \cite{Mondal2016} \\
90401323002 & 2018-05-19 05:31:09 & 11.8 & 198.8$\pm$0.1 & (1.074$\pm$0.001)$\times$10$^{38}$ & \cite{Koliopanos2021} \\
80702318001 & 2021-06-19 10:31:09 & 13.5 & 157.1$\pm$0.1 & (1.213$\pm$0.002)$\times$10$^{38}$ & This work \\
80702318003 & 2021-10-08 19:36:09 & 17.3 & 64.7$\pm$0.1 & (4.06$\pm$0.01)$\times$10$^{37}$ &  This work \\
80702318005 & 2022-03-29 11:26:09 & 18.9 & 115.3$\pm$0.1 & (6.69$\pm$0.01)$\times$10$^{37}$ & This work \\
30802009002 & 2022-04-14 04:06:09 & 15.3 & 149.3$\pm$0.1 & (8.38$\pm$0.01)$\times$10$^{37}$ & \cite{marino2023} \\
30802009004 & 2022-05-31 13:56:09 & 14.6 & 174.9$\pm$0.1 & (1.075$\pm$0.002)$\times$10$^{38}$ & \cite{marino2023} \\
30802009006 & 2022-07-10 22:36:09 & 20.1 & 65.6$\pm$0.1 & (4.24$\pm$0.01)$\times$10$^{37}$ & \cite{marino2023}  \\
90802327002 & 2022-10-12 14:31:09 & 16.5 & 129.0$\pm$0.1 & (8.38$\pm$0.02)$\times$10$^{37}$ & \cite{dimarco2023} \\
90902308002 & 2023-04-15 00:01:09 & 16.8 & 175.6$\pm$0.1 & (9.89$\pm$0.02)$\times$10$^{37}$ & \cite{dimarco2023} \\
90902308004 & 2023-04-16 14:36:09 & 15.2 & 139.3$\pm$0.1 & (9.57$\pm$0.03)$\times$10$^{37}$ & \cite{dimarco2023} \\
\hline
\multicolumn{6}{c}{4U\,1543$-$624} \\
30601006002 & 2020-04-19 07:21:09 & 32.3 & 6.69$\pm$0.01 & (6.95$\pm$0.06)$\times$10$^{36}$ & \cite{ludlam2021} \\
\hline 
\multicolumn{6}{c}{47\,Tuc X$-$9} \\
80001084002$^\dagger$ & 2015-02-02 11:51:07 & 17.7 & 0.017$\pm$0.001$^*$ & (1.10$\pm$0.03)$\times$10$^{34}$ & \cite{bahramian2017} \\
80001084004$^\dagger$ & 2015-02-03 08:46:07 & 78.7 & 0.0141$\pm$0.0004$^*$ & (1.10$\pm$0.03)$\times$10$^{34}$ & \cite{bahramian2017} \\
\hline
\multicolumn{6}{c}{IGR\,J17062$-$6143} \\
30101034002 & 2015-05-06 19:26:07 & 69.6 & 1.027$\pm$0.004 & (8.23$\pm$0.07)$\times$10$^{35}$ & \cite{degenaar2017} \\
30201018002 & 2016-09-13 08:46:08 & 66.6 & 0.496$\pm$0.003 & (3.74$\pm$0.03)$\times$10$^{35}$ & \cite{VanDenEijnden2018}  \\ 
30801032002 & 2022-09-19 12:36:09 & 62.7 & 1.127$\pm$0.004 & (1.07$\pm$0.01)$\times$10$^{36}$ & This work \\
\hline
\multicolumn{6}{c}{4U\,1626$-$67} \\
30101029002 & 2015-05-04 12:26:07 & 64.7 & 12.70$\pm$0.01 & (1.149$\pm$0.002)$\times$10$^{37}$ & \cite{dai2017}  \\
90901318002 & 2023-05-02 19:06:09 & 27.4 & 3.86$\pm$0.01 &  (3.04$\pm$0.02)$\times$10$^{36}$ & \cite{sharma2023}  \\  
90901318004 & 2023-05-19 13:01:09 & 18.9 & 3.47$\pm$0.01 &  (2.64$\pm$0.01)$\times$10$^{36}$ & \cite{sharma2025} \\
90901318006 & 2023-06-04 20:56:09 & 18.4 & 3.23$\pm$0.01 &  (2.48$\pm$0.01)$\times$10$^{36}$ & \cite{mohammed2024} \\
90901318008 & 2023-06-22 00:06:09 & 22.3 & 3.08$\pm$0.01 &  (2.26$\pm$0.04)$\times$10$^{36}$ & \cite{mohammed2024} \\
90901318010 & 2023-07-05 21:36:09 & 18.4 & 2.83$\pm$0.01 &  (2.11$\pm$0.01)$\times$10$^{36}$ & \cite{mohammed2024}  \\
\hline 
\multicolumn{6}{c}{MAXI\,J0911$-$655} \\
90201024002 & 2016-05-24 00:31:08 & 57.8 & 2.43$\pm$0.01 & (2.55$\pm$0.01)$\times$10$^{36}$ & \cite{sanna2017}  \\
90201042002 & 2016-11-23 15:26:08 & 33.4 & 3.73$\pm$0.01 & (4.54$\pm$0.03)$\times$10$^{36}$ & \cite{bult2017}  \\
\hline
\multicolumn{6}{c}{IGR\,J16597$-$3704} \\
 90301324001 & 2017-10-26 12:16:09 & 41.4 & 6.12$\pm$0.01 & (3.63$\pm$0.01)$\times$10$^{36}$ & \cite{sanna2018b} \\
\hline
\multicolumn{6}{c}{4U\,1916$-$053} \\ 
90701325002 & 2021-09-01 02:51:09 & 28.5 & 5.02$\pm$0.01 & (3.68$\pm$0.02)$\times$10$^{36}$ & \cite{sharma2025_4u1916} \\
\hline
\multicolumn{6}{c}{4U\,0614+091} \\
30363002002 & 2017-12-01 15:56:09 & 19.2 & 28.03$\pm$0.04 & (2.64$\pm$0.03)$\times$10$^{36}$ & \cite{ludlam2019}  \\
30702009002 & 2021-10-06 05:36:09 & 28.7 & 14.14$\pm$0.02 & (1.23$\pm$0.01)$\times$10$^{36}$ & \cite{moutard2023}  \\ 
30702009004 & 2021-10-09 09:16:09 & 29.4 & 12.06$\pm$0.02 & (1.10$\pm$0.01)$\times$10$^{36}$ & \cite{moutard2023}  \\
30702009006 & 2021-10-11 17:41:09 & 28.7 & 9.94$\pm$0.02 & (8.24$\pm$0.04)$\times$10$^{35}$ & \cite{moutard2023}  \\
30702009008 & 2021-10-13 17:56:09 & 28.6 & 13.51$\pm$0.02 & (1.19$\pm$0.01)$\times$10$^{36}$ & \cite{moutard2023}  \\
30702009010 & 2022-01-19 06:51:09 & 26.5 & 15.02$\pm$0.02 & (1.34$\pm$0.01)$\times$10$^{36}$ & \cite{moutard2023}  \\
\hline 
\multicolumn{6}{c}{Swift\,J1756.9$-$2508} \\
90402313002 & 2018-04-08 08:31:09 & 39.5 & 6.67$\pm$0.01 & (8.73$\pm$0.07)$\times$10$^{36}$ & \cite{sanna2018a} \\
90402313004 & 2018-04-14 02:56:09 & 60.9 & $<$0.34 & --$^\ddagger$ & \cite{sanna2018a}   \\
90501329001 & 2019-06-22 07:51:09 & 36.8 & 2.82$\pm$0.01 & (2.75$\pm$0.01)$\times$10$^{36}$ & \cite{li2021}  \\
\hline
\multicolumn{6}{c}{IGR\,J17494$-$3030} \\
80601309001 & 2020-10-30 23:06:09 & 81.1 & 0.072$\pm$0.001 & (6.59$\pm$0.1)$\times$10$^{34}$ & This work \\
\hline
\hline
\end{tabular} 
\tablefoot{
\tablefoottext{$*$}{The source net count rate is reported in the 6--15 keV band. For details, see the text and \cite{bahramian2017}.}
\tablefoottext{$^\dagger$}{The observations were merged for the analysis.}
\tablefoottext{$^\ddagger$}{We did not derive the luminosity for this epoch since the spectral parameters in quiescence are unknown.}}
\end{table}

\section{\nus\ spectra and fitted models}
\label{sec:nus_spec}
This section reports the \nus\ spectra with the best-fitting models. In each case, we plot the $E \times f(E)$ unfolded spectra and the model with the single components. Post-fit residuals in units of standard deviations are also plotted at the bottom of each panel. Same symbol/colour code is used for all the figures: black dots for FPMA data, grey circle for FPMB data, dashed blue line for the blackbody component, dotted green line for the disk component, dash‑dotted orange line for the thermally Comptonised continuum component and yellow for the iron line. In a few spectra, one or more components do not appear in the plot, because their contributions fall outside the plotted flux range. 

\begin{figure}[h!]
    \centering
    \includegraphics[width=0.46\textwidth]{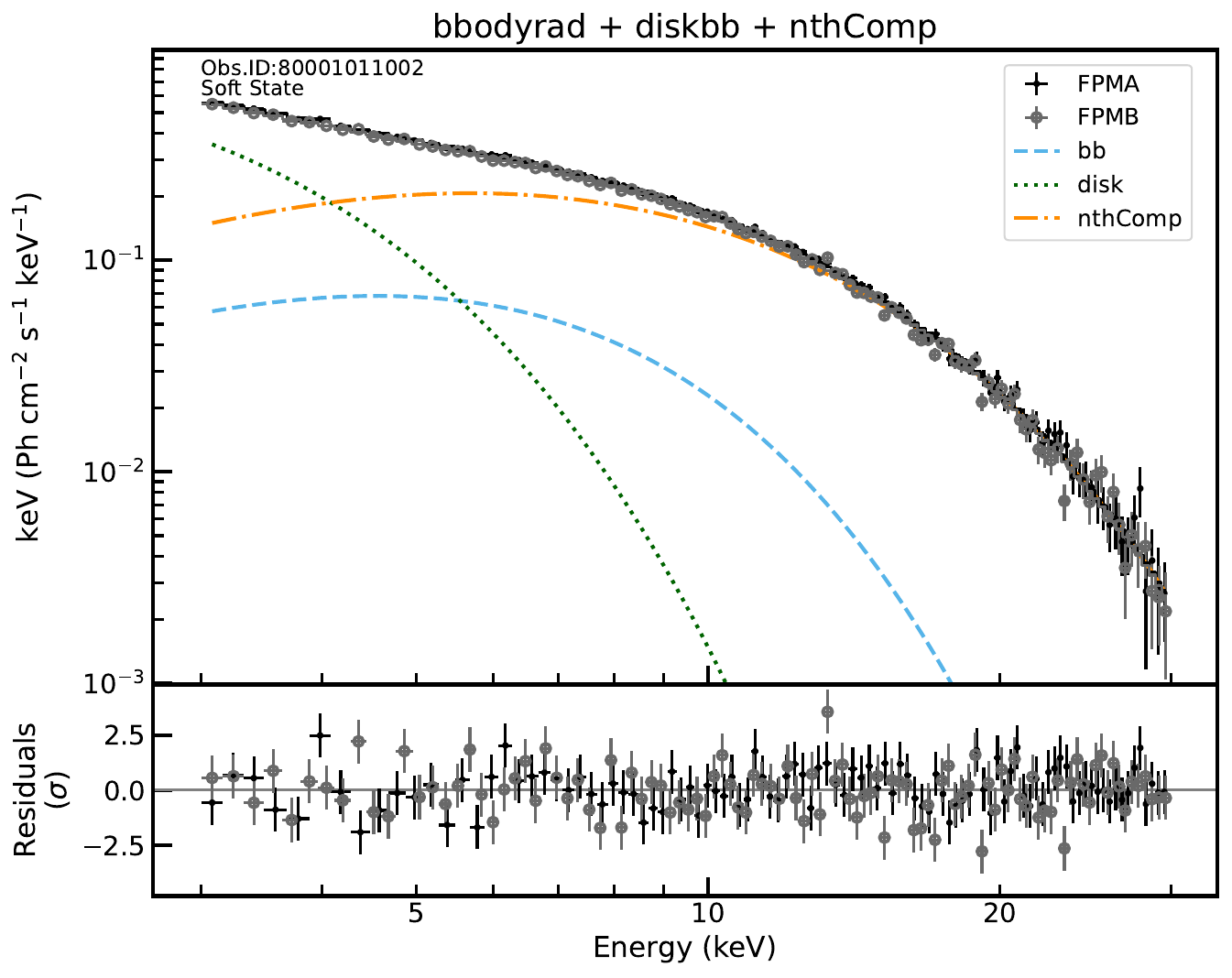}
    \includegraphics[width=0.46\textwidth]{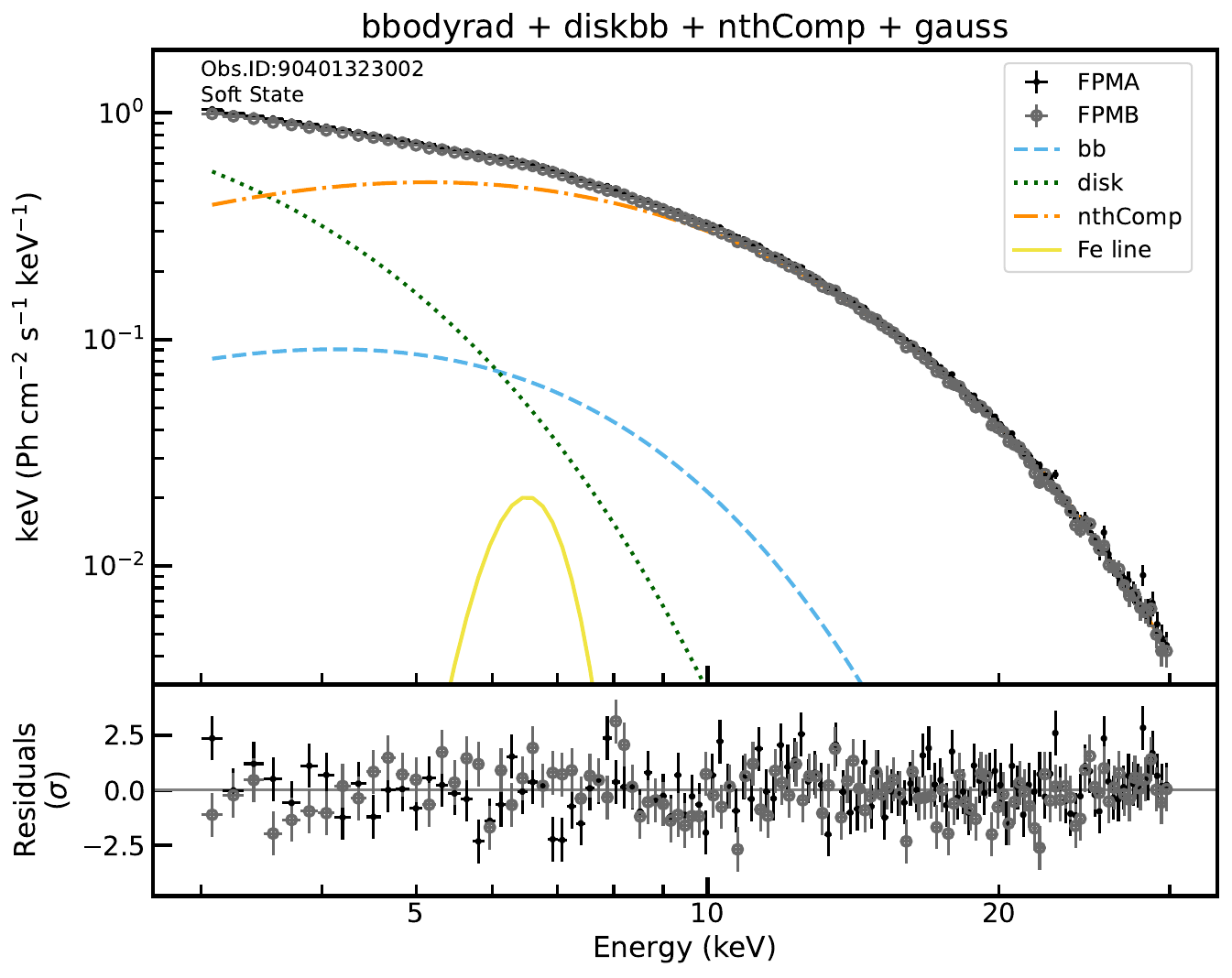}
    \includegraphics[width=0.46\textwidth]{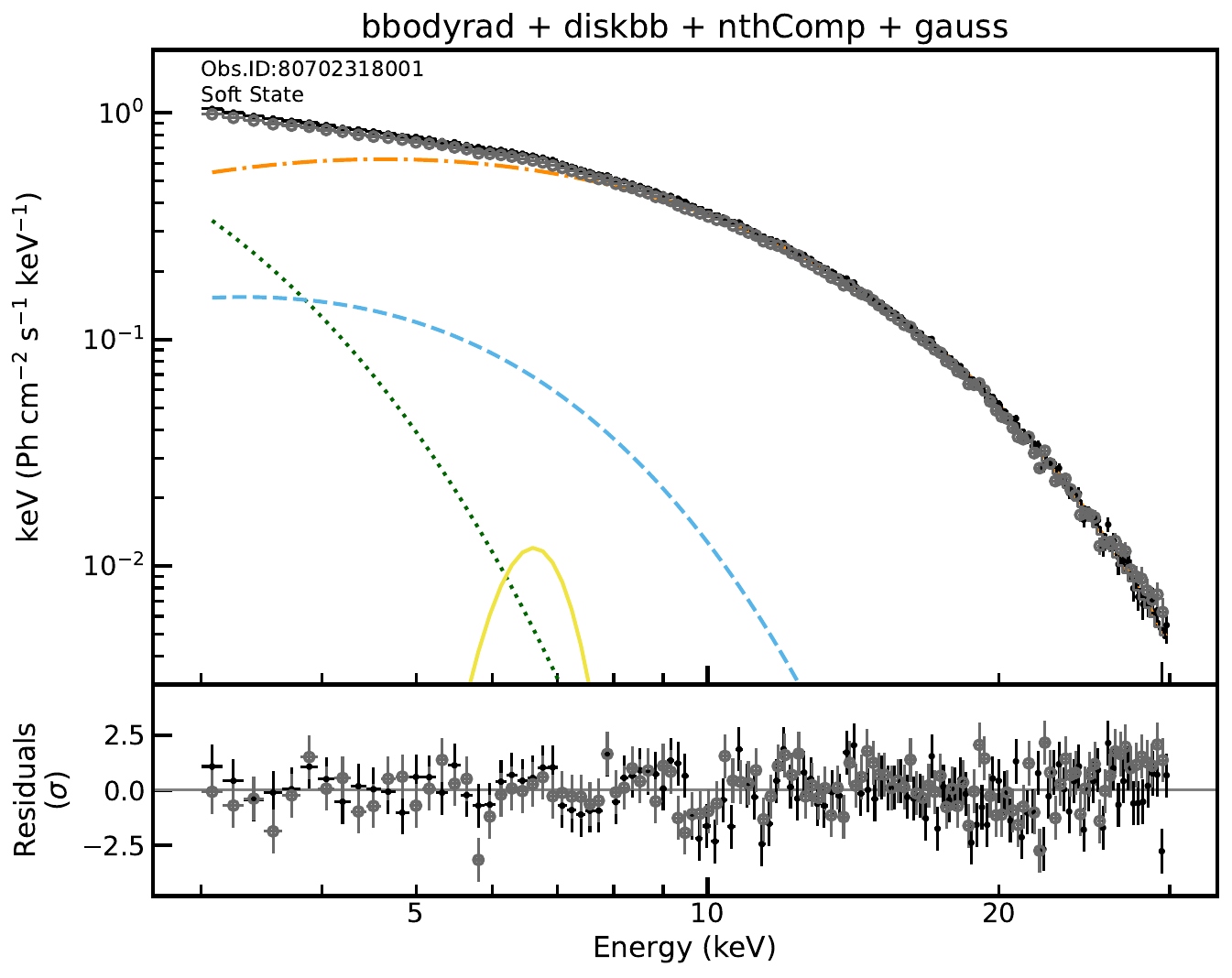}
    \includegraphics[width=0.46\textwidth]{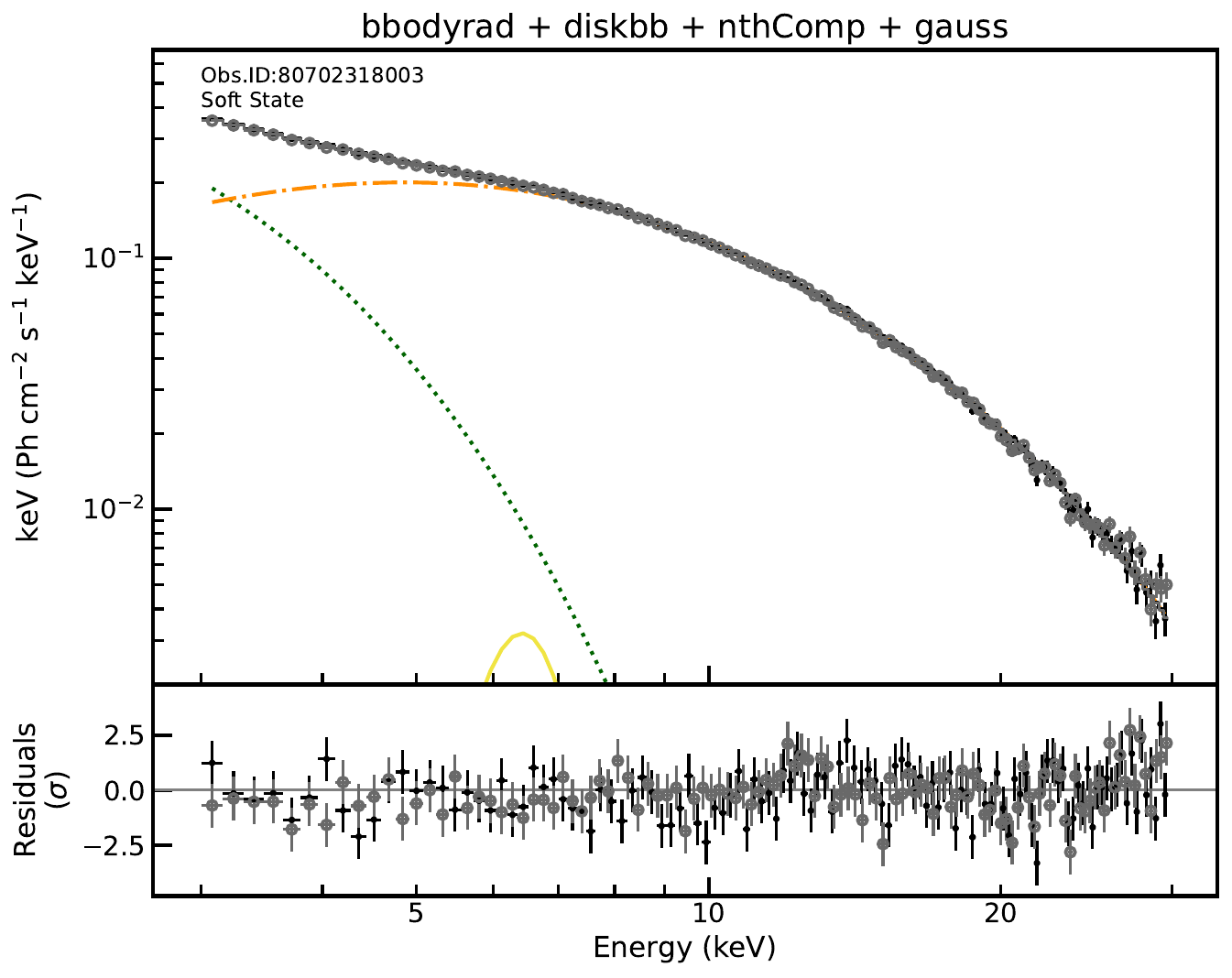}
    \includegraphics[width=0.46\textwidth]{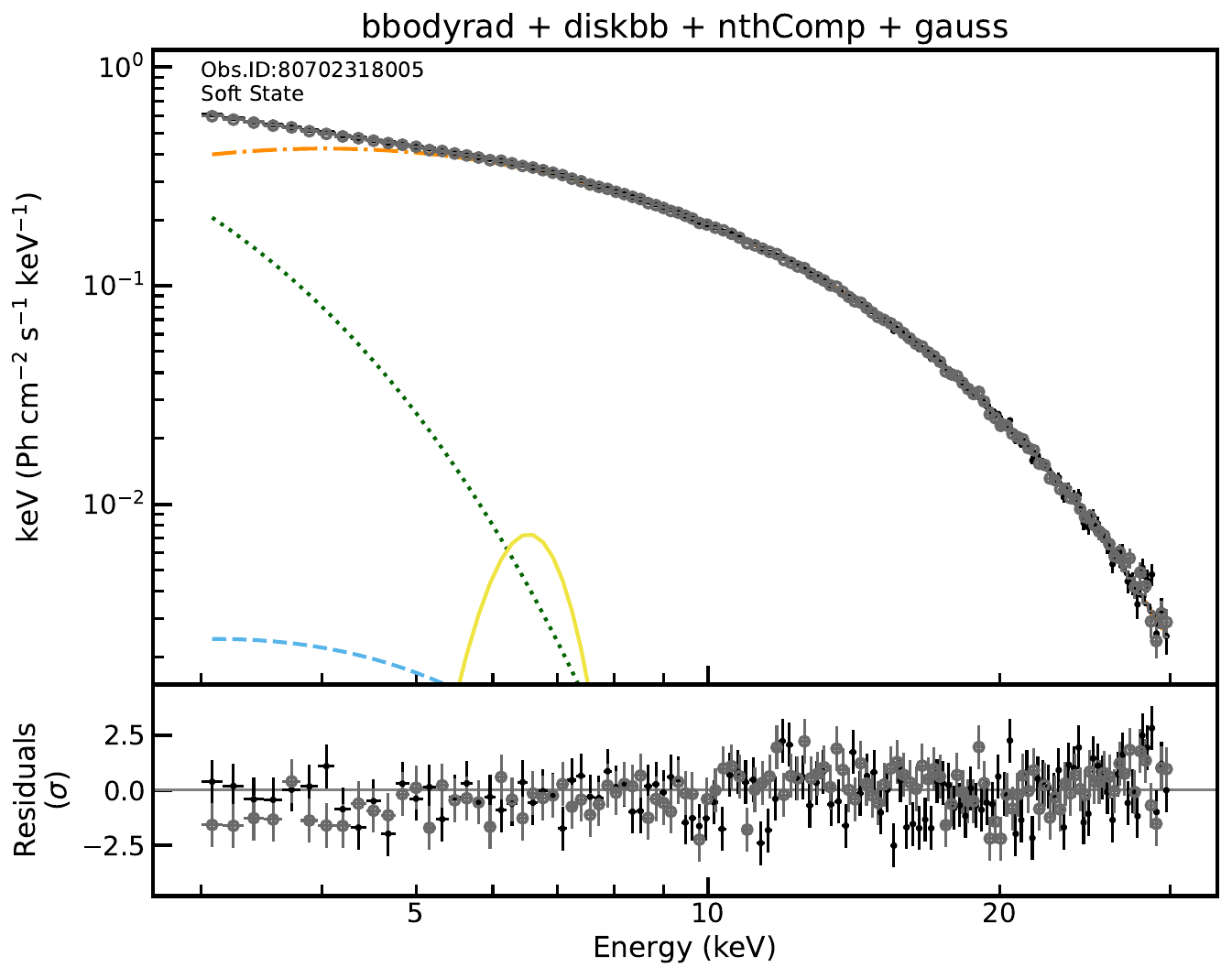}
    \includegraphics[width=0.46\textwidth]{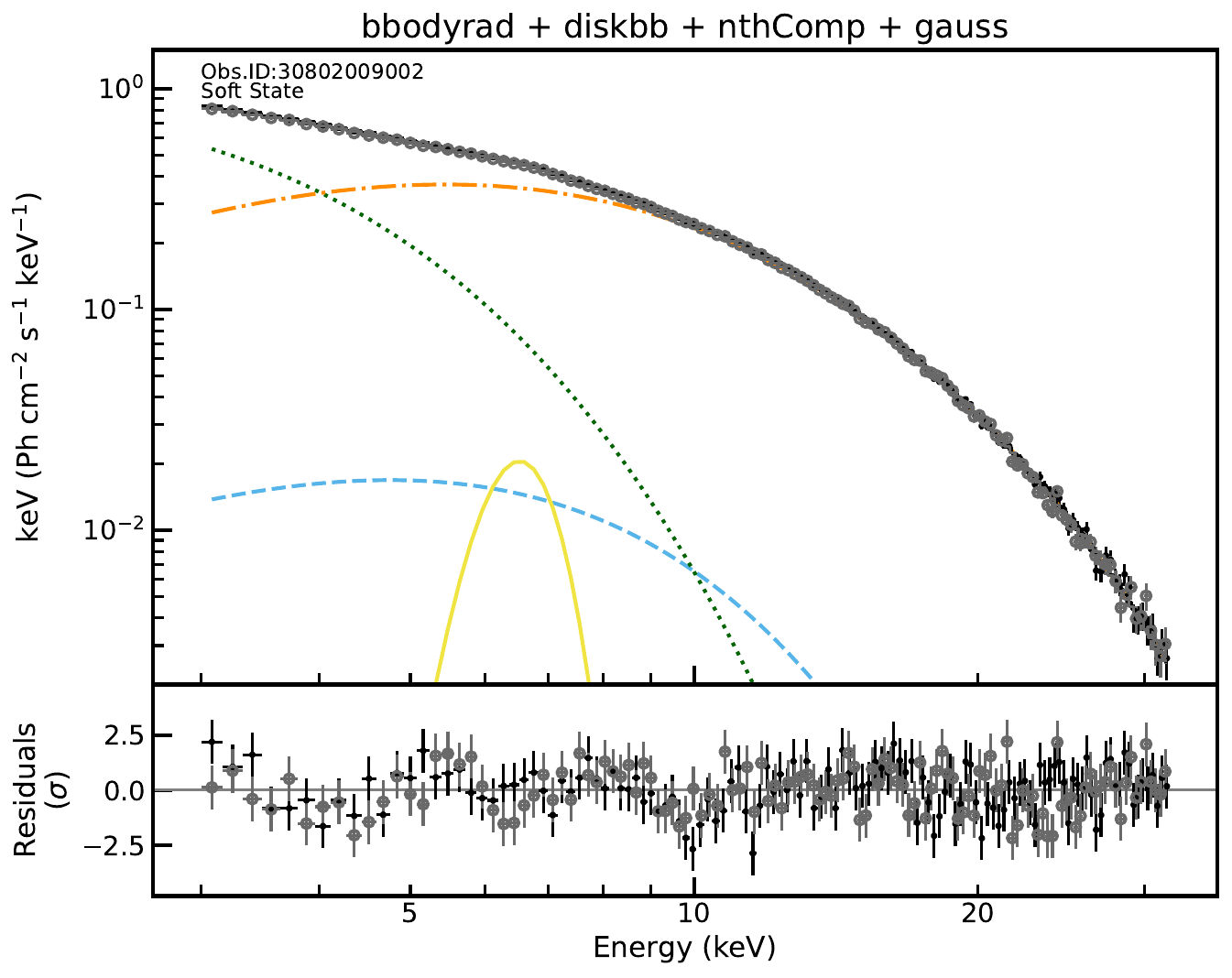}
    \caption{Spectra of 4U\,1820--303.}
    \label{fig:spec_para_1820}
\end{figure}

\begin{figure}
    \ContinuedFloat
    \centering
    \includegraphics[width=0.46\textwidth]{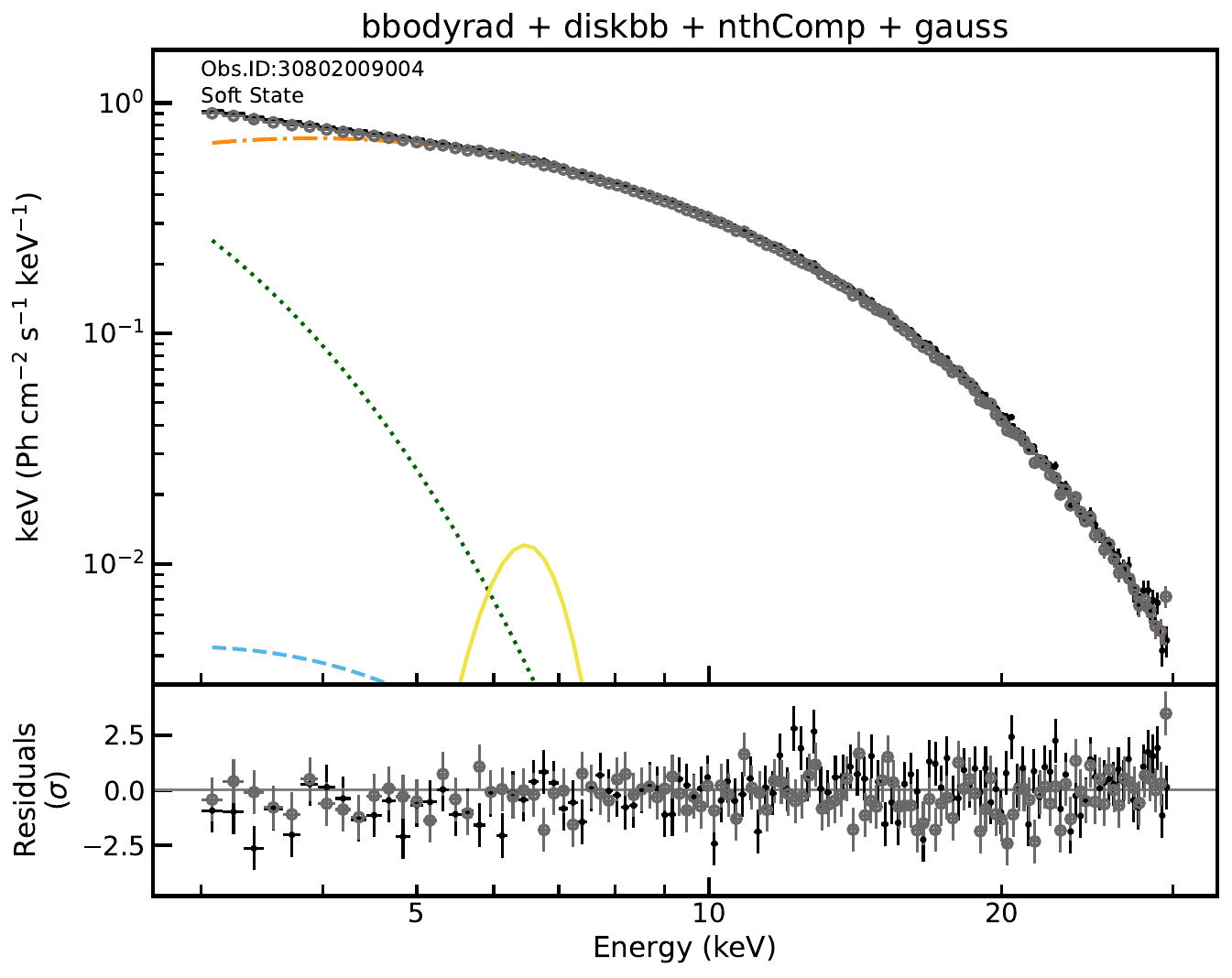}
    \includegraphics[width=0.46\textwidth]{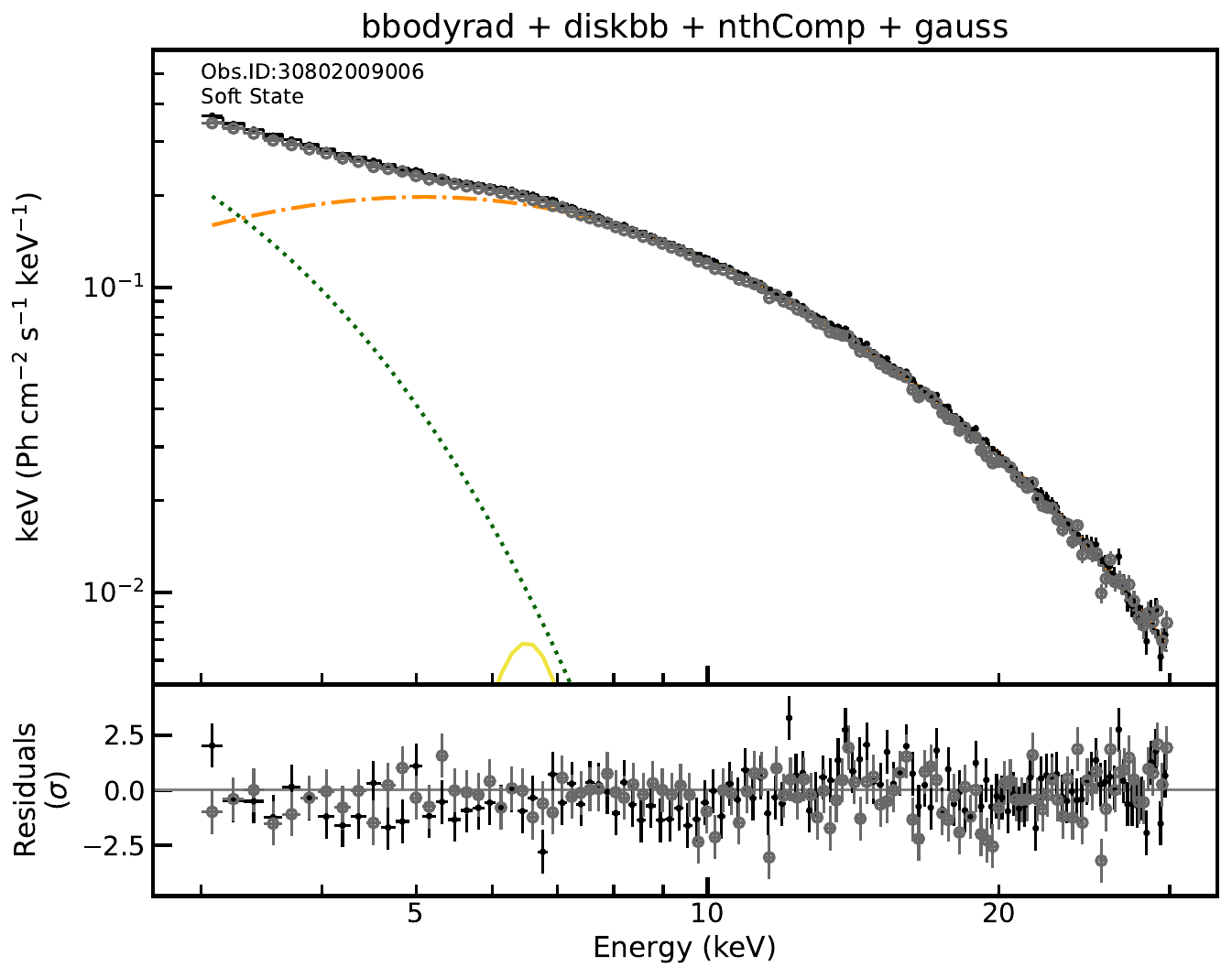}
    \includegraphics[width=0.46\textwidth]{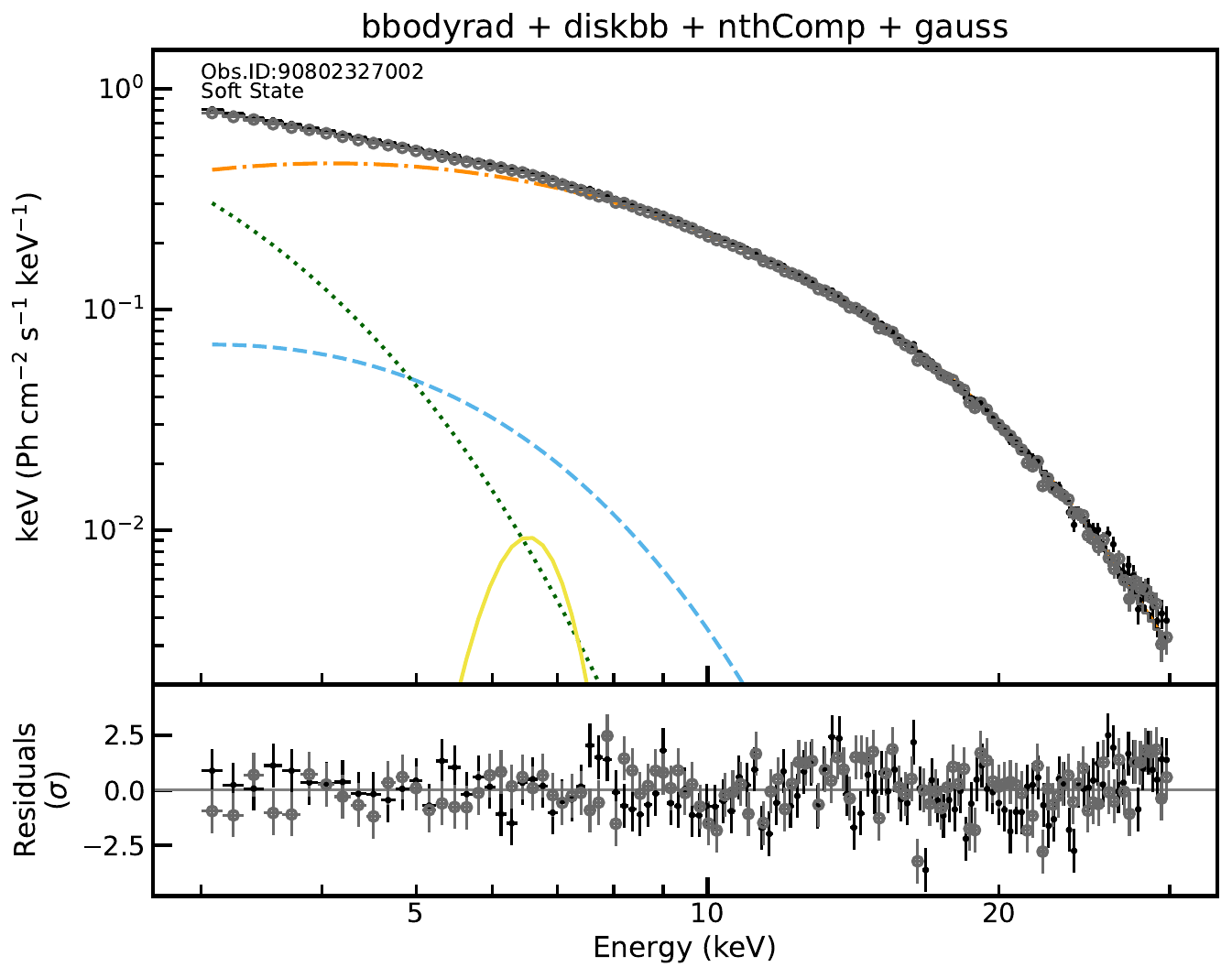}
    \includegraphics[width=0.46\textwidth]{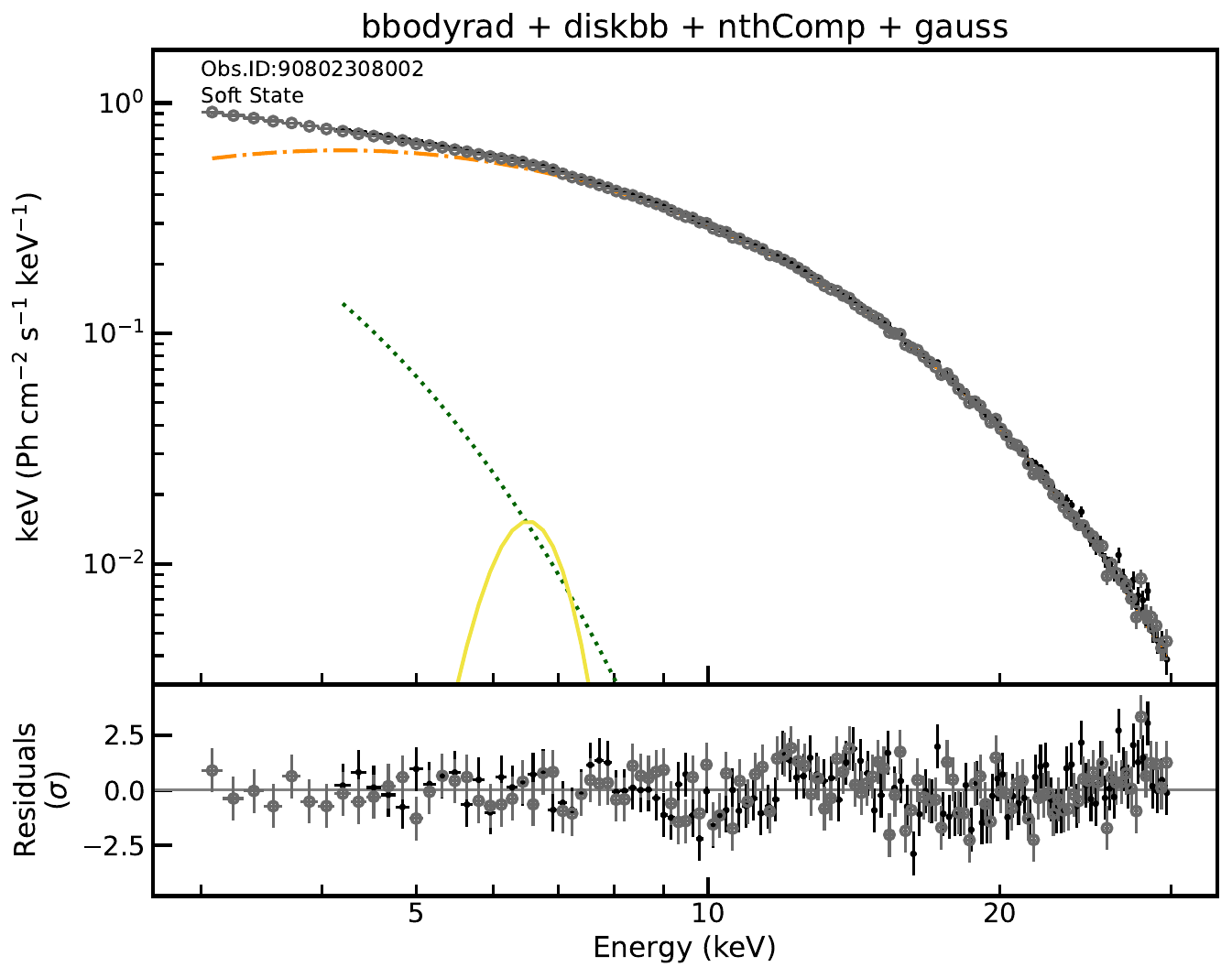}
\includegraphics[width=0.46\textwidth]{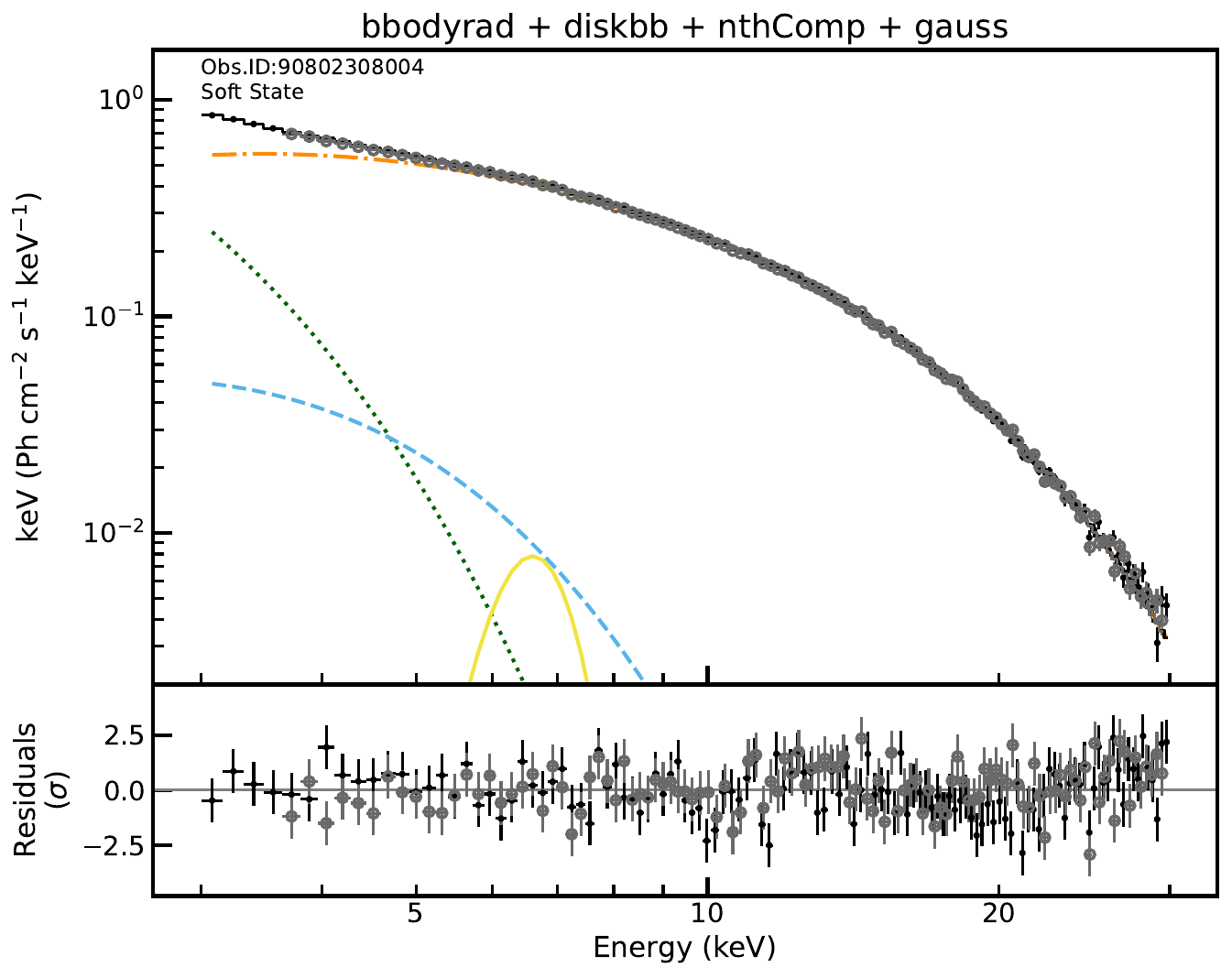}
    \caption{continued.}
\end{figure}

\begin{figure}[th!]
    \centering
    \includegraphics[width=0.46\textwidth]{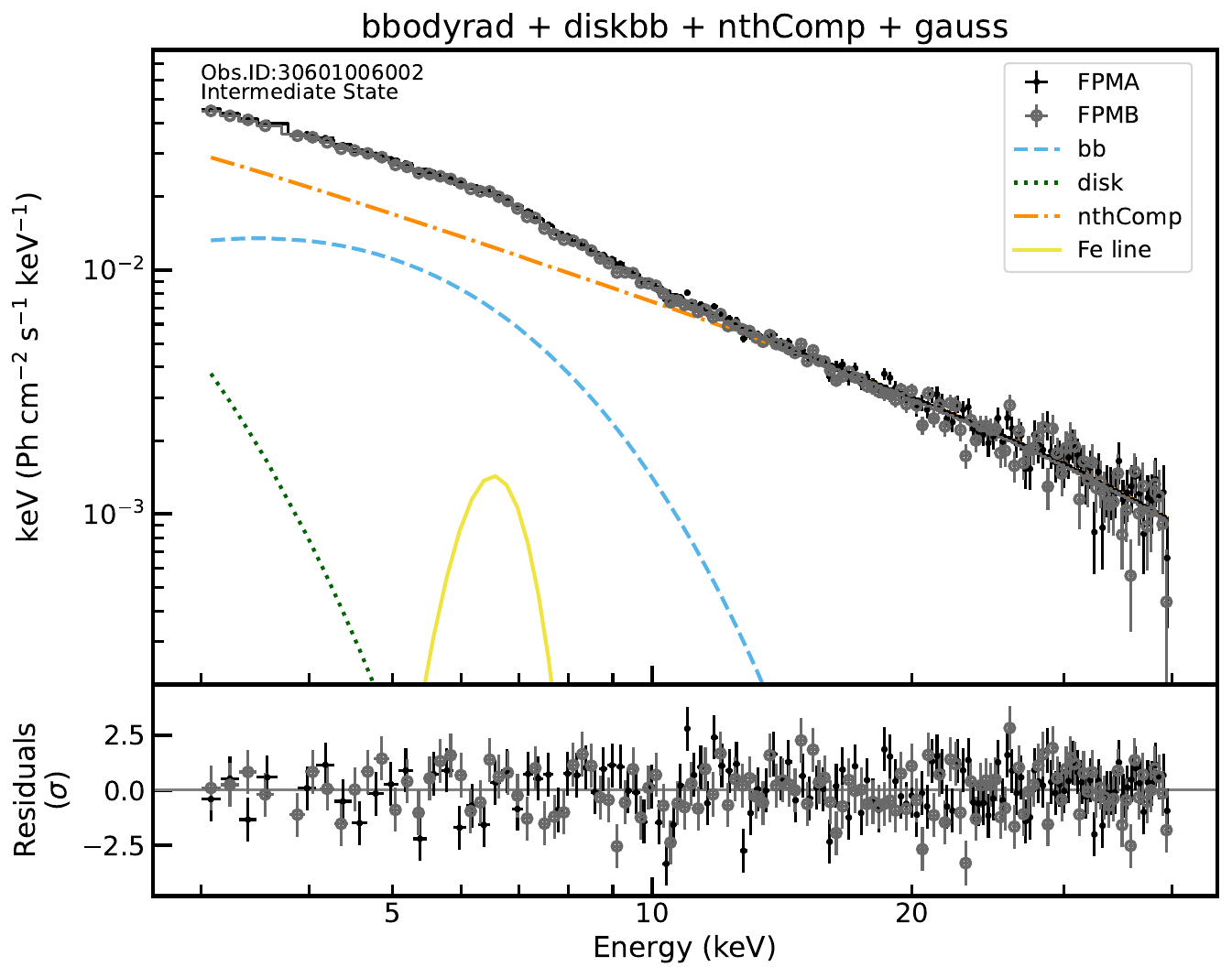}
    \includegraphics[width=0.46\textwidth]{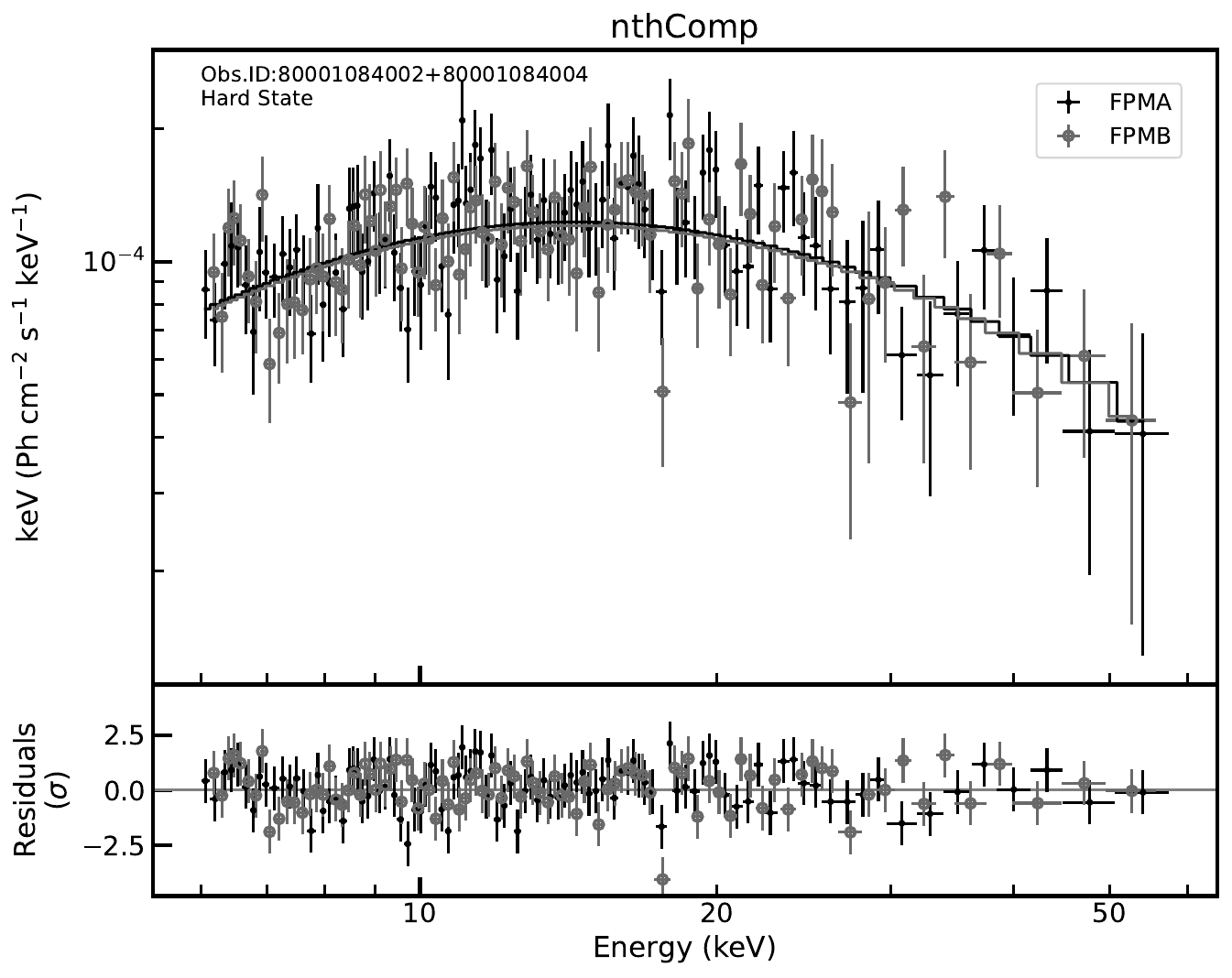}
    \caption{{\it Left:} Spectrum of 4U\,1543--624. {\it Right:} Spectrum of  47\,Tuc X$-$9.}
    \label{fig:spec_para_1543_47tuc}
\end{figure}

\begin{figure}[th!]
    \centering
    \includegraphics[width=0.46\textwidth]{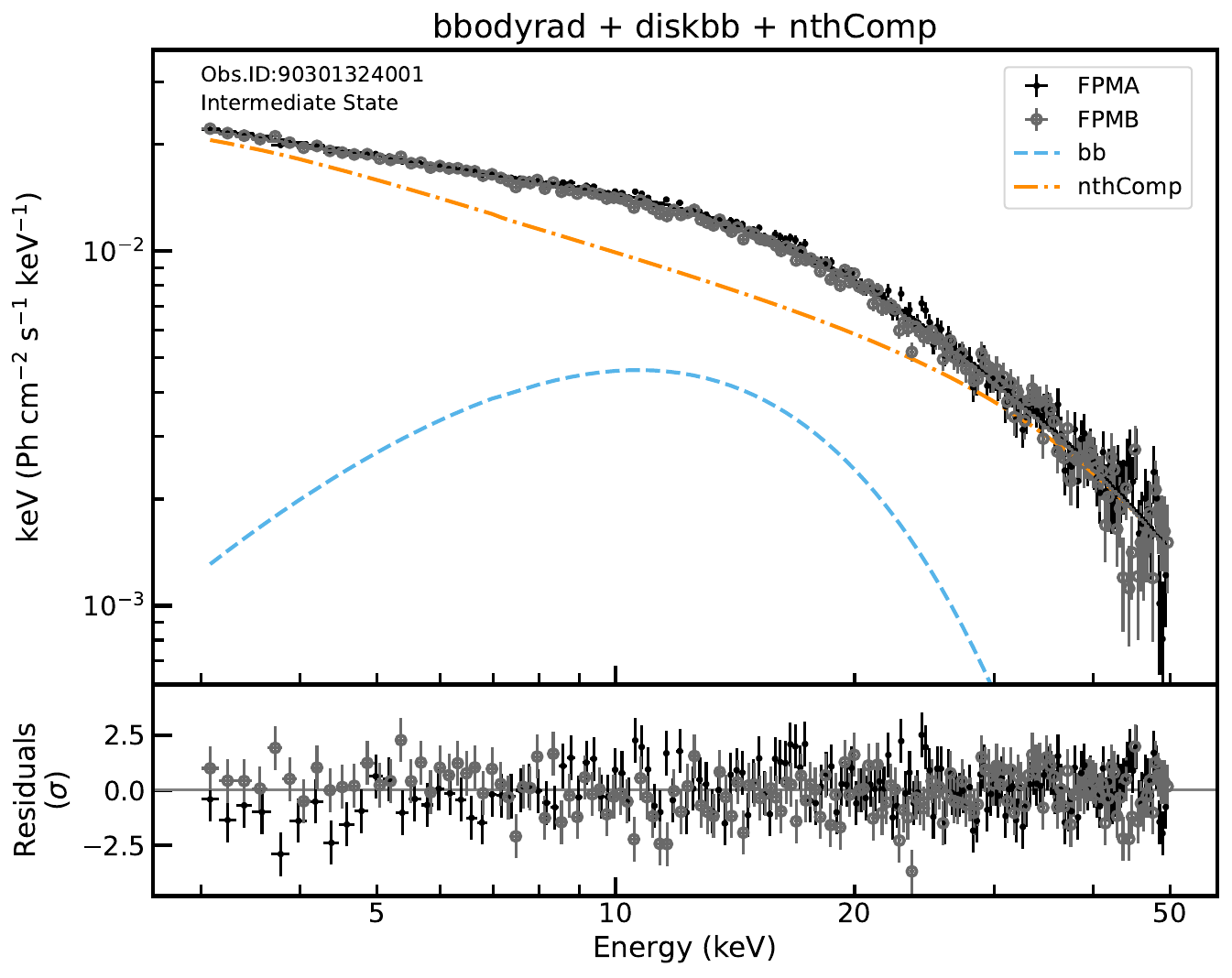}
    \includegraphics[width=0.46\textwidth]{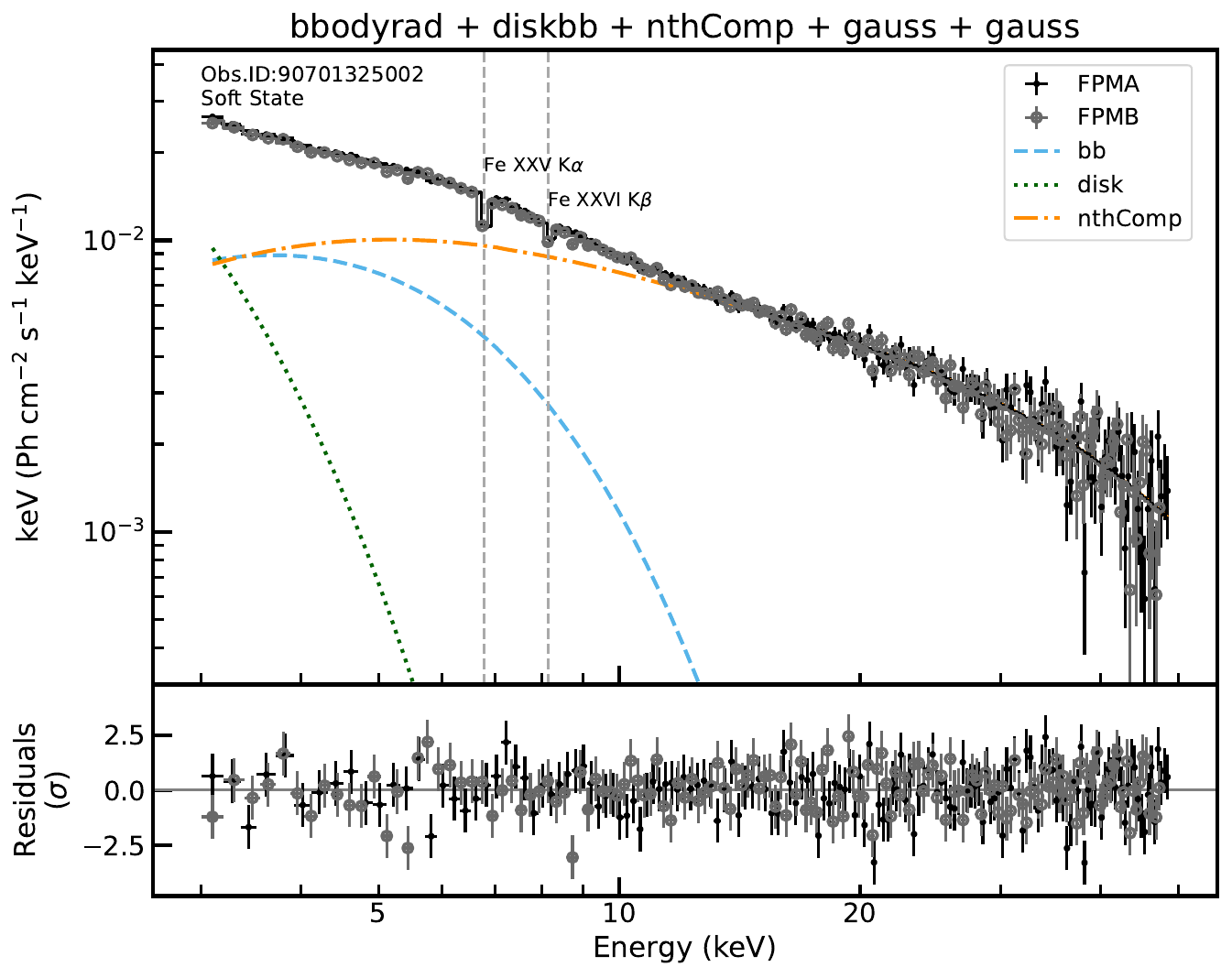}
    \caption{{\it Left:} Spectrum of IGR\,J16597$-$3704. {\it Right:} Spectrum of  4U\,1916$-$053.}
    \label{fig:spec_para_j16597_u1916}
\end{figure}

\begin{figure}[th!]
    \centering
    \includegraphics[width=0.46\textwidth]{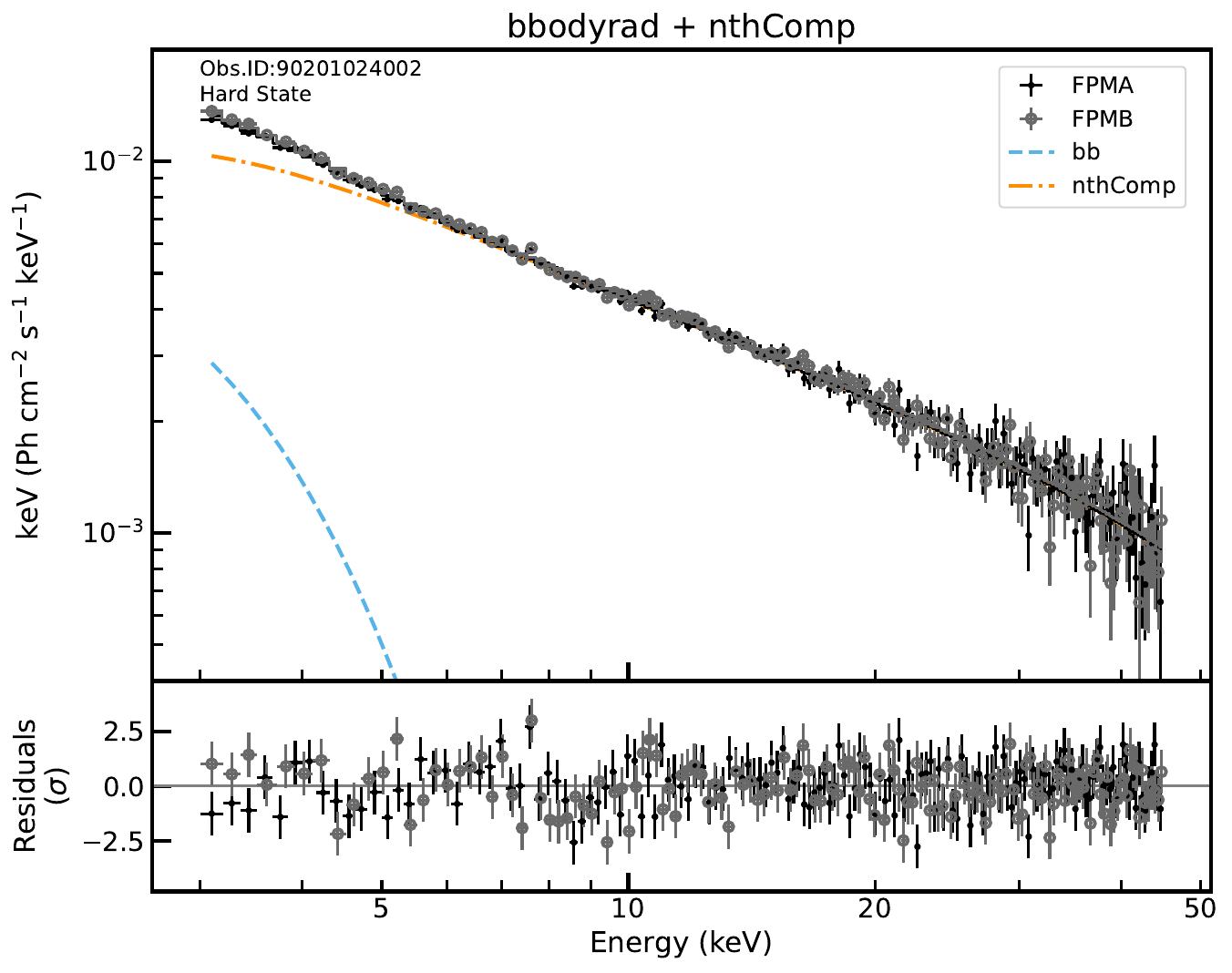}
    \includegraphics[width=0.46\textwidth]{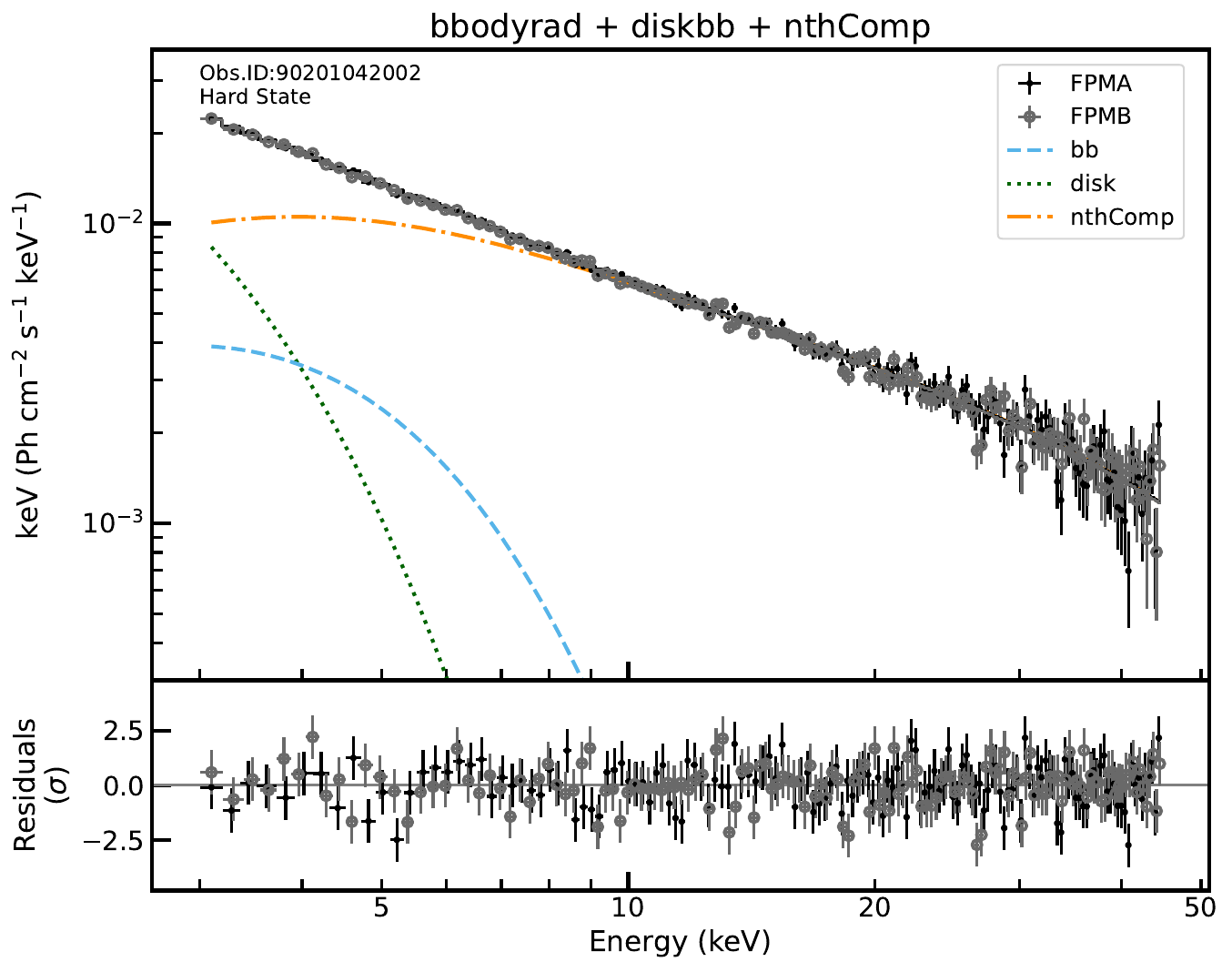}
    \caption{Spectra of MAXI\,J0911$-$655.}
    \label{fig:spec_para_maxi0911}
\end{figure}

\begin{figure}[th!]
    \centering
    \includegraphics[width=0.46\textwidth]{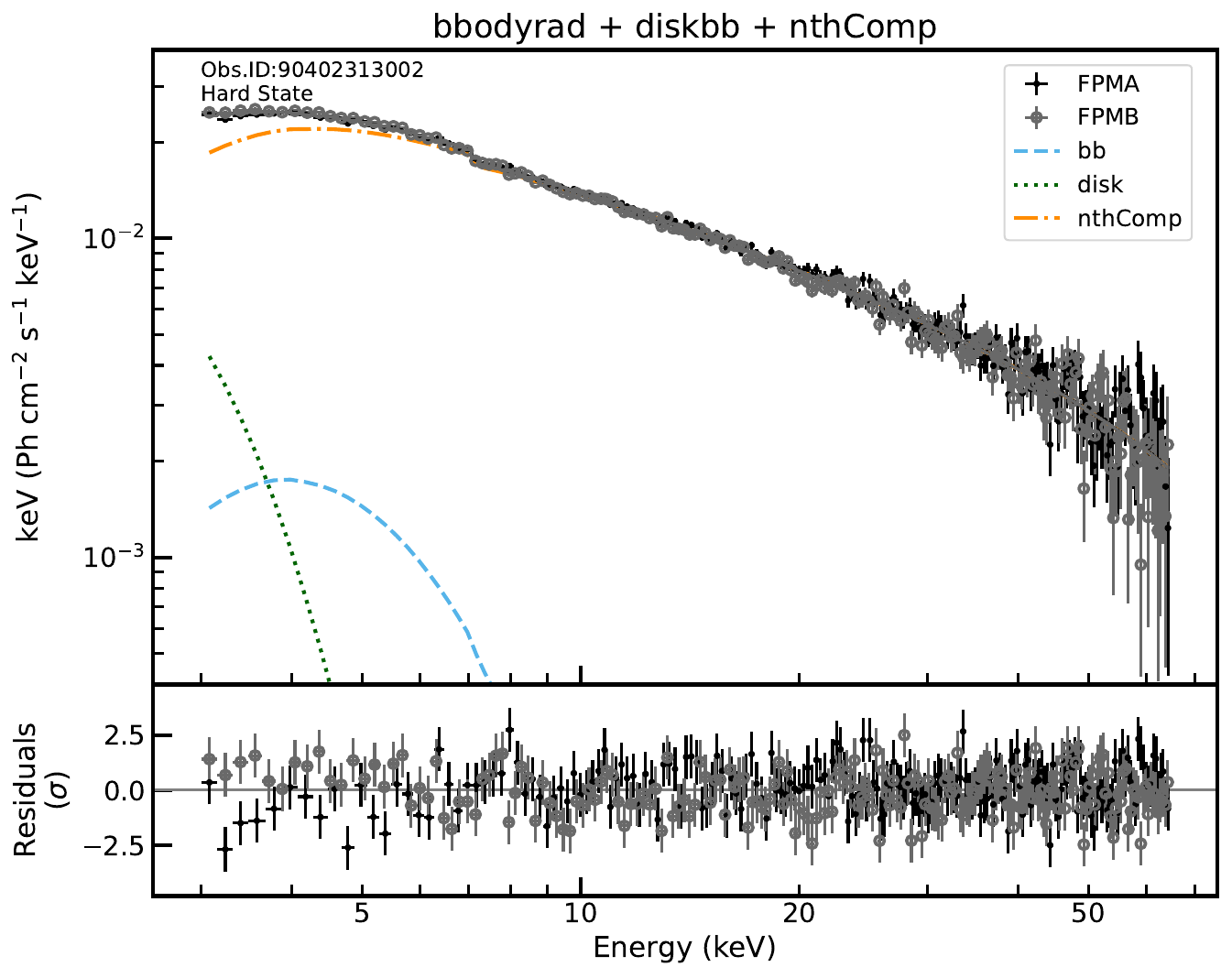}
    \includegraphics[width=0.46\textwidth]{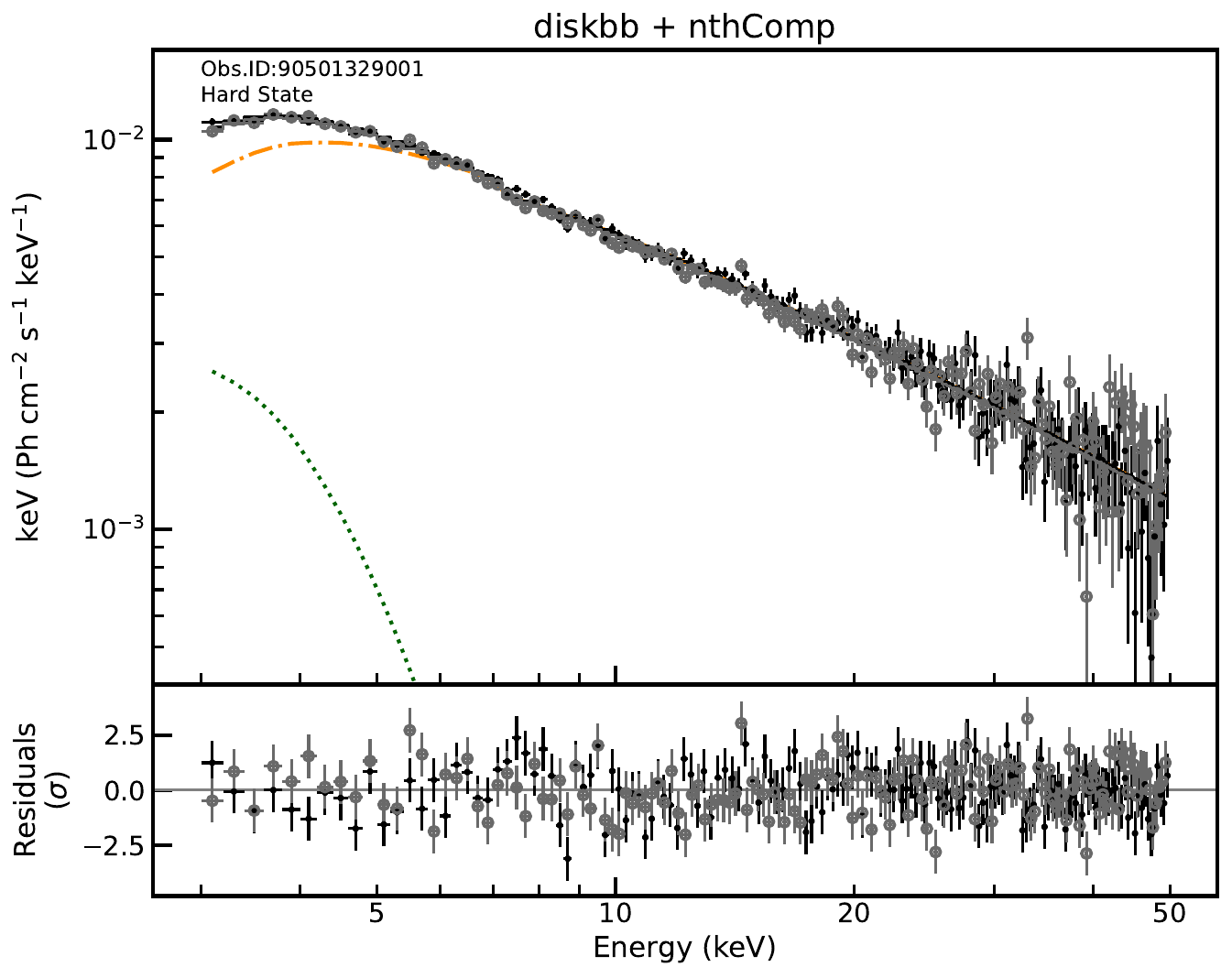}
    \caption{Spectra of Swift\,J1756.9$-$2508.}
    \label{fig:spec_para_swiftj1756}
\end{figure}

\begin{figure}[th!]
    \centering
    \includegraphics[width=0.46\textwidth]{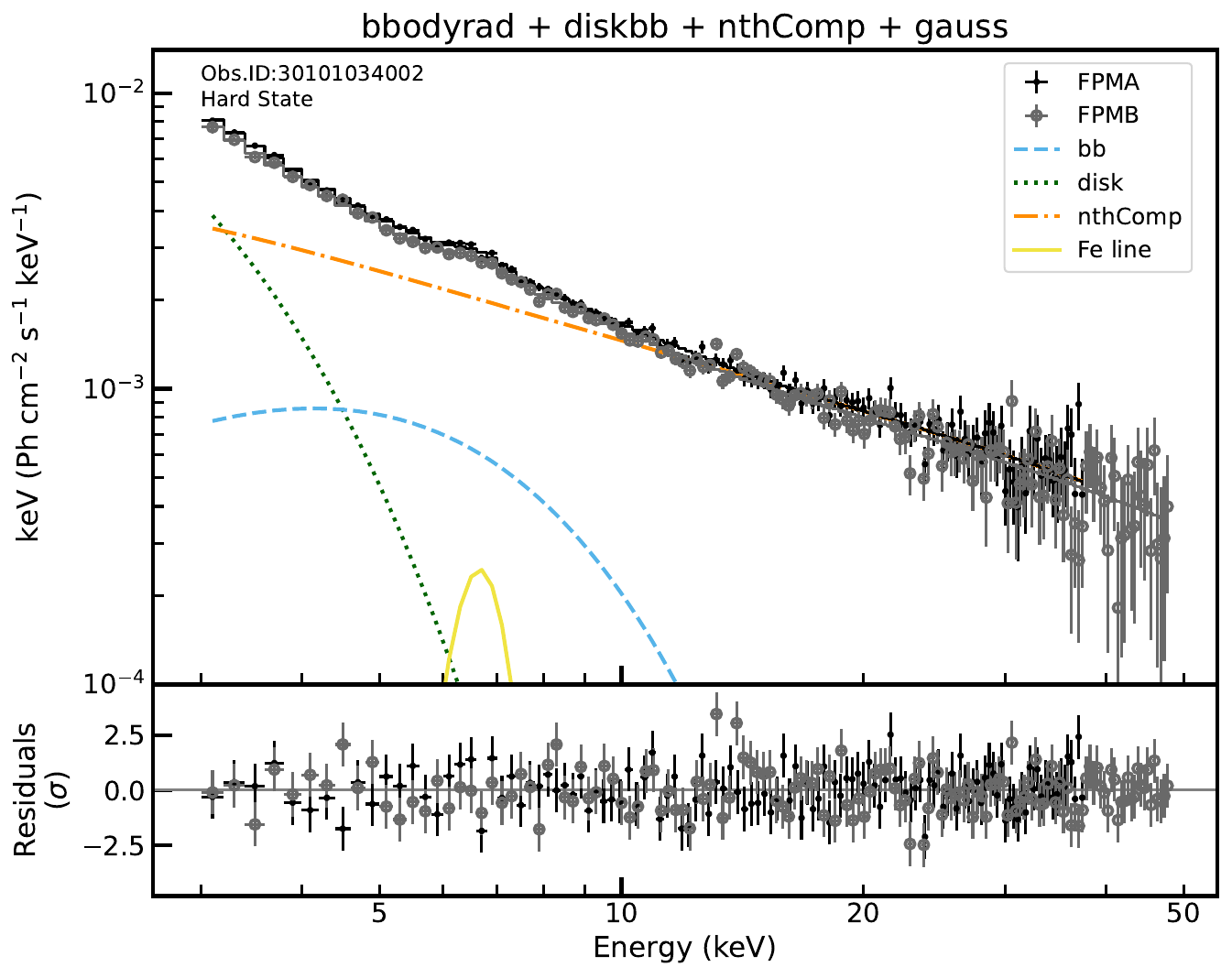}
    \includegraphics[width=0.46\textwidth]{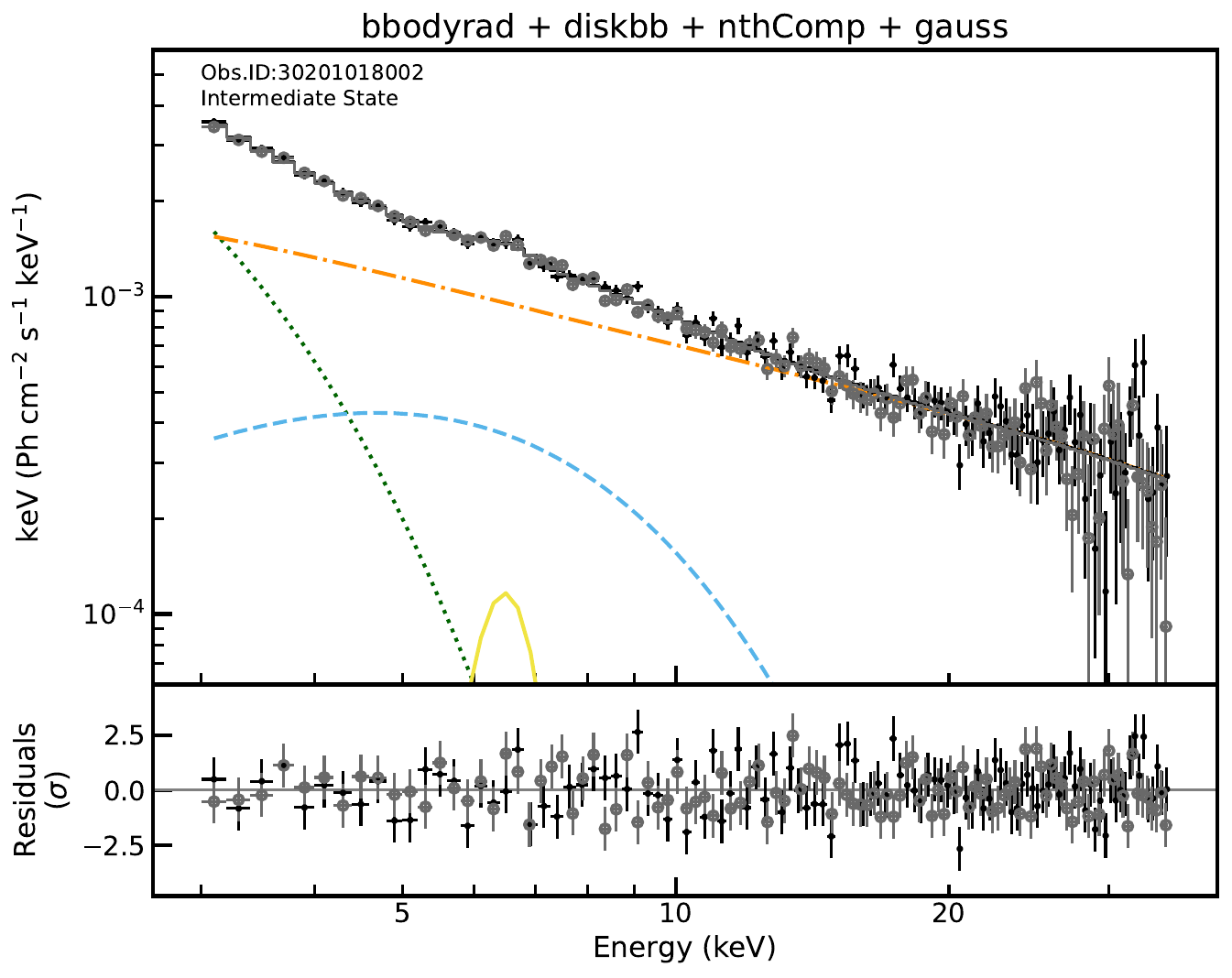}
    \caption{Spectra of IGR\,J17062$-$6143 for the first ({\it left}) and second ({\it right}) epoch.}
    \label{fig:spec_para_j17062}
\end{figure}    
    
\begin{figure}[th!]
    \centering
        \includegraphics[width=0.46\textwidth]{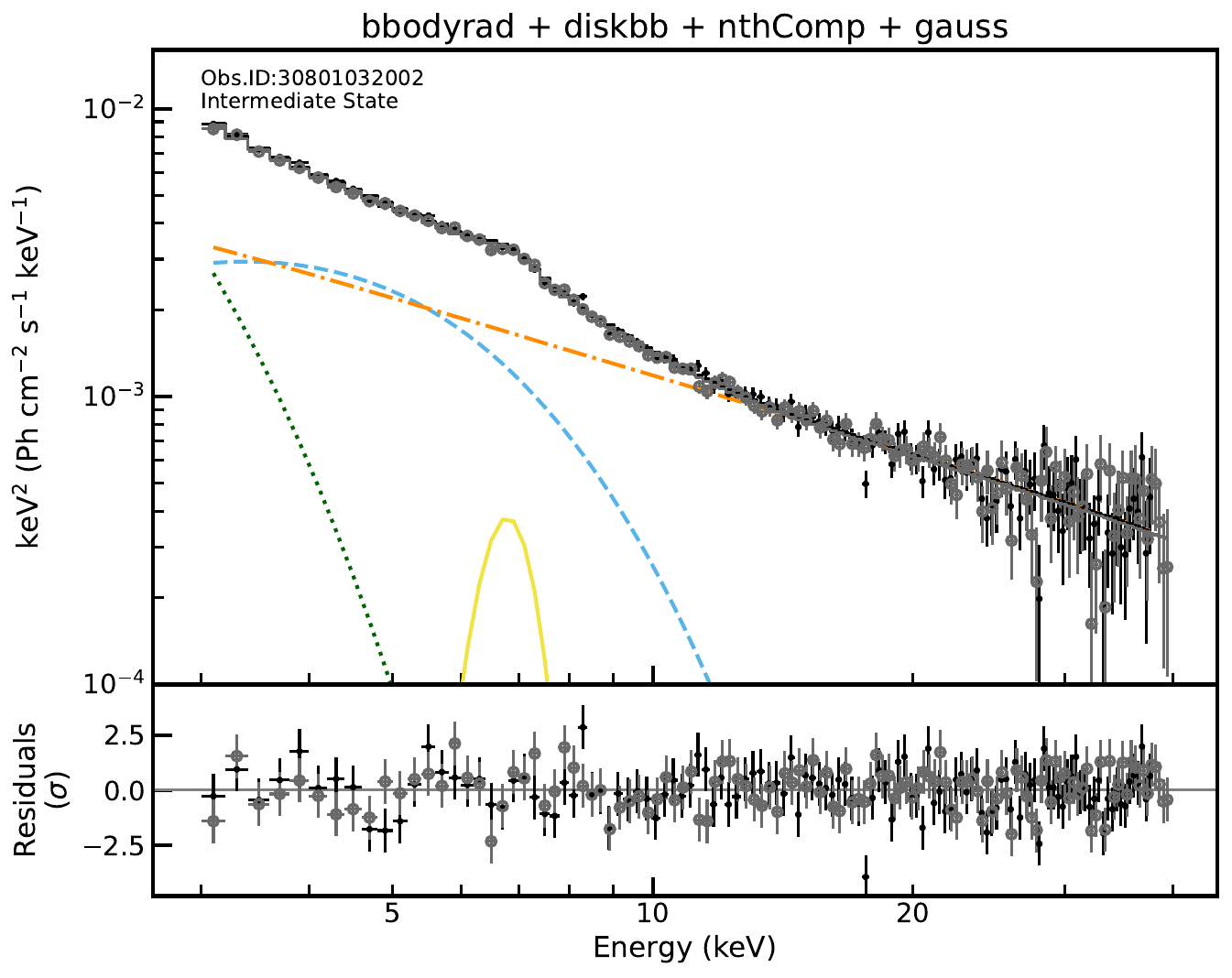}
    \includegraphics[width=0.46\textwidth]{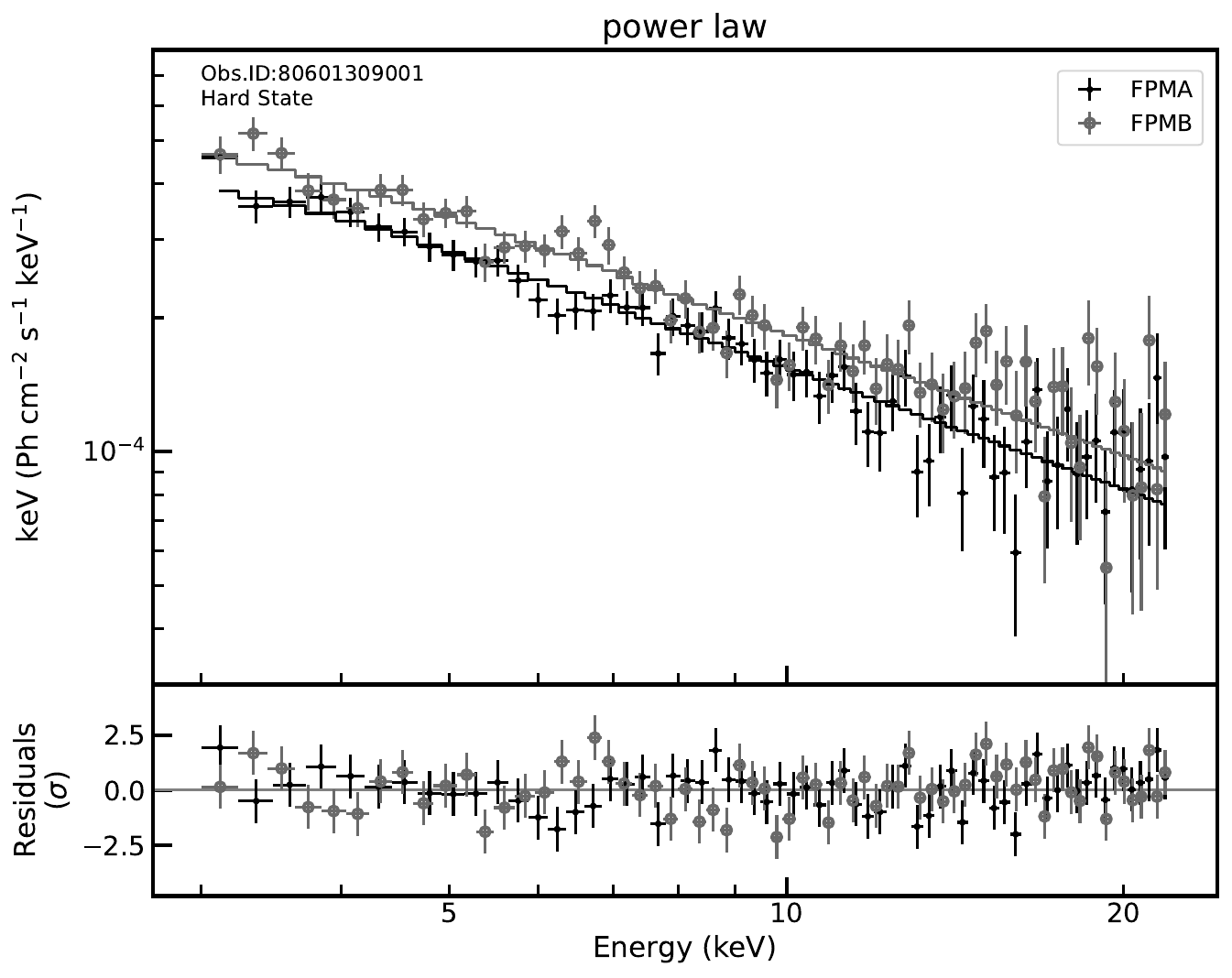}
    \caption{{\it Left}: Spectrum of IGR\,J17062$-$6143 for the third epoch. {\it Right}: Spectrum of IGR\,J17494$-$3030.}
    \label{fig:spec_para_j17494}
\end{figure}

\begin{figure}[th!]
    \centering
    \includegraphics[width=0.46\textwidth]{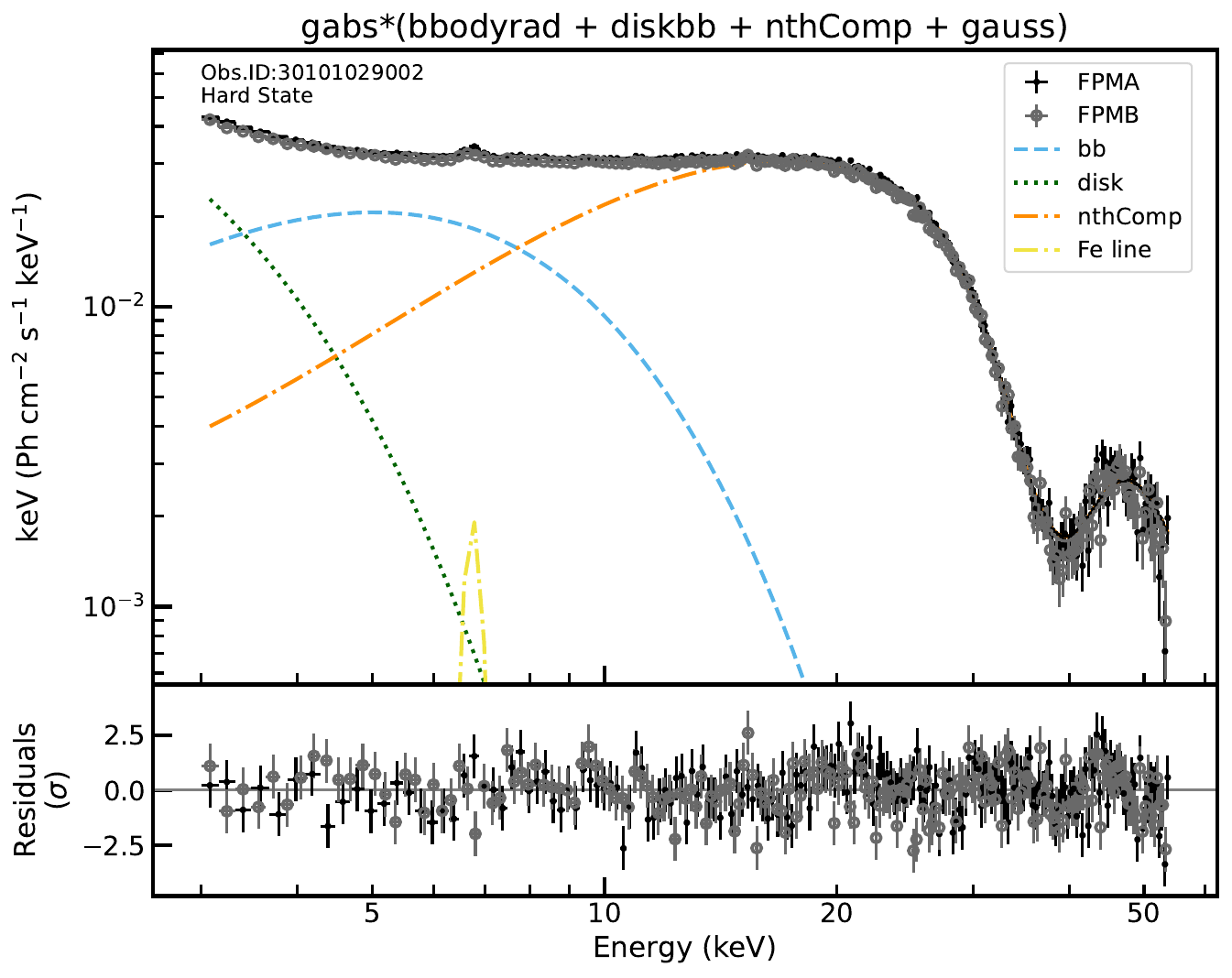}
    \includegraphics[width=0.46\textwidth]{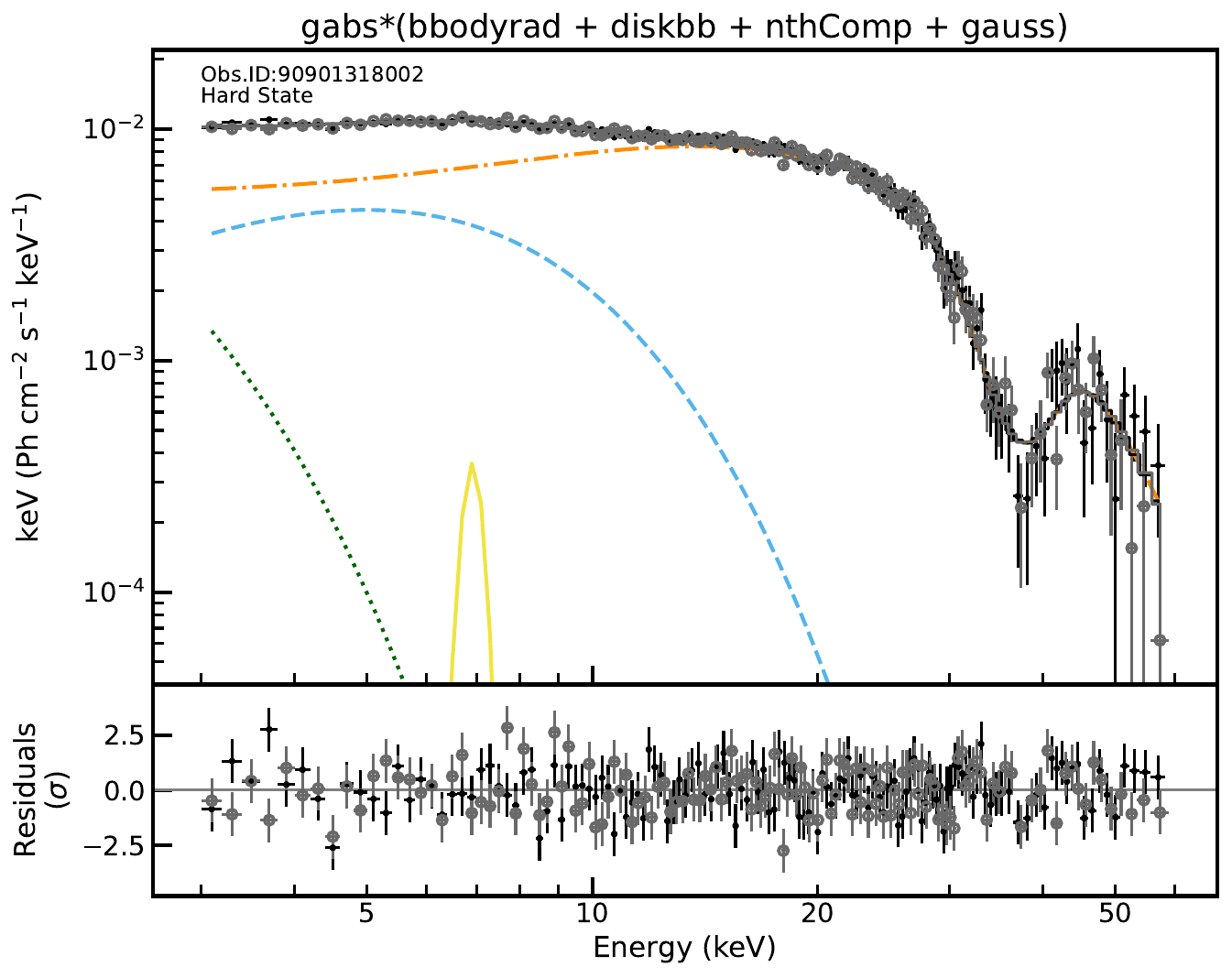}
    \includegraphics[width=0.46\textwidth]{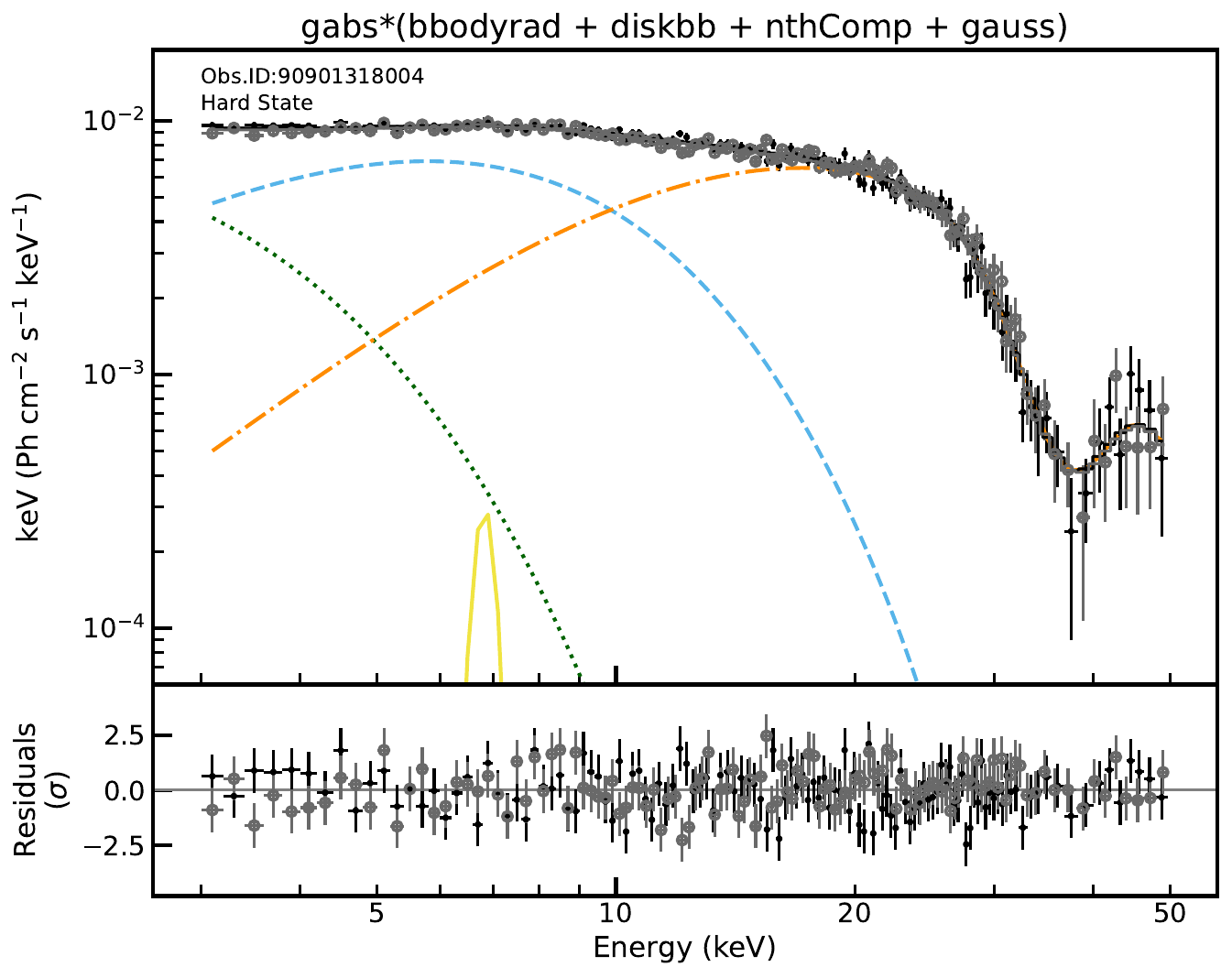}
    \includegraphics[width=0.46\textwidth]{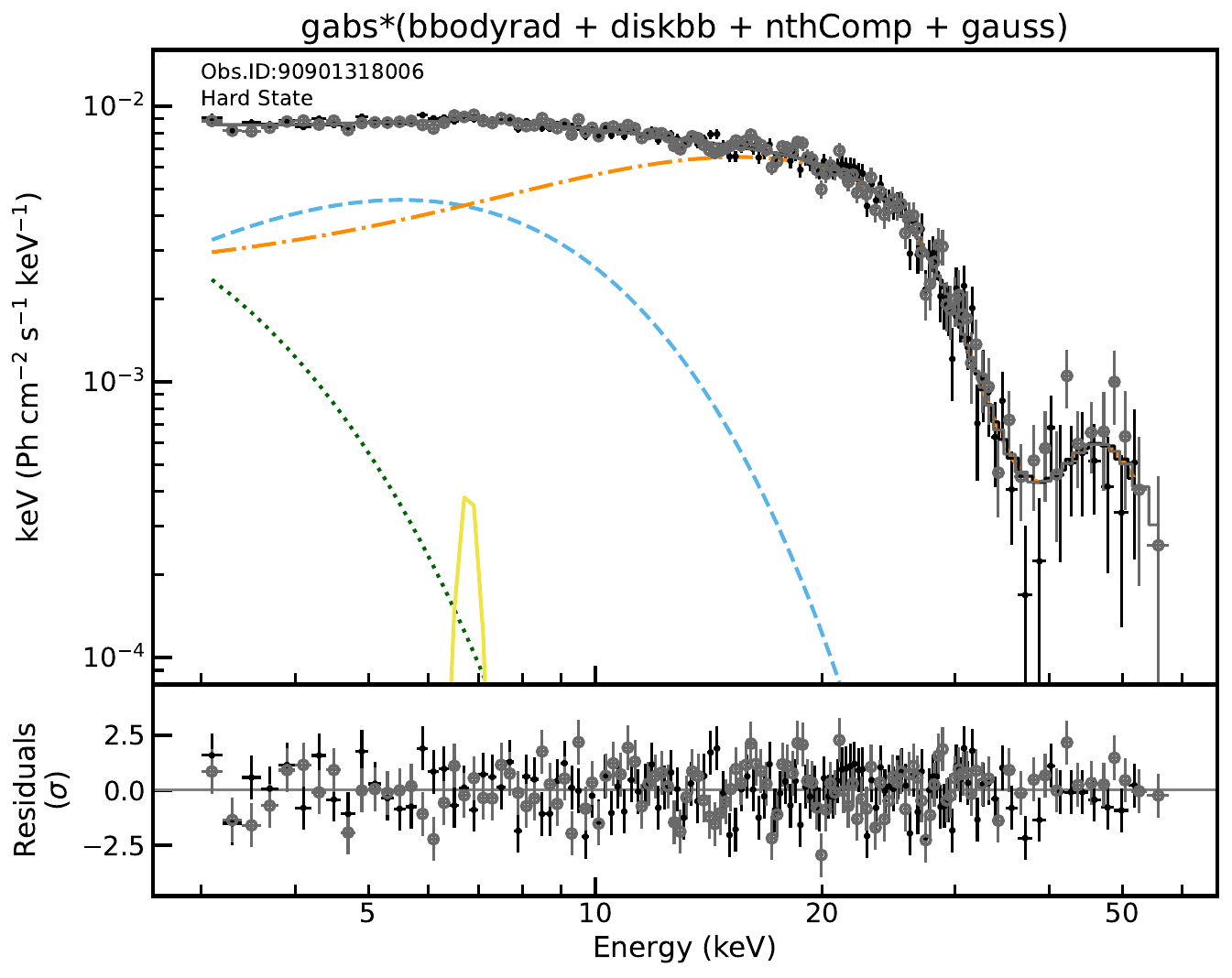}
    \includegraphics[width=0.46\textwidth]{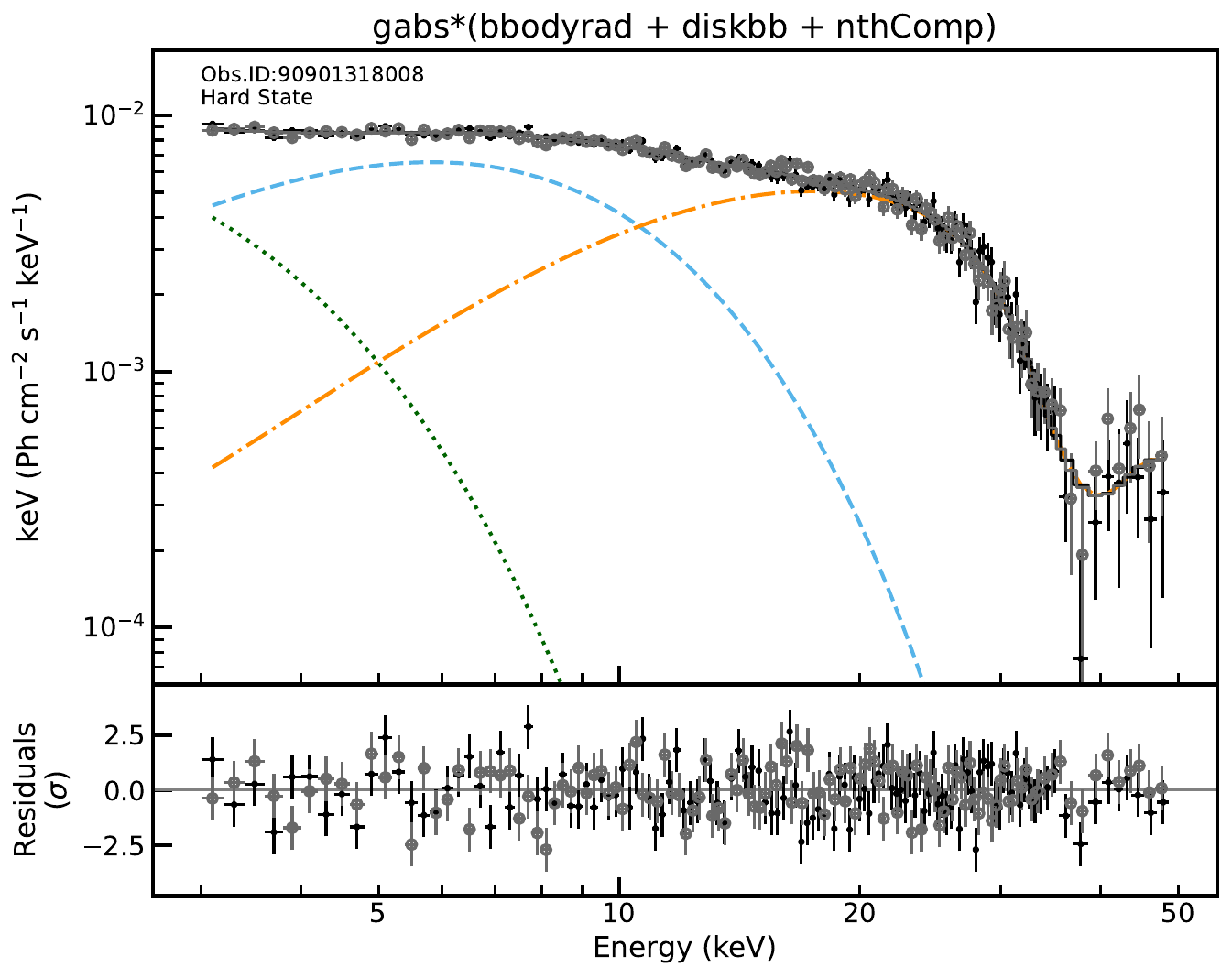}
    \includegraphics[width=0.46\textwidth]{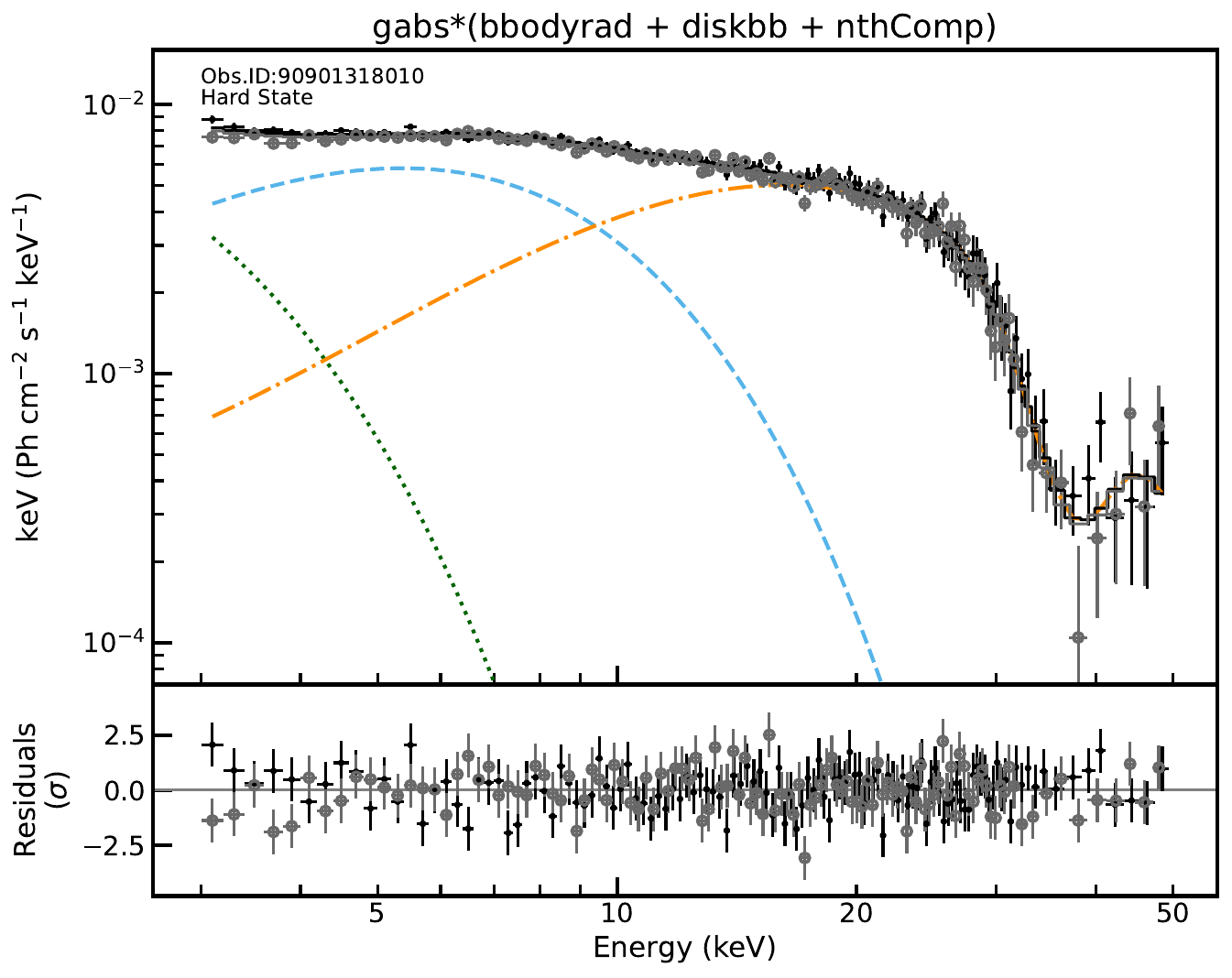}
    \caption{Spectra of 4U\,1626$-$67.}
    \label{fig:spec_para_4u1626}
\end{figure}

\begin{figure}[th!]
    \centering
    \includegraphics[width=0.46\textwidth]{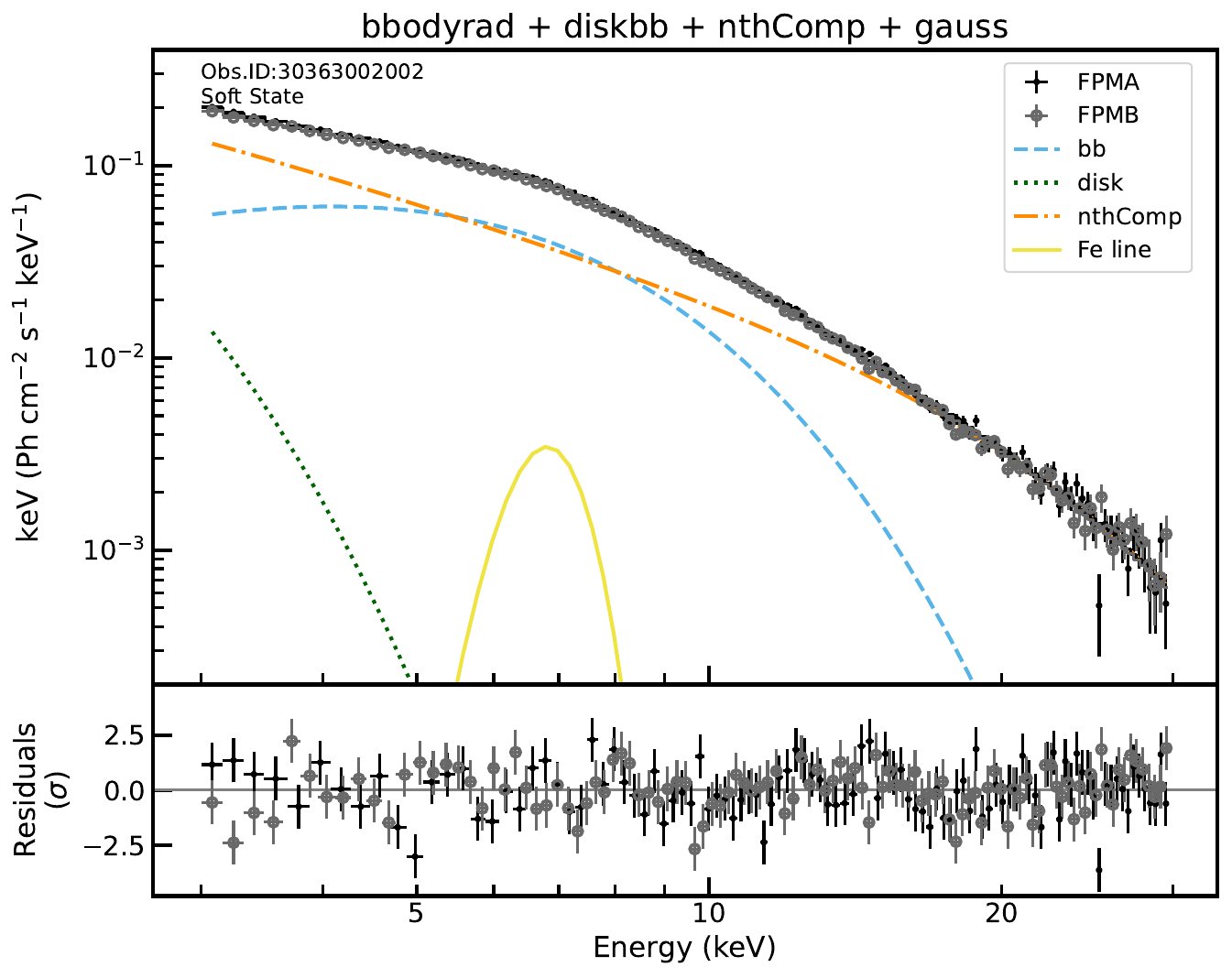}
    \includegraphics[width=0.46\textwidth]{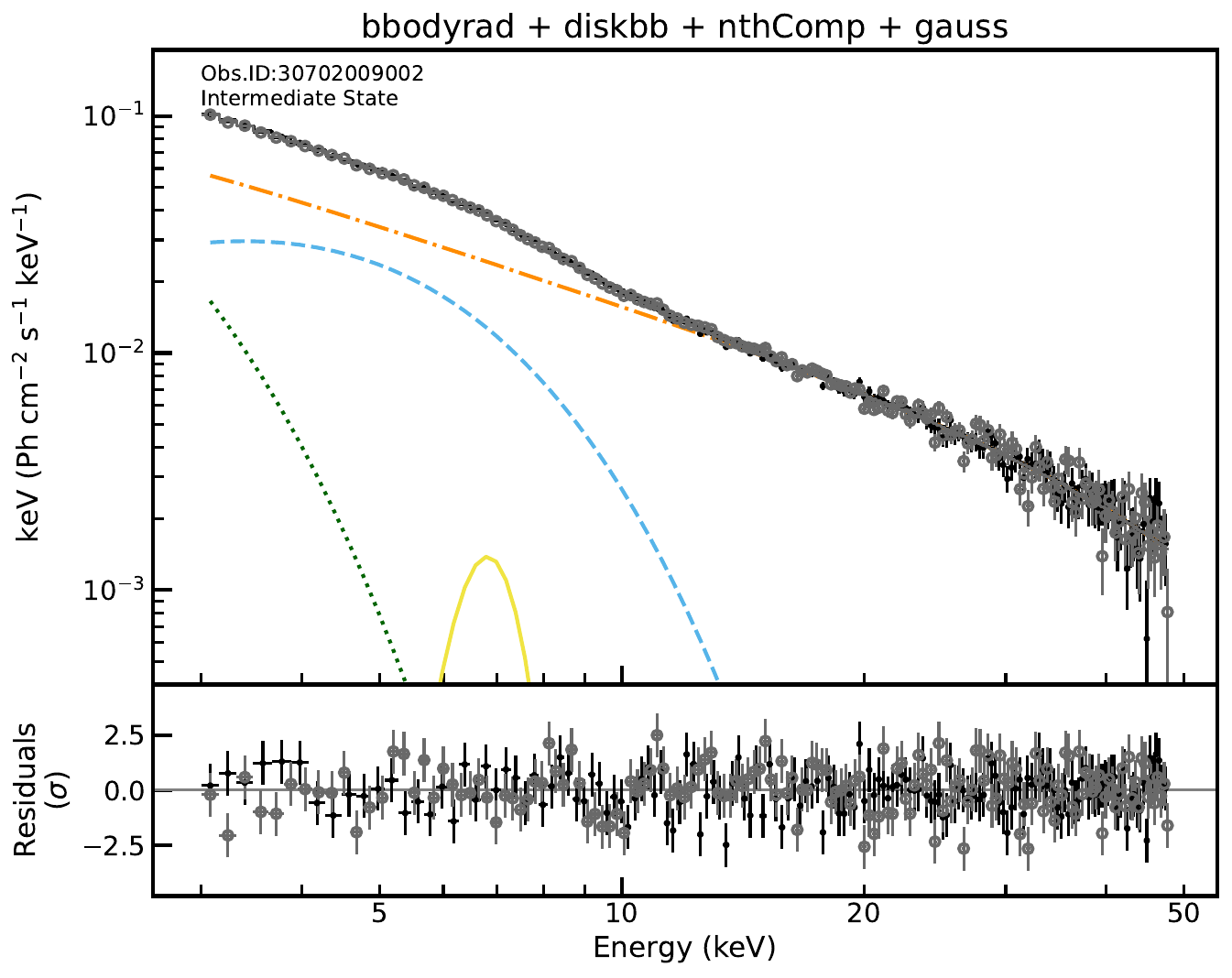}
    \includegraphics[width=0.46\textwidth]{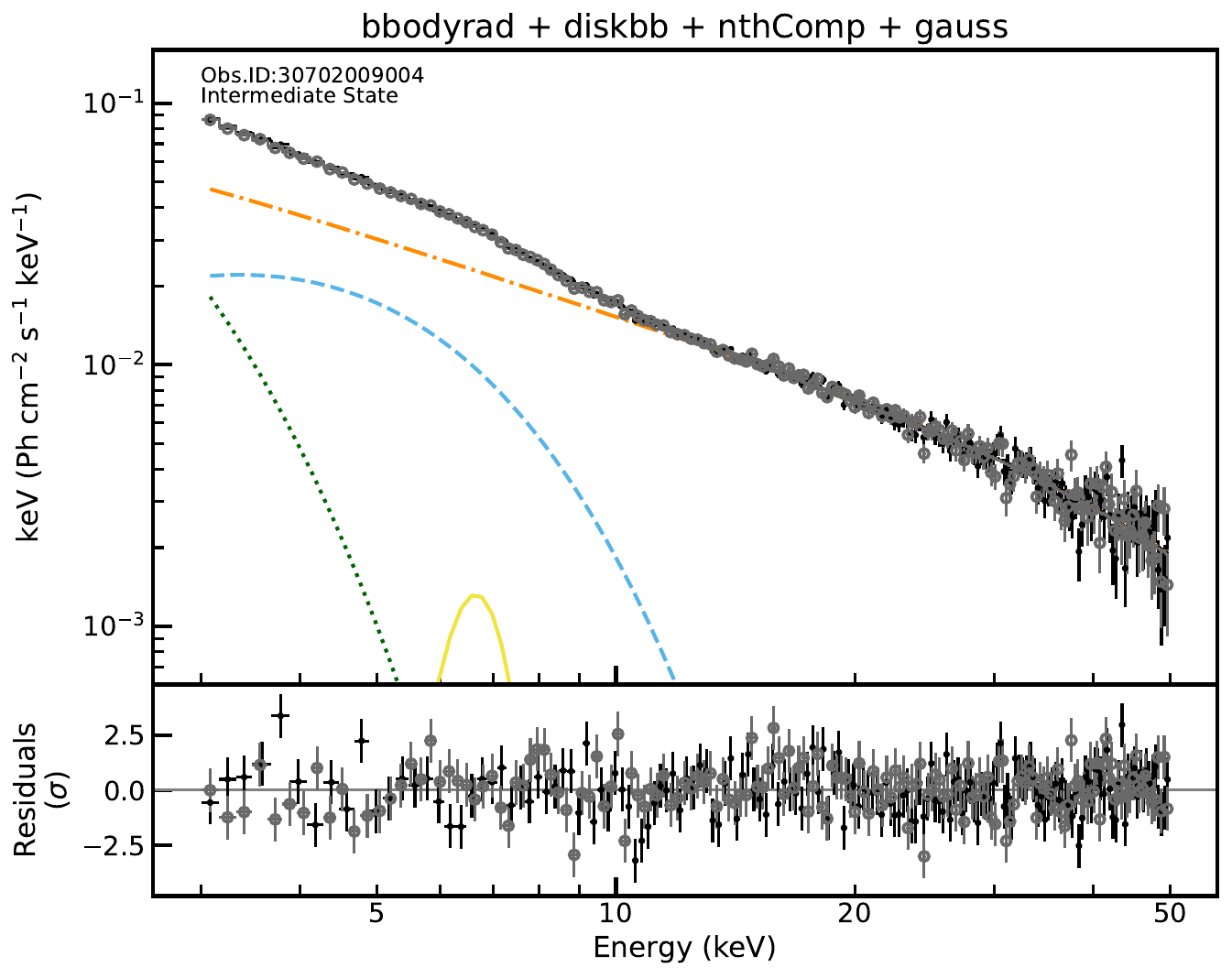}
    \includegraphics[width=0.46\textwidth]{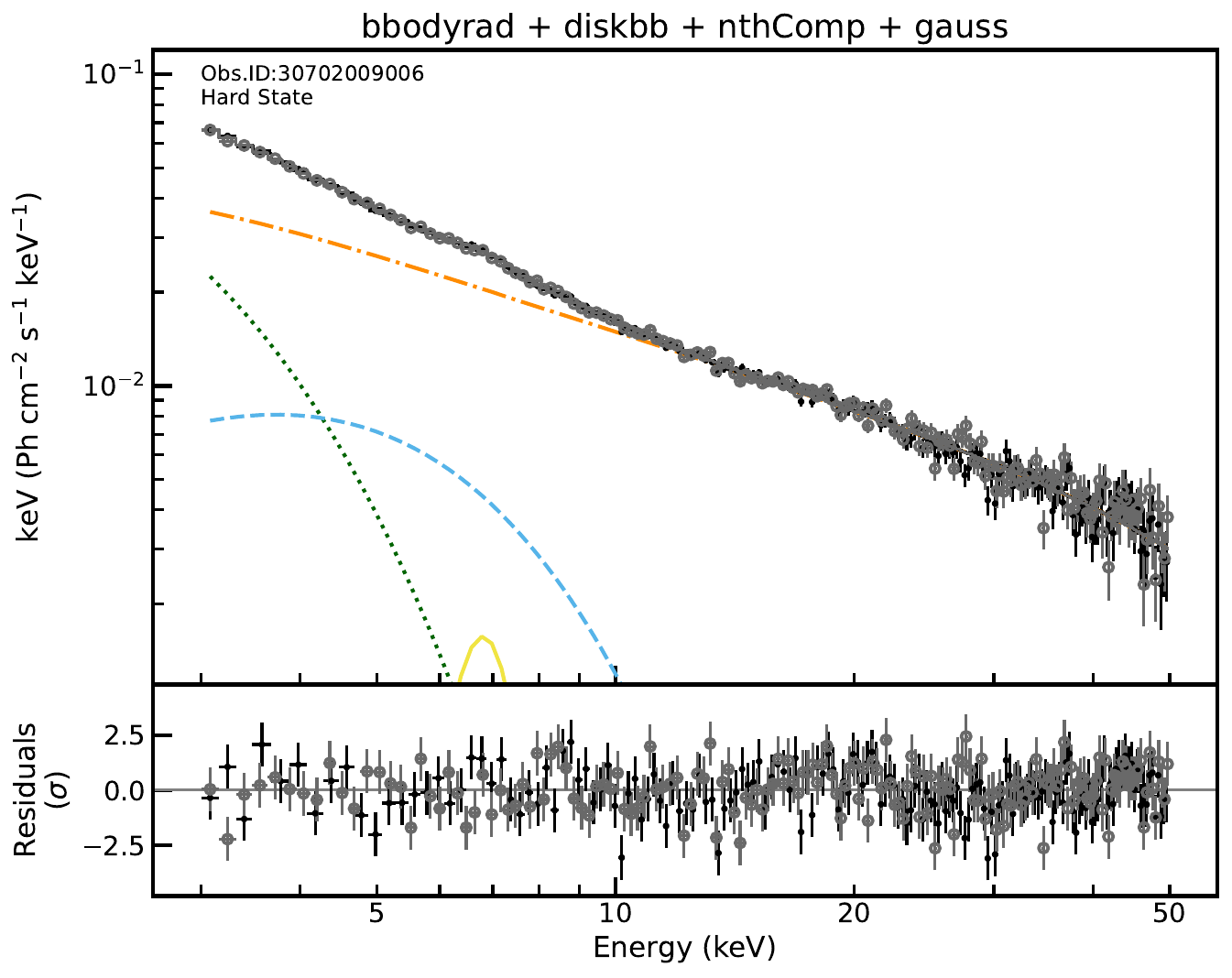}
    \includegraphics[width=0.46\textwidth]{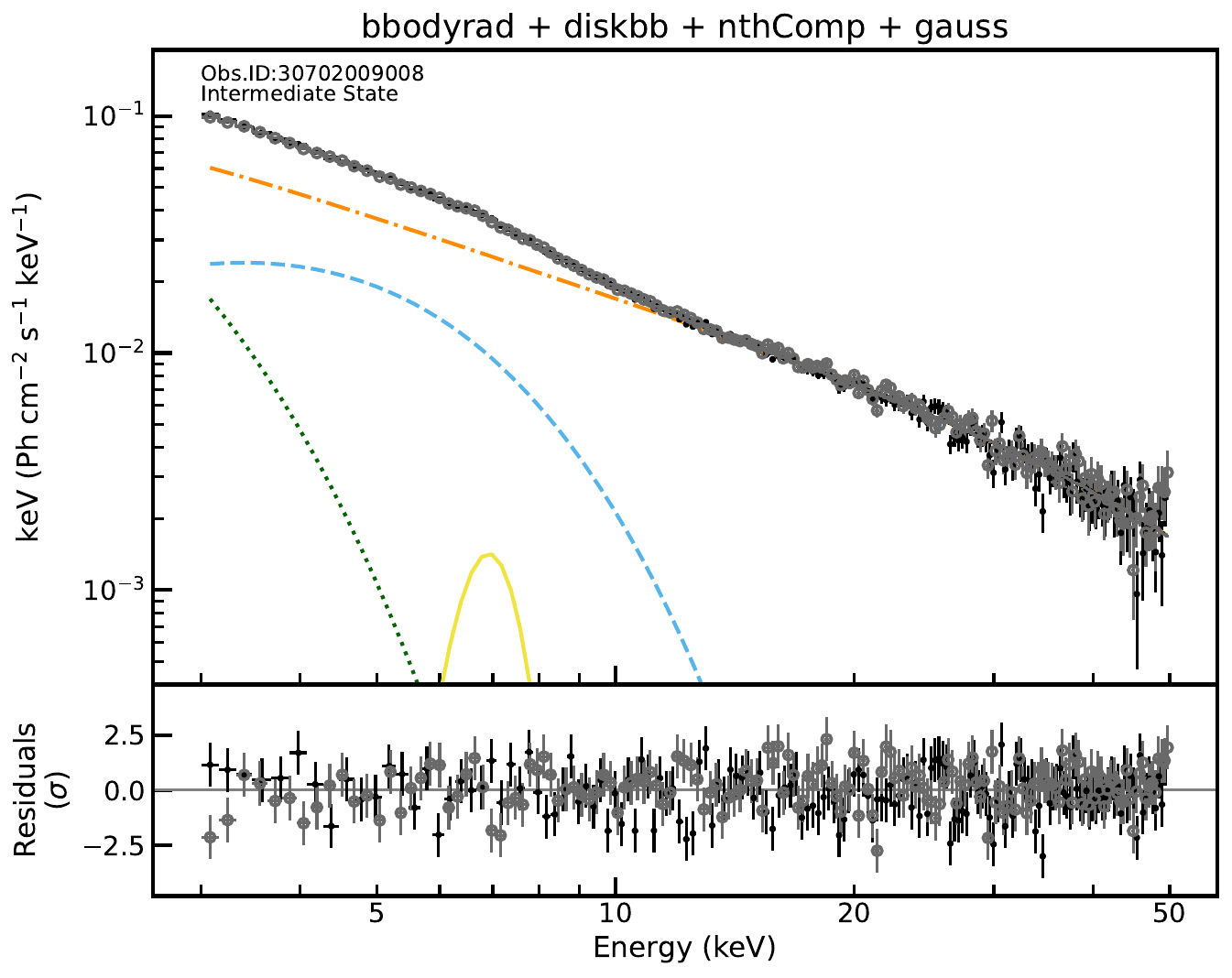}
    \includegraphics[width=0.46\textwidth]{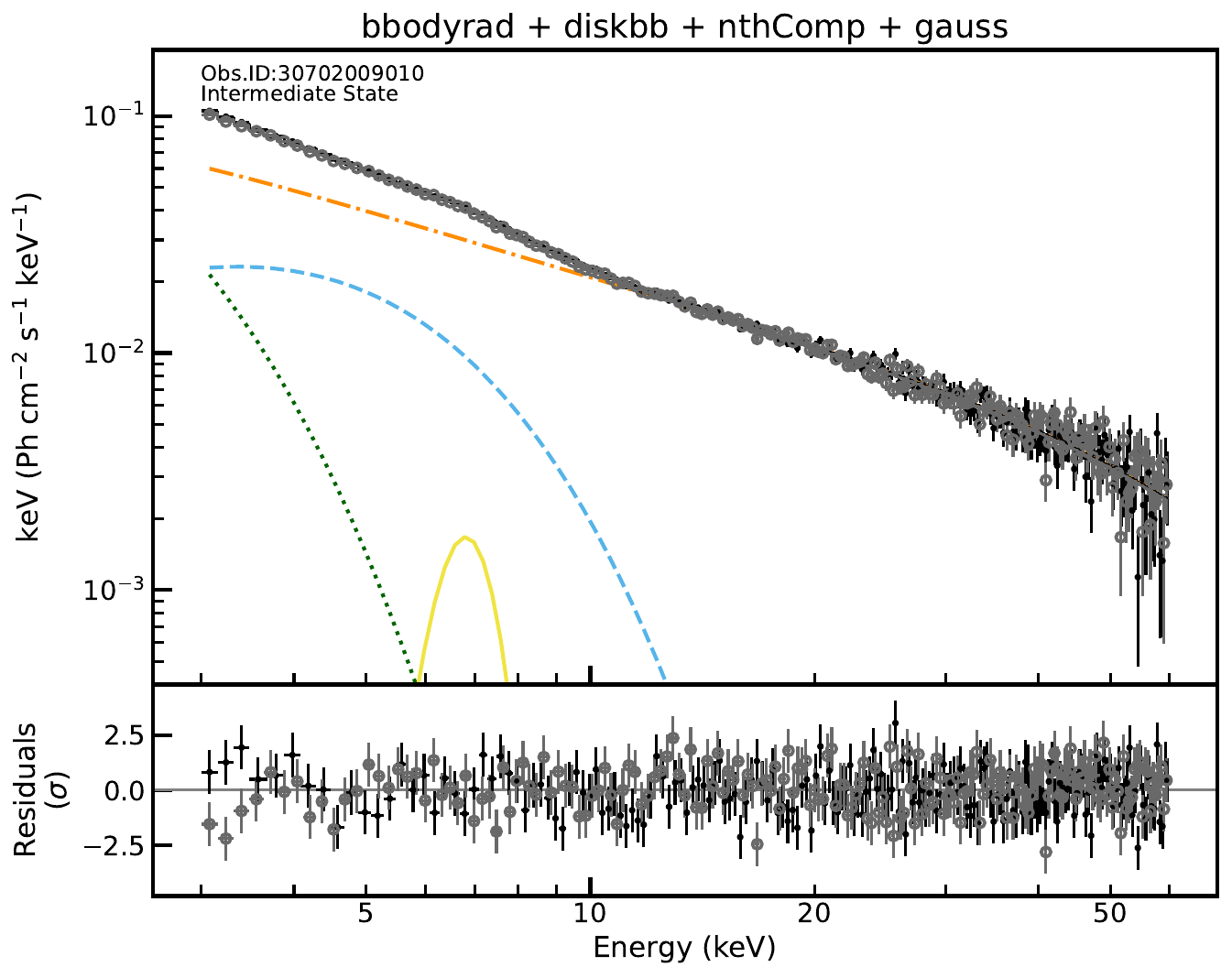}
    \caption{Spectra of 4U\,0614+091.}
    \label{fig:spec_para_4u0614}
\end{figure}

\end{appendix}

\end{document}